\documentclass[journal,twoside,web]{ieeecolor}

\usepackage{generic}
\usepackage{mathtools}
\mathtoolsset{showonlyrefs}

\usepackage{cite}
\usepackage{amsmath,amssymb,amsfonts}
\usepackage[ruled, vlined, linesnumbered]{algorithm2e}
\usepackage{graphicx}
\usepackage{textcomp}
\usepackage[T1]{fontenc}

\usepackage{euscript}
\usepackage{xcolor}
\usepackage{epsfig}
\usepackage{listings}
\usepackage{longtable}
\usepackage{multirow}
\usepackage{epstopdf}
\usepackage{bm}
\usepackage{makecell}
\usepackage{subcaption}
\usepackage{matlab-prettifier}
\usepackage{mathrsfs}

\newtheorem{theorem}{\bf \text{Theorem}}

\newtheorem{assumption}{\bf \text{Assumption}}

\newtheorem{corollary}{\bf \text{Corollary}}[theorem]
\newtheorem{lemma}{\bf \text{Lemma}}
\newtheorem{remark}{\bf \text{Remark}}

\newtheorem{example}{\bf \text{Example}}

\newcommand{\E}{{\mathbb E}}

\newcommand{\R}{{\mathbb R}}

\makeatletter

\newcommand{\Rmnum}[1]{\expandafter \@slowromancap\romannumeral #1@}
\makeatother

\DeclareMathOperator*{\rank}{rank}
\DeclareMathOperator*{\diag}{diag}

\DeclareMathOperator*{\Tr}{Tr}

\DeclareMathOperator*{\EB}{EB}

\DeclareMathOperator*{\tb}{b}

\DeclareMathOperator*{\XMSE}{XMSE}
\DeclareMathOperator*{\ML}{ML}

\DeclareMathOperator*{\MSE}{MSE}

\DeclareMathOperator*{\XBias}{XBias}
\DeclareMathOperator*{\XVar}{XVar}
\DeclareMathOperator*{\XVarHPE}{XVarHPE}

\DeclareMathOperator*{\Bayes}{Bayes}

\DeclareMathOperator*{\FIT}{FIT}

\DeclareMathOperator*\argmin{arg\,min}
\DeclareMathOperator*\argmax{arg\,max}

\def\pf{\textit{Proof. }}
\allowdisplaybreaks[4]

\begin{document}
\title{Bayes Estimators with Performance Comparable to Empirical Bayes Estimators and\\ Improved Local Robustness}
\author{Yue Ju, Ying Wang, Jiabao He, \IEEEmembership{Graduate Student Member, IEEE}, Bo Wahlberg, \IEEEmembership{Life Fellow, IEEE}, and H\r{a}kan Hjalmarsson, \IEEEmembership{Fellow, IEEE}
\thanks{This work was supported by VINNOVA Competence Center for Advanced BioProduction by Continuous Processing (AdBIOPRO) under contract [2016-05181], and by the Swedish Research Council under contract [2019-04956] and contract [2025-04833].}
\thanks{Yue Ju, Ying Wang, Jiabao He, Bo Wahlberg, and H\r{a}kan Hjalmarsson are currently with Department of Decision and Control Systems, School of Electrical Engineering and Computer Science, KTH Royal Institute of Technology, 10044 Stockholm, Sweden (e-mails: yuej@kth.se, yinwang@kth.se, jiabaoh@kth.se, bo@kth.se, hjalmars@kth.se). H\r{a}kan Hjalmarsson is also with the Competence Centre for Advanced BioProduction by Continuous Processing, AdBIOPRO.}}

\maketitle

\begin{abstract} 
Bayes estimation has been extensively studied and widely used in statistics, decision theory, signal processing, machine learning, and system identification. Among its variants, empirical Bayes (EB) estimation has attracted considerable attention due to its favorable estimation performance and computational tractability. However, the direct plug-in dependence of an EB estimator on hyper-parameters can make it locally sensitive to hyper-parameter perturbations. This paper considers the linear regression model and focuses on the EB estimator by employing the marginal maximum likelihood hyper-parameter estimator. For conciseness, this estimator is simply referred to as the EB estimator. Given a family of EB weighting functions, a generalized Bayes estimator is constructed with the same excess mean squared error (XMSE) as the corresponding EB estimator. Here, the XMSE is a second-order asymptotic measure of the mean squared error difference between the estimator of interest and the maximum likelihood estimator. Furthermore, the EB estimator is shown to be at most first-order sensitive to hyper-parameter perturbations, whereas the constructed Bayes estimator is at most second-order sensitive, making it locally more robust. The computational complexities of these two estimators are also analyzed. In some cases, the constructed Bayes estimator can be computationally comparable to, or more efficient than, the EB estimator. These theoretical results are further supported by numerical simulations.
\end{abstract}



\section{Introduction}

\IEEEPARstart{B}{ayes} estimation has its roots in Bayes' theorem and the early inverse-probability ideas \cite{Bayes1763, Laplace1820}, and was later developed into a general framework in statistics and decision theory \cite{Wald1950, Lehmann:06}. Throughout the past several decades, Bayes estimation methods have been extensively studied and widely used in signal processing, machine learning, and system identification, see, e.g., \cite{Lehmann:06, RasmussenW:06, KRE2022, PCCDL2022}. 

In parametric estimation problems, a standard Bayes estimator minimizes the weighted average of a given loss function. These weights are proportional to the product of likelihood and a fixed weighting function (prior) that incorporates available prior information about the parameters to be estimated. The weighting function often induces shrinkage of the maximum likelihood (ML) estimator, and Bayes estimation is therefore closely connected with shrinkage theory \cite{Strawderman1971, CM2012}. The (generalized) Bayes estimators often exhibit favorable admissibility, minimaxity, or domination properties, see, e.g., \cite{Lehmann:06}.

A fundamental challenge in Bayes estimation is how to choose an appropriate weighting function. Several approaches have been developed to address this challenge. One approach adopts a noninformative weighting function when prior information is limited or when one wants to mitigate the influence of subjective weighting specification, e.g., Jeffreys weighting function \cite{Jeffreys1946, SLR25}. Another class of approaches first selects a family of weighting functions indexed by hyper-parameters and then tunes the hyper-parameters using available data. For instance, the EB approach estimates the hyper-parameters from observed data, commonly by marginal likelihood maximization (a standard EB method) or other data-driven estimation methods \cite{M1983, RasmussenW:06}. The hierarchical Bayes approach assigns another weighting function to the hyper-parameters \cite{LS72, PL2023}, which is often called a hyper-weighting function (hyper-prior). The robust Bayes approach first specifies a class of plausible weighting functions and then evaluates the estimation accuracy over the entire class \cite{BB1986, Lehmann:06}.

This paper focuses on the EB estimator using the marginal maximum likelihood hyper-parameter estimator, which is simply referred to as the EB estimator. This estimator has been used in many fields, see, e.g., \cite{WR2004, RasmussenW:06, MCL18asy}, because it often provides a favorable balance between estimation accuracy and computational cost. In particular, given a family of Gaussian weighting functions with parametrized covariance matrices, often known as kernel matrices, the resulting EB estimator reduces to the kernel-based regularized estimator. It has gained considerable attention and been extensively studied in system identification, see, e.g., \cite{PD2010, COL12a, CL:12, PCCDL2022, MCL18asy, JCML21eb, JCCWH25}.  

However, due to the direct dependence on hyper-parameters, the EB estimator can be sensitive to hyper-parameter perturbations. This means that small hyper-parameter estimation errors can result in noticeable performance degradation. To handle this issue, we exploit the XMSE introduced in \cite{JCWH26}, which is a second-order asymptotic measure of the difference in mean squared error (MSE) between the estimator of interest and the ML estimator, to construct a generalized Bayes estimator. Given a family of EB weighting functions, the constructed generalized Bayes estimator has the same XMSE as the EB estimator. Accordingly, these two estimators perform comparably for large sample sizes, while the designed Bayes estimator is more robust to hyper-parameter perturbations. 

This paper \emph{extends} \cite{JWH25csl, JCCWH25}, which constructed generalized Bayes estimators with the same XMSE as the EB estimator for the Gaussian EB weighting families. Matching their XMSEs leads to a partial differential equation (PDE) for the Bayes weighting function, which generally does not admit an explicit solution and may require numerical methods. In particular, closed-form Bayes weighting functions were derived in \cite{JWH25csl} under the identity Gram matrix limit and the scaled identity kernel matrix, and in \cite{JCCWH25} under a general Gram matrix limit and a more general scaled kernel matrix. These closed-form results both rely on the parametrizations that reduce the PDE to a scalar ordinary differential equation (ODE). Such reduction is not possible for the setting in this paper: a \emph{general} Gram matrix limit and a \emph{general} EB weighting family. Rather than reducing the resulting PDE to a scalar ODE, as done in \cite{JWH25csl, JCCWH25}, or solving it numerically, this paper provides a characterization of the Bayes weighting function by matching the XMSE components. It further analyzes the local sensitivity to hyper-parameter perturbations and the computational complexity of the resulting Bayes estimators.




The main contributions of this paper are as follows:
\begin{enumerate}
\item[1)] An explicit expression for the XMSE of the EB estimator is derived by specializing \cite[Theorem~2]{JCWH26} to the case of EB hyper-parameter estimation.
\item[2)] Given a family of EB weighting functions, a weighting function is characterized through a pointwise optimization problem such that the resulting generalized Bayes estimator has the same XMSE as the corresponding EB estimator. This weighting function can be expressed as the upper envelope of the given EB weighting functions. The characterization is further specialized to Gaussian EB weighting functions using scale-shape and diagonal parametrizations of the kernel matrix.
\item[3)] The EB estimator is shown to exhibit at most first-order sensitivity to hyper-parameter perturbations, whereas the constructed generalized Bayes estimator is second-order sensitive and is therefore locally more robust.
\item[4)] The computational complexities of the EB estimator and the constructed Bayes estimator are analyzed. Sufficient conditions are established under which the constructed estimator can be computationally comparable to, or more efficient than, the EB estimator. The theoretical findings are further supported through numerical simulations.
\end{enumerate}

The remainder of this paper is organized as follows. Section~\ref{sec: preliminaries and problem statement} introduces preliminaries and formulates the problem. Section~\ref{sec: design of weighting} derives the XMSE of the EB estimator and constructs a generalized Bayes estimator with the same XMSE as the EB estimator. Section~\ref{sec: local sensitivity} compares its local sensitivity to hyper-parameter perturbations with that of the EB estimator, and Section~\ref{sec: implementation and complexity} analyzes their computational complexities. Section~\ref{sec: numerical simulation} presents numerical simulations. Section~\ref{sec: conclusion} concludes this paper. The proofs of theorems are included in Appendix.






\textbf{Notation.} The set of $m_{1}\times m_{2}$ real matrices is denoted ${\R}^{m_{1}\times m_{2}}$. The $m\times m$ identity matrix is denoted by $\mathbf{I}_{m}$. The $(k,l)$th entry, transpose, and inverse of a matrix $\mathbf{A}$ are denoted $[\mathbf{A}]_{k,l}$, $\mathbf{A}^{\top}$, and $\mathbf{A}^{-1}$, respectively. The partial derivative of $f(\bm{x}):\R^{m}\to\R$ with respect to (w.r.t.) $[\bm{x}]_{k}$ is denoted ${\partial f(\bm{x})}/{\partial [\bm{x}]_{k}}$. Its gradient and Hessian w.r.t. $\bm{x}$ are denoted $\nabla_{x}f(\bm{x})\coloneqq \left[{\partial f(\bm{x})}/{\partial [\bm{x}]_{1}},\ \cdots,\ {\partial f(\bm{x})}/{\partial [\bm{x}]_{m}}\right]^{\top}\in\R^{m}$ and $\nabla^2_{xx}f(\bm{x})\in\R^{m\times m}$ with $[\nabla^2_{xx}f(\bm{x})]_{k,l}={\partial^2 f(\bm{x})}/{\partial [\bm{x}]_{k}\partial [\bm{x}]_{l}}$, respectively. A positive definite (negative definite) matrix is denoted $\mathbf{A}\succ 0$ ($\mathbf{A}\prec 0$). The mathematical expectation of a random vector $\bm{a}$ is denoted $\E(\bm{a})$. A Gaussian distributed random vector $\bm\xi\in\R^{m}$ with mean $\bm{a}\in\R^{m}$ and covariance matrix $\mathbf{A}\in\R^{m\times m}$ is denoted $\mathcal{N}(\bm{a},\mathbf{A})$. The rank of a matrix is denoted $\rank(\cdot)$. The Euclidean norm of a vector is denoted $\|\cdot\|_{2}$. The Frobenius norm of a matrix is denoted $\|\cdot\|_{F}$. The trace and determinant of a square matrix are denoted $\Tr(\cdot)$ and $\det(\cdot)$, respectively. The indicator function of an event $\mathcal{A}$ is denoted $\bm{1}_{\{\mathcal{A}\}}$, which equals one when $\mathcal{A}$ occurs and zero otherwise. For $\bm{a}\in\R^{m}$, $\diag\{\bm{a}\}$ denotes the $m\times m$ diagonal matrix whose diagonal entries are components of $\bm{a}$. 


\vspace{-1mm}

\section{Preliminaries and Problem Statement}\label{sec: preliminaries and problem statement}

Consider the linear regression model, given by
\begin{align}\label{eq: linear regresssion model}
\bm{Y}=\mathbf{\Phi}\bm\theta+\bm{E},
\end{align}
where $\bm{Y}\in\R^{N}$, $\mathbf{\Phi}\in\R^{N\times n_{\theta}}$, and $\bm{E}\in\R^{N}$ represent the measurement output vector, the regression matrix, and the measurement noise vector, respectively, and $\bm{\theta}\in\R^{n_{\theta}}$ denotes the unknown parameter vector to be estimated. The objective is to construct estimators of $\bm\theta$ from $\mathbf{\Phi}$ and $\bm{Y}$. Given an estimator $\hat{\bm\theta}_{N}\in\R^{n_{\theta}}$ of $\bm\theta$, a commonly used criterion for evaluating its performance is MSE, defined as
\begin{align}\label{eq: def of MSE}
\MSE(\hat{\bm\theta}_{N})=\E(\|\hat{\bm\theta}_{N}-\bm\theta_{0}\|_{2}^2),
\end{align}
where the expectation $\E(\cdot)$ is w.r.t. the noise $\bm{E}$, and $\bm\theta_{0}\in\R^{n_{\theta}}$ is the true value of $\bm\theta$. A smaller $\MSE(\hat{\bm\theta}_{N})$ indicates better estimation performance. Before presenting the estimators studied in this paper, we introduce the following assumption.

\begin{assumption}\label{asp:input and noise}
\begin{enumerate}
\item[]
\item[1)] The model order $n_{\theta}$ is fixed, and larger than or equal to the true order.
\item[2)] The regression matrix $\mathbf{\Phi}$ is known and deterministic, and satisfies $\rank(\mathbf{\Phi})=n_{\theta}$ for $N\geq n_{\theta}$.
\item[3)] The measurement noise $\bm{E}$ is Gaussian distributed, i.e., $\bm{E}\sim\mathcal{N}(\bm{0},\sigma^2\mathbf{I}_{N})$, where $\sigma^2>0$ is known. 
\end{enumerate}
\end{assumption}

\vspace{-2mm}

\subsection{Bayes and empirical Bayes estimators}\label{subsec: Bayes estimators}

Under the model in \eqref{eq: linear regresssion model} and Assumption~\ref{asp:input and noise}, we first recall the definitions of several classical estimators. Given a weighting function $\pi(\bm\theta)\geq 0$ and under squared error loss, the generalized Bayes estimator is defined as
\begin{equation}\label{eq: def of Bayes estimator}
\begin{aligned}
\hat{\bm\theta}^{\Bayes}=&\argmin_{\hat{\bm\theta}_{N}\in\R^{n_{\theta}}}{\textstyle\int} \|\hat{\bm\theta}_{N}-\bm\theta\|_{2}^2\tfrac{p(\bm{Y}|\bm\theta)\pi(\bm\theta)}{\int p(\bm{Y}|\bm\theta)\pi(\bm\theta)d\bm\theta}d\bm\theta\\
=&\frac{\int \bm\theta p(\bm{Y}|\bm\theta)\pi(\bm\theta)d\bm\theta}{\int p(\bm{Y}|\bm\theta)\pi(\bm\theta)d\bm\theta},
\end{aligned}
\end{equation}
provided the above integrals are well-defined, where $p(\bm{Y}|\bm\theta)$ denotes the density of $\bm{Y}|\bm\theta\sim\mathcal{N}(\mathbf{\Phi}\bm\theta,\sigma^2\mathbf{I}_{N})$, and the weighting function is allowed to be improper, i.e., $\int\pi(\bm\theta)d\bm\theta=\infty$. 

When the weighting function is flat, i.e., $\pi(\bm\theta)\propto 1$, the Bayes estimator in \eqref{eq: def of Bayes estimator} coincides with the ML estimator
\begin{align}\label{eq: def of ML estimator}
\hat{\bm\theta}^{\ML}=\argmax_{\bm\theta\in\R^{n_{\theta}}}p(\bm{Y}|\bm\theta)=(\mathbf{\Phi}^{\top}\mathbf{\Phi})^{-1}\mathbf{\Phi}^{\top}\bm{Y}.
\end{align}

When the weighting function is parameterized as $\pi(\bm\theta|\bm\eta)$, where $\bm\eta\in\mathcal{D}_{\eta}\subset\R^{n_{\eta}}$ denotes the hyper-parameter vector, an EB estimator can be obtained by substituting a data-dependent hyper-parameter estimate $\hat{\bm\eta}_{N}\in\mathcal{D}_{\eta}$ into the generalized Bayes estimator in \eqref{eq: def of Bayes estimator}. It takes the form
\begin{align}\label{eq: def of EB estimator}
\hat{\bm\theta}^{\EB}(\hat{\bm\eta}_{N})=\frac{\int \bm\theta p(\bm{Y}|\bm\theta)\pi(\bm\theta|\hat{\bm\eta}_{N})d\bm\theta}{\int p(\bm{Y}|\bm\theta)\pi(\bm\theta|\hat{\bm\eta}_{N})d\bm\theta}.
\end{align}
In this paper, we focus on the EB hyper-parameter estimator obtained by maximizing the marginal likelihood, or equivalently, by minimizing the negative log marginal likelihood,
\begin{equation}\label{eq: def of EB hyperparameter estimator}
\begin{aligned}
\hat{\bm\eta}_{\EB}
=&\argmin_{\bm\eta\in\mathcal{D}_{\eta}}\mathscr{F}_{\EB}(\bm\eta),\\
\mathscr{F}_{\EB}(\bm\eta)=&-\log\left[\textstyle{\int} p(\bm{Y}|\bm\theta)\pi(\bm\theta|\bm\eta)d\bm\theta \right].
\end{aligned}
\end{equation} 
Substituting $\hat{\bm\eta}_{N}=\hat{\bm\eta}_{\EB}$ into \eqref{eq: def of EB estimator} yields $\hat{\bm\theta}^{\EB}(\hat{\bm\eta}_{\EB})$, which is simply referred to as the EB estimator throughout this paper. Below, we show that $\hat{\bm\theta}^{\EB}(\hat{\bm\eta}_{\EB})$ depends on the observed data solely through $\hat{\bm\theta}^{\ML}$, which is a minimal sufficient statistic\footnote{A statistic $\bm{T}(\bm{Y})$ is said to be \emph{sufficient} for $\bm\theta$ if the conditional distribution of $\bm{Y}$ given $\bm{T}(\bm{Y})$ does not depend on $\bm\theta$. A sufficient statistic (SS) $\bm{S}(\bm{Y})$ is said to be \emph{minimal} if among all SSs it provides the greatest possible reduction of the data, which means that for any SS $\bm{T}(\bm{Y})$, there exists a function $h(\cdot)$ such that $\bm{S}(\bm{Y})={h}(\bm{T}(\bm{Y}))$ almost surely \cite[Chapter~1.6]{Lehmann:06}.} for $\bm\theta$ under the linear regression model in \eqref{eq: linear regresssion model}.

\begin{lemma}\label{lemma: expression of EB estimator using MSS}
Under Assumption~\ref{asp:input and noise}, we have
\begin{align}\label{eq: rewritten expression for EB estimator}
\hat{\bm\theta}^{\EB}(\hat{\bm\eta}_{N})=&\frac{\int \bm\theta p(\bm\theta|\hat{\bm\theta}^{\ML})\pi(\bm\theta|\hat{\bm\eta}_{N})d\bm\theta}{\int p(\bm\theta|\hat{\bm\theta}^{\ML})\pi(\bm\theta|\hat{\bm\eta}_{N})d\bm\theta},\\
\label{eq: rewritten expression for F_EB}
\mathscr{F}_{\EB}(\bm\eta)=&-\log[C_{N}(\hat{\bm\theta}^{\ML})]-\log\{\E_{\bm\theta|\hat{\bm\theta}^{\ML}}[\pi(\bm\theta|\bm\eta)]\},
\end{align}
where $C_{N}(\hat{\bm\theta}^{\ML})$ denotes a function of $\hat{\bm\theta}^{\ML}$, and $p(\bm\theta|\hat{\bm\theta}^{\ML})$ is the density of $\bm\theta|\hat{\bm\theta}^{\ML}\sim\mathcal{N}(\hat{\bm\theta}^{\ML},\sigma^2(\mathbf{\Phi}^{\top}\mathbf{\Phi})^{-1})$.
\end{lemma}

In particular, for the Gaussian EB weighting families, i.e., 
\begin{align}\label{eq: Gaussian EB weighting families}
\pi(\bm\theta|\bm\eta)\ \text{is}\ \text{the}\ \text{density}\ \text{of}\ \bm\theta|\bm\eta\sim\mathcal{N}(\bm{0},\mathbf{P}(\bm\eta)),
\end{align}
where $\mathbf{P}(\bm\eta)\succ 0$, both the corresponding EB estimator and EB hyper-parameter cost function admit closed-form expressions. From \eqref{eq: def of EB estimator}--\eqref{eq: def of EB hyperparameter estimator} and \cite[Lemma~2]{JCWH26}, we have
\begin{align}\label{eq: expression for regularized estimator}
\hat{\bm\theta}^{\EB}(\hat{\bm\eta}_{N})=&\left[\mathbf{\Phi}^{\top}\mathbf{\Phi}+\sigma^2\mathbf{P}(\hat{\bm\eta}_{N})^{-1}\right]^{-1}\mathbf{\Phi}^{\top}\bm{Y},\\
=&\hat{\bm\theta}^{\ML}-\sigma^2(\mathbf{\Phi}^{\top}\mathbf{\Phi})^{-1}\mathbf{S}(\hat{\bm\eta}_{N})^{-1}\hat{\bm\theta}^{\ML},\\
\label{eq: expression for F_EB}
\mathscr{F}_{\EB}(\bm\eta)=&\tfrac{1}{2}\left[\bm{Y}^{\top}\mathbf{Q}(\bm\eta)^{-1}\bm{Y}+\log\det(\mathbf{Q}(\bm\eta))+N\log(2\pi)\right],\nonumber\\
=&\tfrac{1}{2}\left[(\hat{\bm\theta}^{\ML})^{\top}\mathbf{S}(\bm\eta)^{-1}\hat{\bm\theta}^{\ML}+\log\det(\mathbf{S}(\bm\eta))\right]+\tilde{C}_{N},
\end{align}
where $\mathbf{S}(\bm\eta)=\mathbf{P}(\bm\eta)+\sigma^2(\mathbf{\Phi}^{\top}\mathbf{\Phi})^{-1}$, $\mathbf{Q}(\bm\eta)=\mathbf{\Phi}\mathbf{P}(\bm\eta)\mathbf{\Phi}^{\top}+\sigma^2\mathbf{I}_{N}$, and $\tilde{C}_{N}\in\R$ is independent of $\bm\eta$. This follows from the conjugacy between this Gaussian weighting function and the Gaussian likelihood $p(\bm{Y}|\bm\theta)$. Note that the covariance matrix $\mathbf{P}(\bm\eta)$ is also known as the kernel matrix, and $\hat{\bm\theta}^{\EB}(\hat{\bm\eta}_{N})$ in \eqref{eq: expression for regularized estimator} coincides with the regularized least-squares estimator using a quadratic penalty term, i.e., the minimizer of $\|\bm{Y}-\mathbf{\Phi}\bm\theta\|_{2}^2+\sigma^2\bm\theta^{\top}\mathbf{P}(\bm\eta)^{-1}\bm\theta$. Thus, $\hat{\bm\theta}^{\EB}(\hat{\bm\eta}_{N})$ is also called the kernel-based regularized estimator.

\subsection{Motivation and problem statement}\label{subsec: problem statement}

The EB method was introduced by Robbins~\cite{R1956} and has since been extensively studied and widely applied, see, e.g., \cite{M1983, PRS14, M2018, JCWH26}. It provides a computationally tractable scheme to tune the weighting function from observed data and often achieves favorable estimation accuracy. In particular, in the context of system identification, the EB estimator using the Gaussian weighting families in \eqref{eq: Gaussian EB weighting families}, i.e., $\hat{\bm\theta}^{\EB}(\hat{\bm\eta}_{\EB})$ defined in \eqref{eq: expression for regularized estimator}--\eqref{eq: expression for F_EB}, is appealing for estimating the impulse responses of stable systems, see, e.g., \cite{PD2010, PCCDL2022}. However, due to the direct plug-in dependence on $\hat{\bm\eta}_{\EB}$ in \eqref{eq: def of EB hyperparameter estimator}, the EB estimator $\hat{\bm\theta}^{\EB}(\hat{\bm\eta}_{\EB})$ may be sensitive to local hyper-parameter perturbations. For illustration, we present a numerical example.

\begin{example}\label{example: illustrating example}

We first generate $100$ systems using the MATLAB command \lstinline[style=Matlab-editor]{rss(30,1,1)} 
and the method in \cite[Section 2]{COL12a}, such that the largest pole modulus of each system is smaller than $0.95$. The impulse response of each system is truncated at order $n_{\theta}=20$ to obtain $\tilde{\bm\theta}_{0}\in\R^{20}$, and is then scaled as $\bm\theta_{0}=m_{\theta}\tilde{\bm\theta}_{0}$, where $m_{\theta}\in\R$ is chosen such that $\|\bm\theta_{0}\|_{2}=1$. We generate Gaussian-distributed inputs with zero mean and unit variance, and form $\tilde{\mathbf{\Phi}}\in\R^{N\times n_{\theta}}$ as a lower triangular Toeplitz matrix. We set $\sigma^2=1$ and $N=80$, and choose $m_{u}\in\R$ such that $\mathbf{\Phi}=m_{u}\tilde{\mathbf{\Phi}}$ and the sample signal-to-noise ratio\footnote{\label{footnote: sample snr}The sample SNR is defined as the ratio between the sample variance of the noiseless output $\mathbf{\Phi}\bm\theta_{0}$ and the measurement noise variance $\sigma^2$.} (SNR) is $10$. For each system, we conduct $100$ Monte Carlo (MC) simulations to evaluate the average performance\footnote{To measure the average performance of an estimator $\hat{\bm\theta}_{N}$, we first compute the sample MSE and average FIT \cite{JCWH26} of $\hat{\bm\theta}_{N}$ over all the MC simulations, and then compute their sample means over all the systems, where $\FIT(\hat{\bm\theta}_{N})=100\times(1-{\|\hat{\bm\theta}_{N}-\bm\theta_{0}\|_{2}}/{\|\bm\theta_{0}-\bar{\theta}_{0}\|_{2}})$ with $\bar{\theta}_{0}=\tfrac{1}{n_{\theta}}\sum_{k=1}^{n_{\theta}}[\bm\theta_{0}]_{k}$. The average FIT is a relative quantity based on the normalization of the sample MSE. They often exhibit similar trends, but generally do not have an exact one-to-one correspondence.} of $\hat{\bm\theta}^{\ML}$ in \eqref{eq: def of ML estimator} and $\hat{\bm\theta}^{\EB}(\hat{\bm\eta}_{\EB})$ in \eqref{eq: expression for regularized estimator}--\eqref{eq: expression for F_EB} using the tuned-correlated (TC) kernel \cite{PD2010, COL12a}, i.e., 
\begin{equation}\label{eq: def of TC kernel}
\begin{gathered}
[\mathbf{P}(\bm\eta)]_{k,l}=c\min(\alpha^{k},\alpha^{l}),\\
\bm\eta=[c,\alpha]^{\top},\ \mathcal{D}_{\eta}=\{\underline{c}\leq c\leq \overline{c},\ \underline{\alpha}\leq \alpha\leq \overline{\alpha}\},
\end{gathered}
\end{equation}
where $\underline{c}=\exp(-60)$, $\overline{c}=\exp(60)$, $\underline{\alpha}=10^{-4}$, and $\overline{\alpha}=1-10^{-4}$. For $\hat{\bm\theta}^{\ML}$ and $\hat{\bm\theta}^{\EB}(\hat{\bm\eta}_{\EB})$, the sample means of their sample $\MSE$s are $3.15\times 10^{-2}$ and $2.10\times 10^{-2}$, and the sample means of their average $\FIT$s are $71.24$ and $78.15$, respectively. It shows that $\hat{\bm\theta}^{\EB}(\hat{\bm\eta}_{\EB})$ performs better than $\hat{\bm\theta}^{\ML}$.

Let $\hat{\bm\eta}_{\EB}\coloneqq [\hat{c}_{\EB},\hat{\alpha}_{\EB}]^{\top}$. For each MC simulation, we keep $\hat{c}_{\EB}$ fixed and perturb the shape hyper-parameter estimate as $\tilde{\alpha}_{\EB}=\hat{\alpha}_{\EB}+\delta\alpha$ with $\delta\alpha=-0.06,-0.05,-0.04,\cdots,0.06$. Accordingly, the perturbed EB estimator is denoted $\hat{\bm\theta}^{\EB}(\tilde{\bm\eta}_{\EB})$, where $\tilde{\bm\eta}_{\EB}=[\hat{c}_{\EB},\tilde{\alpha}_{\EB}]^{\top}$. Fig.~\ref{fig: illustrating_example} shows how the shape hyper-parameter perturbation $\delta\alpha$ affects the average performance of $\hat{\bm\theta}^{\EB}(\tilde{\bm\eta}_{\EB})$, where sample $\delta\!\MSE$ is defined as 
\begin{align}
\text{sample}\ \!\MSE(\hat{\bm\theta}^{\EB}(\tilde{\bm\eta}_{\EB}))-\text{sample}\ \!\MSE(\hat{\bm\theta}^{\EB}(\hat{\bm\eta}_{\EB})),
\end{align}
and average $\delta\!\FIT$ is defined analogously. Fig.~\ref{fig: illustrating_example} indicates that even a small perturbation $\delta\alpha$ leads to a noticeable increase in sample $\MSE$ and a decrease in average $\FIT$.

\vspace{-2mm}

\begin{figure}[!htbp]
\centering
 \includegraphics[width=0.86\linewidth]{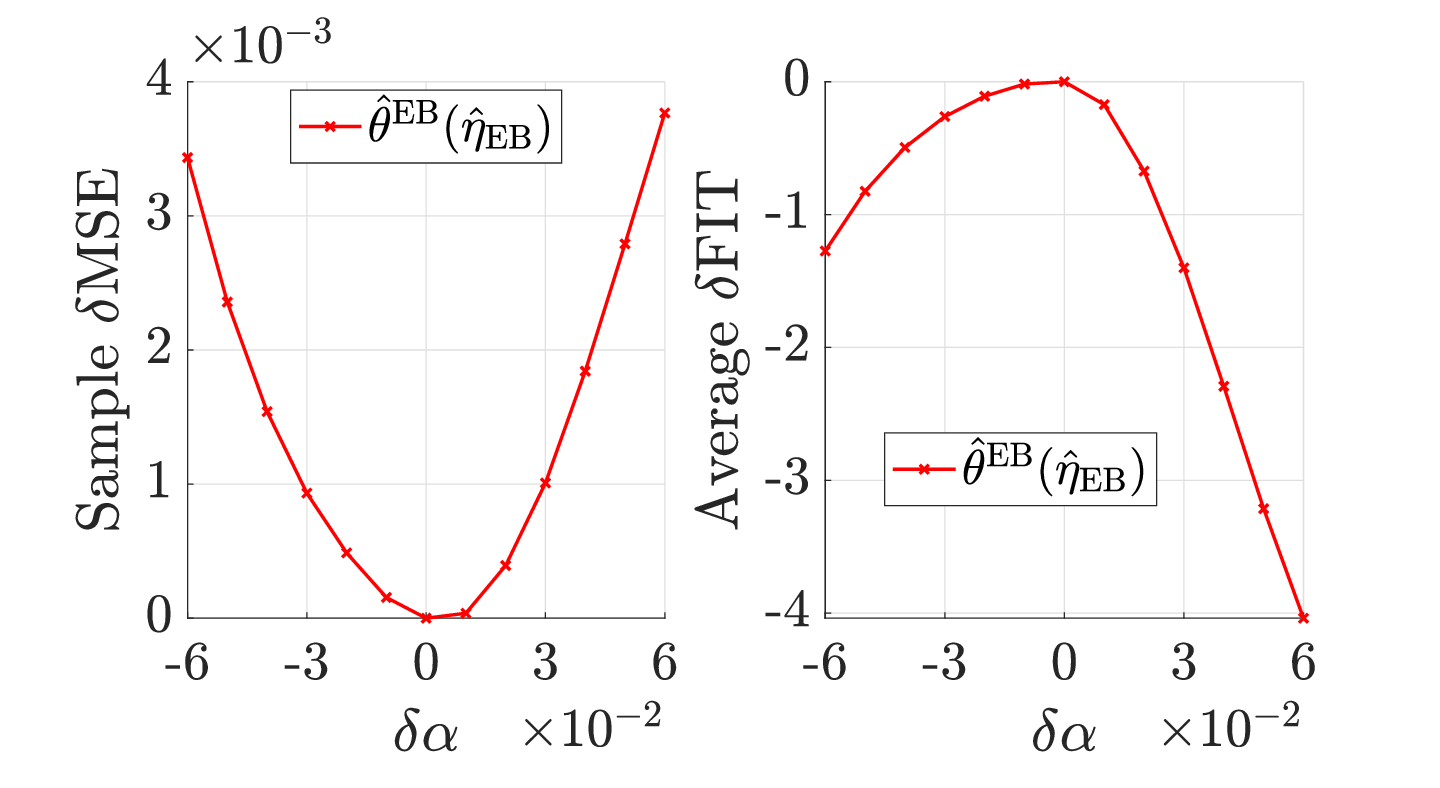}
\caption{Sample means of sample $\delta$MSE and average $\delta$FIT due to the shape hyper-parameter perturbation $\delta\alpha$.}
\label{fig: illustrating_example}
\end{figure}

\end{example}

 \vspace{-2mm}

Example~\ref{example: illustrating example} shows that, although $\hat{\bm\theta}^{\EB}(\hat{\bm\eta}_{\EB})$ can outperform $\hat{\bm\theta}^{\ML}$, it can be sensitive to local perturbations of the estimated hyper-parameters. Such sensitivity may be problematic, because small hyper-parameter estimation errors, caused, e.g., by numerical issues in the used optimization algorithm, may lead to finite-sample performance degradation and numerical fragility. This motivates us to construct an alternative estimator with comparable performance to that of $\hat{\bm\theta}^{\EB}(\hat{\bm\eta}_{\EB})$ in \eqref{eq: def of EB estimator}--\eqref{eq: def of EB hyperparameter estimator}, but reduced sensitivity to hyper-parameter perturbations. To this end, we consider the Bayes estimator $\hat{\bm\theta}^{\Bayes}$ in \eqref{eq: def of Bayes estimator}, which avoids a direct plug-in dependence on hyper-parameters. More concretely, our goal is to design a weighting function such that the resulting $\hat{\bm\theta}^{\Bayes}$ performs comparably to $\hat{\bm\theta}^{\EB}(\hat{\bm\eta}_{\EB})$.

To make this design problem tractable, we need an analytical performance criterion. The finite-sample performance of $\hat{\bm\theta}^{\EB}(\hat{\bm\eta}_{\EB})$ is typically evaluated by its MSE in \eqref{eq: def of MSE}. However, it is in general beyond reach to derive an analytical expression for $\MSE(\hat{\bm\theta}^{\EB}(\hat{\bm\eta}_{\EB}))$. To bypass this difficulty, we shift our focus to the large-sample regime and adopt an asymptotic yet analytical quantity, the XMSE, as an alternative. This quantity was first introduced in \cite{JCWH26}, and requires the following large-sample condition.

\begin{assumption}\label{asp: limit of regression matrix}
We have $\lim_{N\to\infty}\mathbf{\Phi}^{\top}\mathbf{\Phi}/N=\mathbf{\Sigma}\succ 0$.
\end{assumption}

Under Assumptions~\ref{asp:input and noise}--\ref{asp: limit of regression matrix}, the XMSE of an estimator $\hat{\bm\theta}_{N}$ is 
\begin{align}
\XMSE(\hat{\bm\theta}_{N})\coloneqq \lim_{N\to\infty}N^2\left[\MSE(\hat{\bm\theta}_{N})-\MSE(\hat{\bm\theta}^{\ML}) \right]
\end{align}
provided that the limit exists. It is a second-order quantity and measures the difference between the MSEs of $\hat{\bm\theta}_{N}$ and $\hat{\bm\theta}^{\ML}$ for large sample sizes $N$. Under Assumptions~\ref{asp:input and noise}--\ref{asp: limit of regression matrix} and suitable regularity conditions, explicit expressions for the XMSEs of the Bayes and EB estimators were derived in \cite[Theorem~2 and Corollary~2.1]{JCWH26}. This naturally motivates the first problem: 
\begin{itemize}
\item {Given an EB weighting family $\pi(\bm\theta|\bm\eta)$ and the associated EB estimator $\hat{\bm\theta}^{\EB}(\hat{\bm\eta}_{\EB})$ in \eqref{eq: def of EB estimator}--\eqref{eq: def of EB hyperparameter estimator}, can we design a nonnegative weighting function $\pi_{\star}(\bm\theta)$ such that the resulting Bayes estimator, denoted by $\hat{\bm\theta}^{\Bayes}(\pi_{\star})$, satisfies}
\begin{align}\label{eq: XMSE matching condition}
\XMSE(\hat{\bm\theta}^{\Bayes}(\pi_{\star}))=\XMSE(\hat{\bm\theta}^{\EB}(\hat{\bm\eta}_{\EB}))?
\end{align}
\end{itemize}

This problem was partially studied in \cite[Theorem~10 and Corollary~10.1]{JCCWH25} and \cite[Theorem~1]{JWH25csl}, but only under special structural assumptions of $\mathbf{\Sigma}$ and/or the EB weighting families. Whether the weighting function $\pi_{\star}(\bm\theta)$ can be designed so that \eqref{eq: XMSE matching condition} holds for a \emph{general} EB weighting family and a \emph{general} $\mathbf{\Sigma}$ remains open. This is precisely the problem we are addressing.

If a weighting function satisfying \eqref{eq: XMSE matching condition} can be constructed, two further problems arise. Firstly, since this design problem is motivated by the sensitivity of $\hat{\bm\theta}^{\EB}(\hat{\bm\eta}_{\EB})$ to hyper-parameter perturbations, it is natural to ask whether the resulting Bayes estimate $\hat{\bm\theta}^{\Bayes}(\pi_{\star})$ is indeed more locally robust. Secondly, Bayes estimates in \eqref{eq: def of Bayes estimator} often admit no closed-form expressions, and thus the computational complexity of $\hat{\bm\theta}^{\Bayes}(\pi_{\star})$ deserves further analysis. This leads to the next two problems:

\begin{itemize}
\item What is the local sensitivity of $\hat{\bm\theta}^{\Bayes}(\pi_{\star})$ w.r.t. hyper-parameter perturbations in comparison to $\hat{\bm\theta}^{\EB}(\hat{\bm\eta}_{\EB})$?
\item What is the computational complexity of $\hat{\bm\theta}^{\Bayes}(\pi_{\star})$ in comparison to $\hat{\bm\theta}^{\EB}(\hat{\bm\eta}_{\EB})$?
\end{itemize}

\section{Design of Bayes Estimators with the Same XMSE as EB Estimators}\label{sec: design of weighting}

To address the first problem, we first impose an additional assumption on the weighting function $\pi(\bm\theta)$ and then recall the expression for $\XMSE(\hat{\bm\theta}^{\Bayes})$ derived in \cite[Corollary~2.1]{JCWH26}.

\begin{assumption}\label{asp: differentiablity and growth rate conditions}
\begin{enumerate}
\item[]
\item[1)] There exists an open neighborhood of $\bm\theta_{0}$ such that $\pi(\bm\theta)$ is positive and three times continuously differentiable.
\item[2)] There exist $C,a>0$ such that $\pi(\bm\theta)\leq C\exp(a\|\bm\theta\|_{2}^2)$.
\end{enumerate}
\end{assumption}

\begin{remark}\label{rmk: meaning of differentiablity and growth rate condition}
Item $2)$ in Assumption~\ref{asp: differentiablity and growth rate conditions} is a sufficient condition to control the tail behavior of $\pi(\bm\theta)$. Hence, the integrals in \eqref{eq: def of Bayes estimator} are well-defined for sufficiently large $N$, and the tail contributions to them are asymptotically negligible. A related condition on such negligibility was given in \cite[Lemma~2]{GHPZ1990}.
\end{remark}

\begin{lemma}\label{lemma: XMSE expression for Bayes estimator}
Under Assumptions~\ref{asp:input and noise}--\ref{asp: differentiablity and growth rate conditions}, the $\XMSE$ of a generalized Bayes estimator $\hat{\bm\theta}^{\Bayes}$ takes the form
\begin{align*}
\XMSE(\hat{\bm\theta}^{\Bayes})=&\|\XBias(\hat{\bm\theta}^{\Bayes})\|_{2}^2+\Tr[\XVar(\hat{\bm\theta}^{\Bayes})],
\end{align*}
where
\begin{align*}
\XBias(\hat{\bm\theta}^{\Bayes})=&\left.\sigma^2\mathbf{\Sigma}^{-1}\nabla_{\theta}\log\left[\pi(\bm\theta)\right]\right|_{\bm\theta=\bm\theta_{0}},\\
\XVar(\hat{\bm\theta}^{\Bayes})=&\left.2(\sigma^2)^2\mathbf{\Sigma}^{-1}\nabla^2_{\theta\theta}\log\left[\pi(\bm\theta)\right]\mathbf{\Sigma}^{-1}\right|_{\bm\theta=\bm\theta_{0}}.
\end{align*}
\end{lemma}

Lemma~\ref{lemma: XMSE expression for Bayes estimator} shows that the XMSE of $\hat{\bm\theta}^{\Bayes}$ consists of two components. The squared excess bias term $\|\XBias(\hat{\bm\theta}^{\Bayes})\|_{2}^2$ is nonnegative, whereas the sign of the excess variance term $\Tr[\XVar(\hat{\bm\theta}^{\Bayes})]$ is affected by the curvature of $\log(\pi(\bm\theta))$. 



\subsection{XMSE of an EB estimator $\hat{\bm\theta}^{\EB}(\hat{\bm\eta}_{\EB})$}\label{subsec: XMSE of EB}

To make the condition in \eqref{eq: XMSE matching condition} tractable, it remains to derive an explicit expression for $\XMSE(\hat{\bm\theta}^{\EB}(\hat{\bm\eta}_{\EB}))$. Although \cite[Theorem~2]{JCWH26} established a general expression for the XMSE of $\hat{\bm\theta}^{\EB}(\hat{\bm\eta}_{N})$ in \eqref{eq: def of EB estimator}, the explicit specialization to $\hat{\bm\theta}^{\EB}(\hat{\bm\eta}_{\EB})$ in \cite[Corollaries~2.3--2.4]{JCWH26} was limited to Gaussian EB weighting families in \eqref{eq: Gaussian EB weighting families}. The XMSE expression for $\hat{\bm\theta}^{\EB}(\hat{\bm\eta}_{\EB})$ using a general EB weighting family was \emph{not} derived there, which is the focus of this subsection.

We define
\begin{align}\label{eq: def of eta_b_star for theta}
\bm\eta_{\tb\star}(\bm\theta)=\argmin_{\bm\eta\in\mathcal{D}_{\eta}}-\log[\pi(\bm\theta|\bm\eta)],
\end{align}
and make the following additional assumption.



\begin{assumption}\label{asp: regularity conditions for EB hyperparameter estimator}
\begin{enumerate}
\item[]
\item[1)] The set $\mathcal{D}_{\eta}$ is compact, and $-\log[\pi(\bm\theta_{0}|\bm\eta)]$ has a unique minimizer $\bm\eta_{\tb\star}(\bm\theta_{0})$, which locates in the interior of $\mathcal{D}_{\eta}$.
Moreover, we have $\nabla_{\eta\eta}^2\log[\pi(\bm\theta_{0}|\bm\eta)]|_{\bm\eta=\bm\eta_{\tb\star}(\bm\theta_{0})}\prec 0$.
\item[2)] There exists an open neighborhood of $\{\bm\theta_{0}\}\times \mathcal{D}_{\eta}$ such that $\pi(\bm\theta|\bm\eta)>0$ and $\pi(\bm\theta|\bm\eta)$ is three times continuously differentiable jointly in $(\bm\theta,\bm\eta)$.
\item[3)] There exist $C,a>0$ such that 
\begin{gather}
\sup_{\bm\eta\in\mathcal{D}_{\eta}}\left\{\pi(\bm\theta|\bm\eta)+\|\nabla_{\eta}\pi(\bm\theta|\bm\eta)\|_{2}+\|\nabla_{\eta\eta}^2\pi(\bm\theta|\bm\eta)\|_{2}\right\}\\
\leq C\exp(a\|\bm\theta\|_{2}^2).
\end{gather}
\end{enumerate}
\end{assumption}

In the following theorem, we derive an explicit expression for $\XMSE(\hat{\bm\theta}^{\EB}(\hat{\bm\eta}_{\EB}))$.

\begin{theorem}\label{thm: XMSE of EB2 estimator}
Under Assumptions~\ref{asp:input and noise}--\ref{asp: limit of regression matrix} and \ref{asp: regularity conditions for EB hyperparameter estimator}, the XMSE of an EB estimator, i.e., $\hat{\bm\theta}^{\EB}(\hat{\bm\eta}_{\EB})$ in \eqref{eq: def of EB estimator}--\eqref{eq: def of EB hyperparameter estimator}, takes the form
\begin{align}\label{eq: XMSE of EB2 estimator}
&\XMSE(\hat{\bm\theta}^{\EB}(\hat{\bm\eta}_{\EB}))=\|\XBias(\hat{\bm\theta}^{\EB}(\hat{\bm\eta}_{\EB}))\|_{2}^2+\\
&\Tr[\XVar(\hat{\bm\theta}^{\EB}(\hat{\bm\eta}_{\EB})) ]
+\Tr[\XVarHPE(\hat{\bm\theta}^{\EB}(\hat{\bm\eta}_{\EB}))],
\end{align}
where
\begin{align}
&\XBias(\hat{\bm\theta}^{\EB}(\hat{\bm\eta}_{\EB}))=\sigma^2\mathbf{\Sigma}^{-1}\nabla_{\theta}\log\left[\pi(\bm\theta|\bm\eta)\right],\\
&\XVar(\hat{\bm\theta}^{\EB}(\hat{\bm\eta}_{\EB}))=2(\sigma^2)^2\mathbf{\Sigma}^{-1}\nabla_{\theta\theta}^{2}\log\left[\pi(\bm\theta|\bm\eta)\right]\mathbf{\Sigma}^{-1},\nonumber\\
\label{eq: def of XVarHPE of EB2 estimator}
&\XVarHPE(\hat{\bm\theta}^{\EB}(\hat{\bm\eta}_{\EB}))=2(\sigma^2)^2\mathbf{\Sigma}^{-1}\mathcal{V}(\bm\theta,\bm\eta)\mathbf{\Sigma}^{-1}\succeq 0,\quad \\
&\mathcal{V}(\bm\theta,\bm\eta)=-\nabla_{\theta\eta}^{2}\log\left[\pi(\bm\theta|\bm\eta)\right]\left\{\nabla_{\eta\eta}^2\log\left[\pi(\bm\theta|\bm\eta)\right]\right\}^{-1}\\
\label{eq: def of mathcal_V matrix}
&\qquad\qquad\ \times\nabla_{\eta\theta}^2\log\left[\pi(\bm\theta|\bm\eta)\right]\succeq 0,
\end{align}
with $\left(\bm\theta,\bm\eta\right)$ evaluated at $\left(\bm\theta_{0},\bm\eta_{\tb\star}(\bm\theta_{0})\right)$.
\end{theorem}

In contrast to the two-component form of $\XMSE(\hat{\bm\theta}^{\Bayes})$ in Lemma~\ref{lemma: XMSE expression for Bayes estimator}, $\XMSE(\hat{\bm\theta}^{\EB}(\hat{\bm\eta}_{\EB}))$ in Theorem~\ref{thm: XMSE of EB2 estimator} has three components. In addition to its squared excess bias $\|\XBias(\cdot)\|_{2}^2$ and its excess variance $\Tr[\XVar(\cdot)]$, it contains an additional component $\Tr[\XVarHPE(\cdot)]$, representing the excess variance induced by hyper-parameter estimation. Moreover, the condition $\nabla_{\eta\eta}^2\log(\pi(\bm\theta|\bm\eta))|_{\bm\theta=\bm\theta_{0},\bm\eta=\bm\eta_{\tb\star}(\bm\theta_{0})}\prec 0$ in Assumption~\ref{asp: regularity conditions for EB hyperparameter estimator} ensures that the component $\Tr[\XVarHPE(\cdot)]$ is nonnegative, implying that estimating the hyper-parameter using EB comes with a performance penalty.

\subsection{Design of weighting function}\label{subsec: design of weighting function}

Combining Lemma~\ref{lemma: XMSE expression for Bayes estimator} and Theorem~\ref{thm: XMSE of EB2 estimator}, we then analyze the XMSE condition \eqref{eq: XMSE matching condition} to design the weighting function $\pi_{\star}(\bm\theta)$. Direct substitution of these two XMSE expressions into \eqref{eq: XMSE matching condition} gives a nonlinear PDE for $\pi_{\star}(\bm\theta)$, which generally admits no explicit solutions and needs numerical methods. Surprisingly, the particular structures of these two XMSE expressions allow us to derive a characterization of $\pi_{\star}(\bm\theta)$ via $\bm\eta_{\tb\star}(\bm\theta)$ in \eqref{eq: def of eta_b_star for theta}.

\begin{theorem}\label{thm: sufficient condition for general EB2 estimator}
Under Assumptions~\ref{asp:input and noise}--\ref{asp: limit of regression matrix} and \ref{asp: regularity conditions for EB hyperparameter estimator}, if
\begin{align}\label{eq: characterization of weighting function}
\pi_{\star}(\bm\theta)=C\pi(\bm\theta|\bm\eta_{\tb\star}(\bm\theta)),
\end{align}
with $C>0$ and $\bm\eta_{\tb\star}(\bm\theta)$ defined in \eqref{eq: def of eta_b_star for theta}, then \eqref{eq: XMSE matching condition} holds with
\begin{gather}\label{eq: matching XBias}
\XBias(\hat{\bm\theta}^{\EB}(\hat{\bm\eta}_{\EB}))=\XBias(\hat{\bm\theta}^{\Bayes}(\pi_{\star})),\\
\XVar(\hat{\bm\theta}^{\EB}(\hat{\bm\eta}_{\EB}))+
\XVarHPE(\hat{\bm\theta}^{\EB}(\hat{\bm\eta}_{\EB}))\\
\label{eq: matching XVar}
={\XVar}(\hat{\bm\theta}^{\Bayes}(\pi_{\star})).
\end{gather}
\end{theorem}

Theorem~\ref{thm: sufficient condition for general EB2 estimator} establishes a direct correspondence between the EB and Bayes estimation paradigms, implying that every EB estimator admits a Bayes counterpart having the same XMSE. Furthermore, the Bayes weighting function $\pi_{\star}(\bm\theta)$ is obtained by profiling the original EB weighting family over $\bm\eta$, i.e., 
\begin{align}\label{eq: equivalent form of pi_star}
\pi_{\star}(\bm\theta)=C{\max}_{\bm\eta\in\mathcal{D}_{\eta}}\pi(\bm\theta|\bm\eta), 
\end{align}
which is equivalent to \eqref{eq: characterization of weighting function}. Compared with the EB estimator $\hat{\bm\theta}^{\EB}(\hat{\bm\eta}_{\EB})$, the resulting Bayes counterpart $\hat{\bm\theta}^{\Bayes}(\pi_{\star})$ avoids the direct dependence on the hyper-parameter estimate $\hat{\bm\eta}_{\EB}$.


\begin{remark}\label{rmk: high-order merging of EB and Bayes}
A related analysis is provided in \cite[Proposition~3.1]{RRP24} within a regular parametric framework. It focuses on the EB posterior and the Bayes posterior based on the oracle-type weighting function $\pi(\bm\theta|\bm\eta_{\tb\star}(\bm\theta_{0}))$. These two posteriors are shown to merge at a rate faster than ${N}^{-1/2}$. By evaluating $\bm\eta_{\tb\star}(\bm\theta)$ at $\bm\theta=\bm\theta_{0}$, the XMSE condition in \eqref{eq: XMSE matching condition} can be viewed as a risk-level analogue to \cite[Proposition~3.1]{RRP24}. 
\end{remark}

\vspace{-2mm}

\subsection{Case study: Gaussian EB weighting families}\label{subsec: case study}

To gain further insights into the designed weighting function $\pi_{\star}(\bm\theta)$ in \eqref{eq: characterization of weighting function}, we consider the Gaussian EB weighting families in \eqref{eq: Gaussian EB weighting families}, which were studied in \cite[Theorem~10]{JCCWH25}. We recall the relevant result below.

\begin{lemma}\label{lemma: pde for Gaussian weighting families}
Assume that Assumptions~\ref{asp:input and noise}--\ref{asp: limit of regression matrix} and Item~$1)$ in Assumption~\ref{asp: regularity conditions for EB hyperparameter estimator} hold, and $\mathbf{P}(\bm\eta)$ is positive definite and three times continuously differentiable on an open neighborhood of $\mathcal{D}_{\eta}$. Consider the Gaussian weighting families in \eqref{eq: Gaussian EB weighting families}. If $\pi_{\star}(\bm\theta)$ satisfies the following PDE
\begin{align}\label{eq: EB PDE}
\nabla_{\theta}\log[\pi_{\star}(\bm\theta)]=-\mathbf{P}(\bm\eta_{\tb\star}(\bm\theta))^{-1}\bm\theta,
\end{align}
where
\begin{subequations}
\label{eq: def of eta_b_star and Wb}
\noeqref{eq: def of eta_b_star for Gaussian weighting}\noeqref{eq: def of Wb for Gaussian weighting}
\begin{align}
\label{eq: def of eta_b_star for Gaussian weighting}
\bm\eta_{\tb\star}(\bm\theta)=&{\argmin}_{\bm\eta\in\mathcal{D}_{\eta}}W_{\tb}(\bm\eta;\bm\theta),\\
\label{eq: def of Wb for Gaussian weighting}
W_{\tb}(\bm\eta;\bm\theta)=&\bm\theta^{\top}\mathbf{P}(\bm\eta)^{-1}\bm\theta+\log(\det(\mathbf{P}(\bm\eta))),
\end{align}
\end{subequations}
then \eqref{eq: XMSE matching condition} holds.
\end{lemma}

Lemma~\ref{lemma: pde for Gaussian weighting families} provides a sufficient PDE-type condition for \eqref{eq: XMSE matching condition}, but does not by itself lead to a characterization of $\pi_{\star}(\bm\theta)$. In the following corollary, we establish such a characterization, which gives the complete family of solutions to \eqref{eq: EB PDE}.



\begin{corollary}\label{corollary: characterization of weighting function}
Consider the Gaussian EB weighting families in \eqref{eq: Gaussian EB weighting families}. Under the assumptions in Lemma~\ref{lemma: pde for Gaussian weighting families}, if
\begin{align}\label{eq: analytic form}
\pi_{\star}(\bm\theta)=C\exp\left[-\tfrac{1}{2}W_{\tb}(\bm\eta_{\tb\star}(\bm\theta);\bm\theta) \right],
\end{align}
where $C>0$, and $W_{\tb}(\bm\eta;\bm\theta)$ and $\bm\eta_{\tb\star}(\bm\theta)$ are defined in \eqref{eq: def of eta_b_star and Wb}, then \eqref{eq: XMSE matching condition} holds. Moreover, \eqref{eq: analytic form} gives the complete family of solutions to the PDE in \eqref{eq: EB PDE}.
\end{corollary}

Although $\pi_{\star}(\bm\theta)$ in \eqref{eq: analytic form} is constructed to match the XMSE of $\hat{\bm\theta}^{\EB}(\hat{\bm\eta}_{\EB})$ using the Gaussian EB weighting families, it is generally \emph{not} Gaussian since $\bm\eta_{\tb\star}(\bm\theta)$ depends on $\bm\theta$ and is not fixed. More precisely, as shown in \eqref{eq: equivalent form of pi_star}, $\pi_{\star}(\bm\theta)$ is a pointwise upper envelope of the Gaussian families. 

We next consider several representative parametrizations of the covariance matrix $\mathbf{P}(\bm\eta)$ in \eqref{eq: Gaussian EB weighting families}. 

\subsubsection{Scale-shape parametrization}
We start with
\begin{align}\label{eq: kernel structure}
\mathbf{P}(\bm\eta)=c\mathbf{K}(\bm\alpha)\ \text{with}\  \mathbf{K}(\bm\alpha)\succ 0\ \text{and}\ \bm\eta=[c,\bm\alpha^{\top}]^{\top},
\end{align}
where $0<\underline{c}\leq c\leq \overline{c}<\infty$ tunes the overall scale, $\bm\alpha\in\mathcal{D}_{\alpha}\subseteq\R^{n_{\eta}-1}$ controls the covariance shape, and $\mathcal{D}_{\eta}=[\underline{c},\overline{c}]\times \mathcal{D}_{\alpha}$. This structure is widely applied in practice, e.g., the TC kernel in \eqref{eq: def of TC kernel} and the stable spline kernel \cite{PDCDL14}. The corresponding form of $\pi_{\star}(\bm\theta)$ is established in the following result. 

\begin{corollary}\label{corollary: characterization with kernel structure}
Consider the Gaussian EB weighting family in \eqref{eq: Gaussian EB weighting families} with the parametrization in \eqref{eq: kernel structure}. Assume that $\bm\theta_{0}\neq \bm{0}$. Under the assumptions of Lemma~\ref{lemma: pde for Gaussian weighting families}, if
\begin{align}\label{eq: more explicit characterization}
\pi_{\star}(\bm\theta)=C\exp\left[-\tfrac{1}{2}\widetilde{W}_{\tb}(\bm\alpha_{\tb\star};\bm\theta) \right],
\end{align}
where $C>0$, and
\begin{subequations}
\begin{align}\label{eq: def of alpha_star}
\bm\alpha_{\tb\star}=&{\argmin}_{\bm\alpha\in\mathcal{D}_{\alpha}}\widetilde{W}_{\tb}(\bm\alpha;\bm\theta),\\
\label{eq: def of widetilde_Wb}
\widetilde{W}_{\tb}(\bm\alpha;\bm\theta)=&n_{\theta}\log(\bm\theta^{\top}\mathbf{K}(\bm\alpha)^{-1}\bm\theta)+\log\det(\mathbf{K}(\bm\alpha)),\
\end{align}
\end{subequations}
then \eqref{eq: XMSE matching condition} holds. 
\end{corollary}

\begin{remark}\label{rmk: specializations under structures of P_eta}
When $\mathbf{K}(\bm\alpha)$ in \eqref{eq: kernel structure} specializes to fixed $\mathbf{K}\succ 0$, $\pi_{\star}(\bm\theta)$ in \eqref{eq: more explicit characterization} coincides with the weighting function in \cite[Corollary~10.1]{JCCWH25}. When $\mathbf{\Sigma}=\mathbf{K}=\mathbf{I}_{n_{\theta}}$, it further specializes to the weighting function in \cite[Theorem~1]{JWH25csl} with $C_{1}=0$.
\end{remark}

\begin{remark}\label{rmk: uniqueness of scale and shape decomposition}
Since $c$ is a global scale hyper-parameter, the decomposition $\mathbf{P}(\bm\eta)=c\mathbf{K}(\bm\alpha)$ is non-unique under a positive scalar rescaling of $\mathbf{K}(\bm\alpha)$. By the interiority assumption in Item~$1)$ of Assumption~\ref{asp: regularity conditions for EB hyperparameter estimator}, there exists a neighborhood $\mathcal{U}(\bm\theta_{0})$ in which $\bm\eta_{\tb\star}(\bm\theta)$ remains interior. We further implicitly assume that $[\underline{c},\overline{c}]$ is sufficiently wide such that $\underline{c}<\bm\theta^{\top}\mathbf{K}(\bm\alpha)^{-1}\bm\theta/n_{\theta}<\overline{c}$ for all $\bm\alpha\in\mathcal{D}_{\alpha}$ and $\bm\theta\in\mathcal{U}(\bm\theta_{0})$. Under this condition, 
\begin{align}\label{eq: optimal scale hyperparameter}
c_{\tb\star}(\bm\alpha;\bm\theta)\coloneqq \argmin_{c\in[\underline{c},\overline{c}]}W_{\tb}(\bm\eta;\bm\theta)
=\frac{\bm\theta^{\top}\mathbf{K}(\bm\alpha)^{-1}\bm\theta}{n_{\theta}}.\
\end{align}
If $\mathbf{K}(\bm\alpha)$ is multiplied by a positive scalar, $c_{\tb\star}(\bm\alpha;\bm\theta)$ in \eqref{eq: optimal scale hyperparameter} is inversely rescaled. Therefore, this non-uniqueness does not affect the resulting weighting function $\pi_{\star}(\bm\theta)$ in \eqref{eq: more explicit characterization}.
\end{remark}

Since the scale hyper-parameter $c$ that minimizes $W_{\tb}(\bm\eta;\bm\theta)$ is expressed in closed form by \eqref{eq: optimal scale hyperparameter}, it can be profiled out.
This leaves only the optimization over the shape hyper-parameter $\bm\alpha$. Correspondingly, when we use the Gaussian EB weighting family defined in \eqref{eq: Gaussian EB weighting families} and \eqref{eq: kernel structure}, computing $\hat{\bm\eta}_{\EB}$ in \eqref{eq: def of EB hyperparameter estimator} is generally an $n_{\eta}$-dimensional problem, whereas computing $\bm\eta_{\tb\star}(\bm\theta)$ in \eqref{eq: def of eta_b_star for theta} requires solving an $(n_{\eta}-1)$-dimensional problem.

\subsubsection{Diagonal parametrization}
Another structure of interest is
\begin{align}\label{eq: diagonal kernel}
\mathbf{P}(\bm\eta)=\diag\{\bm\eta\}\ \text{with}\ [\bm\eta]_{k}\geq \underline{\eta}>0\ \text{for}\ k=1,\cdots,n_{\eta},\quad
\end{align}
where $\mathcal{D}_{\eta}=[\underline{\eta},\infty)^{n_{\eta}}$ and $n_{\eta}=n_{\theta}$. This is a lower-bounded variant of the diagonal parametrization used in sparse Bayesian learning (SBL) and automatic relevance determination (ARD) \cite{Tipping2001, WR2004}. Standard SBL and ARD formulations often allow some hyper-parameters to exactly attain zero, thereby pruning the corresponding model parameters and inducing sparsity. In contrast, the lower bound $\underline{\eta}$ rules out exact pruning. In spite of this, when $\underline{\eta}$ is sufficiently small, the resulting EB estimator can strongly shrink small model parameters and thus is suitable for near-sparse settings.

Below, we consider the diagonal parametrization in \eqref{eq: diagonal kernel} and derive the closed-form expression for $\pi_{\star}(\bm\theta)$. 



\begin{corollary}\label{corollary: characterization with diagnal kernel}
Consider the Gaussian EB weighting family in \eqref{eq: Gaussian EB weighting families} with the parametrization in \eqref{eq: diagonal kernel}. Assume that Assumptions~\ref{asp:input and noise}--\ref{asp: limit of regression matrix} hold, and $\underline{\eta}<\min_{k}[\bm\theta_{0}]_{k}^2$. If
\begin{gather}\label{eq: weighting function for diagonal kernel}
\pi_{\star}(\bm\theta)=C{\textstyle\prod}_{k=1}^{n_{\theta}}\pi_{\underline{\eta}}([\bm\theta]_{k}),\ C>0,\\
{\pi}_{\underline{\eta}}([\bm\theta]_{k})=
\left\{\begin{array}{ll} \sqrt{\frac{\mathrm{e}}{\underline{\eta}}}\exp\left\{-\frac{[\bm\theta]_{k}^2}{2\underline{\eta}}\right\}, & [\bm\theta]_{k}^2\leq \underline{\eta},\\
\big|[\bm\theta]_{k}\big|^{-1}, & [\bm\theta]_{k}^2> \underline{\eta}, \end{array}\right.
\end{gather}
then \eqref{eq: XMSE matching condition} holds.
\end{corollary}

\begin{remark}\label{rmk: compactness and coercivity of D_eta}
Note that $\mathcal{D}_{\eta}=[\underline{\eta},\infty)^{n_{\eta}}$ is closed but noncompact, and thus does not satisfy Item~$1)$ of Assumption~\ref{asp: regularity conditions for EB hyperparameter estimator}. In this case, the role of compactness is replaced by the coercivity of the criteria that define $\bm\eta_{\tb\star}(\bm\theta)$ and $\hat{\bm\eta}_{\EB}$, see Appendix~\ref{subsec: proof of diagonal parametrization}.
\end{remark}

As shown in Corollary~\ref{corollary: characterization with diagnal kernel}, when $[\bm\theta]_{k}^2> \underline{\eta}$ for all $k$, \eqref{eq: weighting function for diagonal kernel} specializes to $\pi_{\star}(\bm\theta)=C\prod_{k=1}^{n_{\theta}}|[\bm\theta]_{k}|^{-1}$. It coincides with the normal-Jeffreys prior \cite{F2003, BM2004}, which has been widely used for sparse estimation and variable selection. This prior combines the Gaussian weighting family in \eqref{eq: Gaussian EB weighting families} with $\mathbf{P}(\bm\eta)=\diag\{\bm\eta\}\succ 0$ and the Jeffreys hyper-prior $p([\bm\eta]_{k})\propto [\bm\eta]_{k}^{-1}$, i.e.,
\begin{align*}
\pi_{\text{NJ}}(\bm\theta)={\textstyle\prod}_{k}{\textstyle\int}_{0}^{\infty}p([\bm\theta]_{k}|[\bm\eta]_{k})p([\bm\eta]_{k})d[\bm\eta]_{k}\propto {\textstyle\prod}_{k=1}^{n_{\theta}}\left|[\bm\theta]_{k}\right|^{-1}.
\end{align*}
However, as pointed out in \cite{ADL2013}, $\pi_{\text{NJ}}(\bm\theta)$ is locally nonintegrable around $[\bm\theta]_{k}=0$, and consequently, the integrals in \eqref{eq: def of Bayes estimator} diverge. Consequently, the normal-Jeffreys prior is commonly used for maximum a posteriori (MAP) estimation rather than the Bayes estimation in \eqref{eq: def of Bayes estimator}. To avoid this issue, we impose the positive lower bound $\underline{\eta}$ in \eqref{eq: diagonal kernel}, which removes the singularity at zero and yields a well-defined Bayes estimator $\hat{\bm\theta}^{\Bayes}(\pi_{\star})$. 

\begin{remark}\label{rmk: smooth approximation}
As an alternative to imposing the lower bound $\underline{\eta}$, one may consider $\mathbf{P}(\bm\eta)=\diag\{\bm\eta\}$ and $\mathcal{D}_{\eta}=(0,\infty)^{n_{\eta}}$, and adopt a smooth approximation to the normal-Jeffreys prior,
\begin{align}
\pi_{\star,\epsilon}(\bm\theta)=C{\textstyle\prod}_{k=1}^{n_{\theta}}([\bm\theta]_{k}^2+\epsilon^2)^{-1/2}.
\end{align}
For sufficiently small $\epsilon$, the XMSE of $\hat{\bm\theta}^{\Bayes}(\pi_{\star,\epsilon})$ is expected to be close to that of $\hat{\bm\theta}^{\EB}(\hat{\bm\eta}_{\EB})$. 
\end{remark}

\section{Local Sensitivity of EB and Bayes Estimates to Hyper-parameter Perturbations}\label{sec: local sensitivity}

In the previous section, we have designed a weighting function $\pi_{\star}(\bm\theta)$ such that the resulting Bayes estimator $\hat{\bm\theta}^{\Bayes}(\pi_{\star})$ has the same XMSE as the EB estimator $\hat{\bm\theta}^{\EB}(\hat{\bm\eta}_{\EB})$. We now turn to the second problem and compare their local sensitivity to hyper-parameter perturbations. Throughout this section, the observed data are kept fixed, and local sensitivity is quantified at the resulting EB and Bayes estimates.

For certain parametrizations of the EB weighting families, $\bm\eta_{\tb\star}(\bm\theta)$ in \eqref{eq: def of eta_b_star for theta} can be obtained analytically, yielding a closed-form $\pi_{\star}(\bm\theta)$, e.g., \cite[Corollary~10.1]{JCCWH25}, \cite[Theorem~1]{JWH25csl}, and Corollary~\ref{corollary: characterization with diagnal kernel}. Hence, the resulting Bayes estimate $\hat{\bm\theta}^{\Bayes}(\pi_{\star})$ avoids local sensitivity to perturbations in $\bm\eta_{\tb\star}(\bm\theta)$. For general parametrizations, the EB and Bayes estimates are both affected by perturbations in hyper-parameter quantities. However, the perturbations enter in different ways.
\begin{itemize} 
\item For $\hat{\bm\theta}^{\EB}(\hat{\bm\eta}_{\EB})$, the perturbation enters through $\hat{\bm\eta}_{\EB}$ in \eqref{eq: def of EB hyperparameter estimator} and is represented by $\delta\bm\eta\in\R^{n_{\eta}}$, whose magnitude is measured by $\|\delta\bm\eta\|_{2}$.
\item For $\hat{\bm\theta}^{\Bayes}(\pi_{\star})$, the perturbation enters through $\bm\eta_{\tb\star}(\bm\theta)$ in \eqref{eq: def of eta_b_star for theta} and is represented by $\delta\bm\eta(\cdot):\R^{n_{\theta}}\mapsto\R^{n_{\eta}}$, whose magnitude is measured by  
\begin{align}\label{eq: uniform norm}
\|\delta\bm\eta\|_{\infty}\coloneqq \sup_{\bm\theta}\|\delta\bm\eta(\bm\theta)\|_{2}.
\end{align}
\end{itemize}
By following Example~\ref{example: illustrating example}, we let $\hat{\bm\theta}^{\EB}(\tilde{\bm\eta}_{\EB})$ and $\hat{\bm\theta}^{\Bayes}(\tilde{\pi}_{\star})$ denote the perturbed EB and Bayes estimates, respectively, where $\tilde{\bm\eta}_{\EB}\coloneqq \hat{\bm\eta}_{\EB}+\delta\bm\eta$ and
\begin{align}
\tilde{\pi}_{\star}(\bm\theta) \coloneqq C\pi\left(\bm\theta|\tilde{\bm\eta}_{\tb\star}(\bm\theta)\right)\ \text{with}\ \tilde{\bm\eta}_{\tb\star}(\bm\theta)\coloneqq\bm\eta_{\tb\star}(\bm\theta)+\delta\bm\eta(\bm\theta).\nonumber
\end{align}

We first give an additional assumption on perturbations, and then analyze local sensitivity of $\hat{\bm\theta}^{\EB}(\hat{\bm\eta}_{\EB})$ and $\hat{\bm\theta}^{\Bayes}(\pi_{\star})$.

\begin{assumption}\label{asp: admissible perturbations}
\begin{enumerate}
\item[]
\item[1)] The perturbation segments $\mathcal{S}_{\EB}\coloneqq \{\hat{\bm\eta}_{\EB}+t\delta\bm\eta: t\in[0,1]\}$ and $\mathcal{S}_{\Bayes}(\bm\theta)\coloneqq \{\bm\eta_{\tb\star}(\bm\theta)+t\delta\bm\eta(\bm\theta):t\in[0,1]\}$ satisfy $\mathcal{S}_{\EB}\subseteq \mathcal{D}_{\eta}$ and $\mathcal{S}_{\Bayes}(\bm\theta)\subseteq \mathcal{D}_{\eta}$.
\item[2)] The EB weighting family $\pi(\bm\theta|\bm\eta)$ is twice continuously differentiable in $\bm\eta$ on an open neighborhood of both $\mathcal{S}_{\EB}$ and $\mathcal{S}_{\Bayes}(\bm\theta)$.
\item[3)] The Bayes perturbation function $\delta\bm\eta(\bm\theta)$ satisfies 
\begin{align}\label{eq: directional stationarity}
\left.[\nabla_{\eta}\pi(\bm\theta|\bm\eta)]^{\top}\delta\bm\eta(\bm\theta)\right|_{\bm\eta=\bm\eta_{\tb\star}(\bm\theta)}=0.
\end{align}
\end{enumerate}
\end{assumption}

\begin{remark}\label{rmk: perturbation assumption meaning}
In Assumption~\ref{asp: admissible perturbations}, Item~$1)$ ensures the feasibility of the perturbations and Item~$2)$ provides the smoothness of $\pi(\bm\theta|\bm\eta)$ required for the corresponding Taylor expansions. As for Item~$3)$, when $\bm\eta_{\tb\star}(\bm\theta)$ in \eqref{eq: def of eta_b_star for theta} is an interior optimizer, \eqref{eq: directional stationarity} follows directly from the first-order optimality condition; when $\bm\eta_{\tb\star}(\bm\theta)$ is on the boundary of $\mathcal{D}_{\eta}$, e.g., the case where $[\bm\theta]_{k}^2\leq \underline{\eta}$ in Corollary~\ref{corollary: characterization with diagnal kernel}, condition \eqref{eq: directional stationarity} restricts the perturbation to a critical direction along which the first-order variation vanishes.
\end{remark}

\begin{theorem}\label{thm: sensitivity analysis}
Let Assumptions~\ref{asp:input and noise}--\ref{asp: limit of regression matrix} and \ref{asp: regularity conditions for EB hyperparameter estimator}--\ref{asp: admissible perturbations} hold. For any sufficiently large fixed $N$ and for fixed observed data, we have\footnote{For a perturbation $\delta\bm{x}$ in a normed space, we write $\|R(\delta\bm{x})\|=O(\|\delta\bm{x}\|^{q})$ with $q>0$ as $\|\delta\bm{x}\|\to 0$ means that there exist constants $C>0$ and $r>0$, independent of $\delta\bm{x}$, such that $\|R(\delta\bm{x})\|\leq C\|\delta\bm{x}\|^{q}$ for any $\|\delta\bm{x}\|<r$.}
\begin{align}\label{eq: sensitivity of EB2 estimator}
\|\hat{\bm\theta}^{\EB}(\tilde{\bm\eta}_{\EB})-\hat{\bm\theta}^{\EB}(\hat{\bm\eta}_{\EB})\|_{2}=&O(\|\delta\bm\eta\|_{2}), \\
\label{eq: sensitivity of Bayes estimator}
\|\hat{\bm\theta}^{\Bayes}(\tilde{\pi}_{\star})-\hat{\bm\theta}^{\Bayes}(\pi_{\star})\|_{2}=&O(\|\delta\bm\eta\|_{\infty}^2), 
\end{align}
as $\|\delta\bm\eta\|_{2}\to 0$ and $\|\delta\bm\eta\|_{\infty}\to 0$, respectively.
\end{theorem}

Theorem~\ref{thm: sensitivity analysis} implies that $\hat{\bm\theta}^{\EB}(\hat{\bm\eta}_{\EB})$ is at most first-order sensitive to $\delta\bm\eta$, while the constructed Bayes estimate $\hat{\bm\theta}^{\Bayes}(\pi_{\star})$ is at most second-order sensitive to $\delta\bm\eta(\bm\theta)$. The reduced sensitivity of $\hat{\bm\theta}^{\Bayes}(\pi_{\star})$ suggests reduced performance degradation under small hyper-parameter perturbations. 
This property also allows $\bm\eta_{\tb\star}(\bm\theta)$ to be numerically computed more coarsely than $\hat{\bm\eta}_{\EB}$, thereby potentially lowering the computational cost of $\hat{\bm\theta}^{\Bayes}(\pi_{\star})$. This point will be further used in the next section.

\section{Implementation and Computational Complexity of EB and Bayes Estimates}\label{sec: implementation and complexity}

In this section, we discuss the numerical implementation of the EB estimate $\hat{\bm\theta}^{\EB}(\hat{\bm\eta}_{\EB})$ and the Bayes estimate $\hat{\bm\theta}^{\Bayes}(\pi_{\star})$, respectively. We first analyze the general computational structures of the two estimators, and then specialize their complexity analysis to the Gaussian and Student-$t$ weighting families.


\subsection{General computational structures}

The general computation of $\hat{\bm\theta}^{\EB}(\hat{\bm\eta}_{\EB})$ consists of 
\begin{itemize}
\item[-] computation of the basic data-dependent quantities, e.g., $\mathbf{\Phi}^{\top}\mathbf{\Phi}$, $\mathbf{\Phi}^{\top}\bm{Y}$, and the required matrix factorizations;
\item[-] computation of the data-dependent hyper-parameter estimate $\hat{\bm\eta}_{\EB}$, which generally involves repeated evaluations of its cost function, i.e., $\mathscr{F}_{\EB}(\bm\eta)$ defined in \eqref{eq: def of EB hyperparameter estimator};
\item[-] evaluation of the EB estimate, i.e., $\hat{\bm\theta}^{\EB}(\bm\eta)$ in \eqref{eq: def of EB estimator}.
\end{itemize}
For a given $\bm\eta$, when the weighting family $\pi(\bm\theta|\bm\eta)$ is conjugate to the likelihood $p(\bm{Y}|\bm\theta)$, both $\mathscr{F}_{\EB}(\bm\eta)$ and $\hat{\bm\theta}^{\EB}(\bm\eta)$ admit closed-form expressions, e.g., the Gaussian weighting families in \eqref{eq: Gaussian EB weighting families}. In the non-conjugate cases, these quantities typically require numerical approximations, e.g., the importance sampling (IS) and Markov chain MC (MCMC) methods \cite{RCC2004, MRRTT1953}.

The general computation of $\hat{\bm\theta}^{\Bayes}(\pi_{\star})$ consists of
\begin{itemize}
\item[-] computation of the basic data-dependent quantities;
\item[-] computation of the weighting function $\pi_{\star}(\bm\theta)$ defined in \eqref{eq: characterization of weighting function}, which involves the computation of $\bm\eta_{\tb\star}(\bm\theta)$ in \eqref{eq: def of eta_b_star for theta};
\item[-] evaluation of the Bayes estimate $\hat{\bm\theta}^{\Bayes}(\pi_{\star})$ defined in \eqref{eq: def of Bayes estimator}, which generally requires numerical approximations.
\end{itemize}
For a given $\bm\theta$, $\bm\eta_{\tb\star}(\bm\theta)$ is independent of the observed data and its computation depends only on the chosen weighting family. It is computed either in closed form or by numerically solving the problem in \eqref{eq: def of eta_b_star for theta} for each integration point.

The two estimates exhibit different computational structures, and \emph{neither} estimate is uniformly computationally preferable. Their relative computational costs will depend strongly on the conjugacy and parametrizations of the EB weighting families. Below, we consider two specific weighting families.

\subsection{Gaussian EB weighting family with \eqref{eq: def of TC kernel}}\label{subsec: Gaussian weighting with TC kernel}
 
The first case studied is the Gaussian EB weighting family in \eqref{eq: Gaussian EB weighting families} under the parametrization in \eqref{eq: def of TC kernel}. 

\subsubsection{Implementation and computational complexity of EB estimate $\hat{\bm\theta}^{\EB}(\hat{\bm\eta}_{\EB})$} 

Since the Gaussian weighting family is conjugate to the Gaussian likelihood, both $\mathscr{F}_{\EB}(\bm\eta)$ and $\hat{\bm\theta}^{\EB}(\bm\eta)$ admit closed-form expressions, see \eqref{eq: expression for regularized estimator}--\eqref{eq: expression for F_EB}, thereby enabling efficient computation of $\hat{\bm\theta}^{\EB}(\hat{\bm\eta}_{\EB})$. 

Here, we use the implementation in \cite[Algorithm~2]{CL:12} and let $M_{\EB}$ denote the required number of evaluations of $\mathscr{F}_{\EB}(\bm\eta)$. The computational costs of its main steps are given below\footnote{In the computational complexity analysis, we let $\mathcal{O}(\cdot)$ denote the order of the number of arithmetic operations.},
\begin{itemize}
\item[-] computation of basic data-dependent quantities: $\mathcal{O}(Nn_{\theta}^2)$,
\item[-] $M_{\EB}$ evaluations of $\mathscr{F}_{\EB}(\bm\eta)$ in \eqref{eq: expression for F_EB}: $\mathcal{O}(M_{\EB}n_{\theta}^3)$,
\item[-] evaluation of $\hat{\bm\theta}^{\EB}(\bm\eta)$ in \eqref{eq: expression for regularized estimator}: $\mathcal{O}(n_{\theta}^2)$.
\end{itemize}
Therefore, the total computational complexity of $\hat{\bm\theta}^{\EB}(\hat{\bm\eta}_{\EB})$ is $\mathcal{O}(Nn_{\theta}^2+M_{\EB}n_{\theta}^3)$.

\subsubsection{Implementation and computational complexity of Bayes estimate $\hat{\bm\theta}^{\Bayes}(\pi_{\star})$} 

The Gaussian EB weighting family in \eqref{eq: Gaussian EB weighting families} and the scale-shape parametrization in \eqref{eq: def of TC kernel} together yield the explicit representation of $\pi_{\star}(\bm\theta)$ in \eqref{eq: more explicit characterization}. By \cite[Proposition~\Rmnum{4}.2 and Remark~\Rmnum{4}.2]{CCL17}, we can rewrite \eqref{eq: def of widetilde_Wb} as
\begin{align}
&\widetilde{W}_{\tb}(\alpha;\bm\theta)=n_{\theta}\log\left\{{\sum}_{j=1}^{n_{\theta}-1}\frac{([\bm\theta]_{j}-[\bm\theta]_{j+1})^2}{\alpha^{j}(1-\alpha)}+\frac{[\bm\theta]_{n_{\theta}}^2}{\alpha^{n_{\theta}}}\right\}\\
\label{eq: rewritten formulation of Wb_tilde_TC}
&\quad+[{n_{\theta}(n_{\theta}+1)}/{2}]\log(\alpha)+(n_{\theta}-1)\log(1-\alpha).
\end{align}
Given $\pi_{\star}(\bm\theta)$, we then utilize the IS method to approximate the Bayes estimate. For the proposal density, we adopt $q_{G}(\bm\theta)\propto p(\bm{Y}|\bm\theta)\pi(\bm\theta|\bm\eta_{\tb\star}(\hat{\bm\theta}^{\ML}))$, where $\bm\eta_{\tb\star}(\hat{\bm\theta}^{\ML})$ is just \eqref{eq: def of alpha_star} and \eqref{eq: optimal scale hyperparameter} evaluated at $\bm\theta=\hat{\bm\theta}^{\ML}$. Equivalently, $q_{G}(\bm\theta)$ is the density of $\bm\theta\sim\mathcal{N}({\bm\mu}_{G},\mathbf{\Sigma}_{G})$, where
\begin{equation}\label{eq: def of mu_q and Sigma_q}
\begin{aligned}
{\bm\mu}_{G}=&\hat{\bm\theta}^{\EB}(\bm\eta_{\tb\star}(\hat{\bm\theta}^{\ML})),\\
\mathbf{\Sigma}_{G}=&\sigma^2[\mathbf{\Phi}^{\top}\mathbf{\Phi}+\sigma^2\mathbf{P}(\bm\eta_{\tb\star}(\hat{\bm\theta}^{\ML}))^{-1}]^{-1}.
\end{aligned}
\end{equation}
The implementation of $\hat{\bm\theta}^{\Bayes}(\pi_{\star})$ is presented in Algorithm~\ref{alg: bayes_is Gaussian}.

\begin{remark}\label{rmk: vectorized batch operations of IS implementation}
The IS samples $\{\bm\theta^{(k)}\}_{k=1}^{M_{\text{IS}}^{\text{B}}}$ are generated independently, and the associated pointwise computations are mutually independent. Thus, they can be implemented efficiently using vectorized batch operations.

\end{remark}

\begin{remark}\label{rmk: choices for proposal distributions}
In contrast with the likelihood-based proposal $\bm\theta\sim\mathcal{N}(\hat{\bm\theta}^{\ML},\sigma^2(\mathbf{\Phi}^{\top}\mathbf{\Phi})^{-1})$ in \cite[Section~\Rmnum{3}]{JWH25csl}, the proposal $q_{G}(\bm\theta)$ exploits the structure of the Gaussian weighting family in \eqref{eq: Gaussian EB weighting families} evaluated at $\bm\eta=\bm\eta_{\tb\star}(\hat{\bm\theta}^{\ML})$. When $q_{G}(\bm\theta)$ approximates the target distribution proportional to $p(\bm{Y}|\bm\theta)\pi_{\star}(\bm\theta)$ well, this proposal can improve the numerical accuracy and stability.
\end{remark}

\begin{algorithm}[!htbp]
\caption{IS implementation of
$\hat{\bm{\theta}}^{\mathrm{Bayes}}(\pi_{\star})$ using the Gaussian EB weighting family in \eqref{eq: Gaussian EB weighting families} with \eqref{eq: def of TC kernel}}
\label{alg: bayes_is Gaussian}
\KwIn{
Observed data $(\bm{Y},\mathbf{\Phi})$,
noise variance $\sigma^2$,
number of IS samples $M_{\text{IS}}^{\text{B}}$.
}
\KwOut{
$\hat{\bm{\theta}}^{\mathrm{Bayes}}(\pi_{\star})$.
}

Compute $\mathbf{\Phi}^{\top}\mathbf{\Phi}$, $\mathbf{\Phi}^{\top}\bm{Y}$, and $\hat{\bm{\theta}}^{\ML}$ in \eqref{eq: def of ML estimator}.
\nllabel{line: basic_ML_Bayes_Gaussian}

Compute \eqref{eq: optimal scale hyperparameter} and \eqref{eq: def of alpha_star} by minimizing $\widetilde W_{\tb}(\alpha;\hat{\bm{\theta}}^{\ML})$ in \eqref{eq: rewritten formulation of Wb_tilde_TC} to obtain $\bm\eta_{\tb\star}(\hat{\bm{\theta}}^{\mathrm{ML}})$.
\nllabel{line: basic_ML_proposal_Bayes_Gaussian}

Compute \eqref{eq: def of mu_q and Sigma_q}, construct the proposal $q_{G}(\bm{\theta})$ and generate
$\{\bm{\theta}^{(k)}\}_{k=1}^{M_{\text{IS}}^{\text{B}}}
\sim q_G(\bm{\theta})$.
\nllabel{line: computation of proposal}

\For{$k=1,\cdots,M_{\mathrm{IS}}^{\mathrm{B}}$}{
    Compute
    $\bm\eta_{\tb\star}(\bm{\theta}^{(k)})$
    by minimizing
    $\widetilde W_{\tb}(\alpha;\bm{\theta}^{(k)})$.

    Compute the IS importance weight $w_{\star}(\bm\theta^{(k)})= {\pi_{\star}(\bm\theta^{(k)})}/{\pi(\bm\theta^{(k)}|\bm\eta_{\tb\star}(\hat{\bm\theta}^{\ML}))}$
    \nllabel{line: computation of IS weight}
}

Compute
\begin{align}\label{eq: Bayes estimate approximation}
\hat{\bm\theta}^{\Bayes}(\pi_{\star})\approx \sum_{m=1}^{M_{\text{IS}}^{\text{B}}}\bm\theta^{(m)}\frac{w_{\star}(\bm\theta^{(m)})}{\sum_{k=1}^{M_{\text{IS}}^{\text{B}}}w_{\star}(\bm{\theta}^{(k)})}.
\end{align}
\nllabel{line: approximation of Bayes Gaussian}

\end{algorithm}

Let $M_{\tb}$ denote the number of evaluations of $\widetilde{W}_{\tb}(\alpha;\bm\theta)$ in \eqref{eq: rewritten formulation of Wb_tilde_TC} required to compute $\bm\eta_{\tb\star}(\bm\theta)$ given $\bm\theta$. The computational cost of Algorithm~\ref{alg: bayes_is Gaussian} is decomposed as follows,
\begin{itemize}
\item[-] computation of basic data-dependent quantities (\emph{Lines~\ref{line: basic_ML_Bayes_Gaussian}--\ref{line: basic_ML_proposal_Bayes_Gaussian} in Algorithm~\ref{alg: bayes_is Gaussian}}): $\mathcal{O}(Nn_{\theta}^2 + n_{\theta}^3 + M_{\tb}n_{\theta})$,
\item[-] computation of IS weights $\{w_{\star}(\bm\theta^{(k)})\}_{k=1}^{M_{\text{IS}}^{\text{B}}}$ (\emph{Lines~\ref{line: computation of proposal}--\ref{line: computation of IS weight} in Algorithm~\ref{alg: bayes_is Gaussian}}): $\mathcal{O}(M_{\text{IS}}^{\text{B}}n_{\theta}^2+M_{\text{IS}}^{\text{B}}M_{\tb}n_{\theta})$,
\item[-] evaluation of $\hat{\bm\theta}^{\Bayes}(\pi_{\star})$ in \eqref{eq: Bayes estimate approximation} (\emph{Line~\ref{line: approximation of Bayes Gaussian} in Algorithm~\ref{alg: bayes_is Gaussian}}): $\mathcal{O}(M_{\text{IS}}^{\text{B}}n_{\theta})$.
\end{itemize}
The derivation is given in Appendix~\ref{subsec: computational analysis of Bayes under Gaussian weighting families}. The overall complexity of Algorithm~\ref{alg: bayes_is Gaussian} is $\mathcal{O}(M_{\text{IS}}^{\text{B}}M_{\tb}n_{\theta}+[N+M_{\text{IS}}^{\text{B}}]n_{\theta}^2 + n_{\theta}^3)$.

The above results suggest that, for the EB weighting family defined in \eqref{eq: Gaussian EB weighting families} and \eqref{eq: def of TC kernel}, and for fixed $N$ and $n_{\theta}$, $\hat{\bm\theta}^{\Bayes}(\pi_{\star})$ can be computationally comparable to, or more efficient than, $\hat{\bm\theta}^{\EB}(\hat{\bm\eta}_{\EB})$ when\footnote{We write $N_{\text{A}}\lesssim N_{\text{B}}$ if there exists a constant $\mathcal{C}>0$ such that $N_{\text{A}}\leq \mathcal{C}N_{\text{B}}$.} 
\begin{align*}
M_{\text{IS}}^{\text{B}}M_{\tb}n_{\theta}+M_{\text{IS}}^{\text{B}}n_{\theta}^2 + n_{\theta}^3 \lesssim M_{\EB}n_{\theta}^3.
\end{align*}
In particular, this condition holds when $M_{\tb}\lesssim n_{\theta}$ and $M_{\text{IS}}^{\text{B}}\lesssim M_{\EB}n_{\theta}$. As discussed after Theorem~\ref{thm: sensitivity analysis}, the local robustness of $\hat{\bm\theta}^{\Bayes}(\pi_{\star})$ permits the use of a relatively small $M_{\tb}$.


\vspace{-3mm}

\subsection{Student-t EB weighting family}\label{subsec: student t weighting family}

The second case considered here is the Student-$t$ EB weighting family \cite{PS1992, Tipping2001, AAMY2008}, given by
\begin{align}\label{eq: student t weighting family}
\pi(\bm\theta|\eta)=\prod_{k=1}^{n_{\theta}}\frac{\Gamma((\nu+1)/2)}{\Gamma(\nu/2)\sqrt{\nu\pi}\eta}\left( 1+\frac{[\bm\theta]_{k}^2}{\nu\eta^2} \right)^{-(\nu+1)/2},
\end{align}
where $\eta>0$ denotes the common scale hyper-parameter, and $\nu>0$ is the degrees of freedom and is assumed to be known. It is heavy-tailed and imposes less shrinkage on large model parameters than a Gaussian weighting family. 

\subsubsection{Implementation and computational complexity of EB estimate $\hat{\bm\theta}^{\EB}(\hat{\eta}_{\EB})$} Since the Student-$t$ weighting family in \eqref{eq: student t weighting family} is non-conjugate to the Gaussian likelihood, we compute both $\hat{\bm\theta}^{\EB}(\eta)$ in \eqref{eq: def of EB estimator} and $\mathscr{F}_{\EB}(\eta)$ in \eqref{eq: def of EB hyperparameter estimator} by using the IS method. For each fixed $\eta$, we use the proposal $\bm\theta|\eta\sim\mathcal{N}(\bm\mu_{S}(\eta),\mathbf{\Sigma}_{S}(\eta))$ and its density is denoted $q_{S}(\bm\theta;\eta)$, where
\begin{align}\label{eq: def of mu_S}
\bm\mu_{S}(\eta)=&{\argmin}_{\bm\theta}\mathcal{J}(\bm\theta;\eta),\\
\mathcal{J}(\bm\theta;\eta)=&\tfrac{\|\bm{Y}-\mathbf{\Phi}\bm\theta\|_{2}^2}{2\sigma^2}+\tfrac{\nu+1}{2}\operatorname{\sum}_{k=1}^{n_{\theta}}\log\left(1+\tfrac{[\bm\theta]_{k}^2}{\nu\eta^2} \right),\nonumber\\
\label{eq: def of Sigma_S}
\mathbf{\Sigma}_{S}(\eta)=&\left[\nabla_{\theta\theta}^{2}\mathcal{J}(\bm\theta;\eta)|_{\bm\theta=\bm\mu_{S}(\eta)}\right]^{-1}.
\end{align}
Note that $\mathcal{J}(\bm\theta;\eta)$ is equivalent to $-\log[p(\bm{Y}|\bm\theta)\pi(\bm\theta|\eta)]$ up to constants independent of $\bm\theta$, and the Gaussian density $q_{S}(\bm\theta;\eta)$ is a Laplace approximation to $p(\bm{Y}|\bm\theta)\pi(\bm\theta|\eta)$. 

The numerical construction of $\bm{\mu}_{S}(\eta)$ and $\mathbf{\Sigma}_{S}(\eta)$ is based on the analytic gradient and Hessian of $\mathcal{J}(\bm\theta;\eta)$, i.e.,
\begin{align}\label{eq: explicit expression for mu_S}
&\nabla_{\theta}\mathcal{J}(\bm\theta;\eta)=
({1}/{\sigma^2})\mathbf{\Phi}^{\top}(\mathbf{\Phi}\bm\theta-\bm{Y})\nonumber\\
&+(\nu+1)\left[ \begin{array}{ccc} \frac{[\bm\theta]_{1}}{\nu\eta^2+[\bm\theta]_{1}^2} & \cdots & \frac{[\bm\theta]_{n_{\theta}}}{\nu\eta^2+[\bm\theta]_{n_{\theta}}^2} \end{array} \right]^{\top},\\
\label{eq: explicit expression for Hessian}
&\nabla_{\theta\theta}^2\mathcal{J}(\bm\theta;\eta)=(1/\sigma^2)\mathbf{\Phi}^{\top}\mathbf{\Phi}+\nonumber\\
&(\nu+1)\diag\left\{\tfrac{\nu\eta^2-[\bm\theta]_{1}^2}{(\nu\eta^2+[\bm\theta]_{1}^2)^2},\cdots,\tfrac{\nu\eta^2-[\bm\theta]_{n_{\theta}}^2}{(\nu\eta^2+[\bm\theta]_{n_{\theta}}^2)^2} \right\}.
\end{align} 
The implementation of $\hat{\bm\theta}^{\EB}(\hat{\eta}_{\EB})$ is given in Algorithm~\ref{alg:EB_IS_Student_t}.


\begin{algorithm}[!htbp]
\caption{IS implementation of $\hat{\bm\theta}^{\EB}(\hat{\eta}_{\EB})$ using the Student-$t$
weighting family in \eqref{eq: student t weighting family}}
\label{alg:EB_IS_Student_t}

\KwIn{Observed data $(\bm Y,\bm\Phi)$, noise variance $\sigma^{2}$,
number of IS samples
$M_{\mathrm{IS}}^{\EB}$.}

\KwOut{$\hat{\bm\theta}^{\EB}(\hat{\eta}_{\EB})$.}

Compute $\mathbf{\Phi}^{\top}\mathbf{\Phi}$, $\mathbf{\Phi}^{\top}\bm{Y}$ and $\hat{\bm{\theta}}^{\ML}$ in \eqref{eq: def of ML estimator}.
\nllabel{line: basic ML Bayes student t}

Initialize the hyper-parameter $\eta$.
\nllabel{line: hyper-parameter initialization}

\While{the stopping criterion is not satisfied}{

For each candidate $\eta$, compute $\bm\mu_{S}(\eta)$ by minimizing
$\mathcal{J}(\bm\theta;\eta)$ in \eqref{eq: def of mu_S} and using a quasi-Newton method with gradient \eqref{eq: explicit expression for mu_S}.
\nllabel{line:EB_Student_t_mu}

Compute $\bm\Sigma_{S}(\eta)$ using \eqref{eq: def of Sigma_S} and \eqref{eq: explicit expression for Hessian}.
\nllabel{line:EB_Student_t_Sigma}

Construct the Gaussian proposal $q_{S}(\bm\theta;\eta)$ and generate $\{\bm\theta^{(k)}\}_{k=1}^{M_{\mathrm{IS}}^{\EB}}
\sim q_{S}(\bm\theta;\eta)$.
\nllabel{line:EB_Student_t_samples}

\For{$k=1,\ldots,M_{\mathrm{IS}}^{\EB}$}{
    Compute the IS weight $w_{\EB}(\bm\theta^{(k)};\eta)= {p(\bm{Y}|\bm\theta^{(k)})\pi(\bm\theta^{(k)}|\eta)}/{q_{S}(\bm\theta^{(k)};\eta)}$.
    \nllabel{line:EB_Student_t_weight}
}

Evaluate the EB hyper-parameter cost 
\begin{align}\label{eq: approximation of EB hp cost function using student t}
\mathscr{F}_{\EB}(\eta)\approx -\log\left[\tfrac{1}{M_{\text{IS}}^{\EB}}{\sum}_{k}w_{\EB}(\bm\theta^{(k)};\eta)\right].\ \ \
\end{align}
\nllabel{line:EB_Student_t_cost}
}

Reuse the IS samples and IS weights to compute
\begin{align}\label{eq: approximation of EB estimate using student t}
\hat{\bm\theta}^{\EB}(\hat{\eta}_{\EB})\approx {\sum}_{m=1}^{M_{\text{IS}}^{\EB}}\bm\theta^{(m)}\frac{w_{\EB}(\bm\theta^{(m)};\hat{\eta}_{\EB})}{\sum_{k}w_{\EB}(\bm\theta^{(k)};\hat{\eta}_{\EB})}.\ \ 
\end{align}
\nllabel{line:EB_Student_t_estimate}

\end{algorithm}

Let $M_{L}$ denote the number of evaluations of $\mathcal{J}(\bm\theta;\eta)$ in \eqref{eq: def of mu_S} required to compute $\bm{\mu}_{S}(\eta)$ for a fixed $\eta$. The computational costs of $\hat{\bm\theta}^{\EB}(\hat{\eta}_{\EB})$ can be decomposed as
\begin{itemize}
\item[-] computation of basic data-dependent quantities (\emph{Line~\ref{line: basic ML Bayes student t} in Algorithm~\ref{alg:EB_IS_Student_t}}): $\mathcal{O}(Nn_{\theta}^2+n_{\theta}^3)$,
\item[-] $M_{\EB}$ evaluations of the cost $\mathscr{F}_{\EB}(\eta)$ in \eqref{eq: approximation of EB hp cost function using student t} (\emph{Lines~\ref{line: hyper-parameter initialization}--\ref{line:EB_Student_t_cost} in Algorithm~\ref{alg:EB_IS_Student_t}}): $\mathcal{O}(M_{\EB}[M_{L}n_{\theta}^2+n_{\theta}^3+M_{\text{IS}}^{\EB}n_{\theta}^2])$,
\item[-] evaluation of $\hat{\bm\theta}^{\EB}(\hat{\eta}_{\EB})$ in \eqref{eq: approximation of EB estimate using student t} (\emph{Line~\ref{line:EB_Student_t_estimate} in Algorithm~\ref{alg:EB_IS_Student_t}}): $\mathcal{O}(M_{\text{IS}}^{\EB}n_{\theta})$.
\end{itemize}
The computational complexities of these steps are analyzed in Appendix~\ref{subsec: computational complexity of EB using student t}.
The total computational complexity of $\hat{\bm\theta}^{\EB}(\hat{\eta}_{\EB})$ is $\mathcal{O}([N+M_{\EB}M_{L}+M_{\EB}M_{\text{IS}}^{\EB}]n_{\theta}^2+M_{\EB}n_{\theta}^3)$.

\subsubsection{Implementation and computational complexity of Bayes estimate $\hat{\bm\theta}^{\Bayes}(\pi_{\star})$} Here, $\eta_{\tb\star}(\bm\theta)$ in \eqref{eq: def of eta_b_star for theta} satisfies an explicit first-order optimality condition $-\nabla_{\eta}\log(\pi(\bm\theta|\eta))=0$, i.e.,
\begin{align}\label{eq: student_t optimality condition}
(\nu+1)\sum_{k=1}^{n_{\theta}}\Big\{{[\bm\theta]_{k}^2}/{[\nu \eta^2 + [\bm\theta]_{k}^2]}\Big\}=n_{\theta}.
\end{align}
Since the left-hand side of \eqref{eq: student_t optimality condition} is monotonically decreasing in $\eta>0$, we solve it by using the bisection method. We still use the IS method to approximate the Bayes estimate. The implementation follows Algorithm~\ref{alg: bayes_is Gaussian}, except that $\eta_{\tb\star}(\bm\theta)$ is obtained by solving \eqref{eq: student_t optimality condition}, the proposal is replaced by $q_{S}(\bm\theta;\eta_{\tb\star}(\hat{\bm\theta}^{\ML}))$, and $w_{\star}(\bm\theta^{(k)})= p(\bm{Y}|\bm\theta^{(k)})\pi_{\star}(\bm\theta^{(k)})/q_{S}(\bm\theta^{(k)};\eta_{\tb\star}(\hat{\bm\theta}^{\ML}))$. We refer to this implementation as ``the modified Algorithm~\ref{alg: bayes_is Gaussian}''.



 Here, we slightly abuse the notation $M_{\tb}$ and let it denote the number of evaluations of the left-hand side of \eqref{eq: student_t optimality condition} required to compute $\eta_{\tb\star}(\bm\theta)$ for each fixed $\bm\theta$. The computational cost of implementing $\hat{\bm\theta}^{\Bayes}(\pi_{\star})$ can be decomposed as
\begin{itemize}
\item[-] computation of basic data-dependent quantities (\emph{Lines~\ref{line: basic_ML_Bayes_Gaussian}--\ref{line: basic_ML_proposal_Bayes_Gaussian} in modified Algorithm~\ref{alg: bayes_is Gaussian}}): $\mathcal{O}(M_{\tb}n_{\theta}+[N+M_{L}]n_{\theta}^2+n_{\theta}^3)$,
\item[-] computation of IS weights $\{w_{\star}(\bm\theta^{(k)})\}_{k=1}^{M_{\text{IS}}^{\text{B}}}$ (\emph{Lines~\ref{line: computation of proposal}--\ref{line: computation of IS weight} in the modified Algorithm~\ref{alg: bayes_is Gaussian}}): $\mathcal{O}(M_{\text{IS}}^{\text{B}}n_{\theta}^2+M_{\text{IS}}^{\text{B}}M_{\tb}n_{\theta})$,
\item[-] evaluation of $\hat{\bm\theta}^{\Bayes}(\pi_{\star})$ (\emph{Line~\ref{line: approximation of Bayes Gaussian} in the modified Algorithm~\ref{alg: bayes_is Gaussian}}): $\mathcal{O}(M_{\text{IS}}^{\text{B}}n_{\theta})$.
\end{itemize}
Therefore, the total computational complexity of $\hat{\bm\theta}^{\Bayes}(\pi_{\star})$ is $\mathcal{O}(M_{\text{IS}}^{\text{B}}M_{\tb}n_{\theta}+[N+M_{L}+M_{\text{IS}}^{\text{B}}]n_{\theta}^2+n_{\theta}^3)$.

As noted in Appendix~\ref{subsec: computational complexity of EB using student t}, constructing the proposal $q_{S}(\bm\theta;\eta)$ costs $\mathcal{O}(M_{L}n_{\theta}^2+n_{\theta}^3)$. Here, $\hat{\bm\theta}^{\EB}(\hat{\eta}_{\EB})$ requires constructing the proposal $M_{\EB}$ times with complexity $\mathcal{O}(M_{\EB}[M_{L}n_{\theta}^2+n_{\theta}^3])$, whereas $\hat{\bm\theta}^{\Bayes}(\pi_{\star})$ constructs it only once. Hence, the computational advantage of $\hat{\bm\theta}^{\Bayes}(\pi_{\star})$ becomes more evident for a larger $M_{\EB}$. For the remaining computations, $\hat{\bm\theta}^{\Bayes}(\pi_{\star})$ is computationally comparable or more efficient when
\begin{align*}
M_{\text{IS}}^{\text{B}}M_{\tb}n_{\theta}+M_{\text{IS}}^{\text{B}}n_{\theta}^2\lesssim M_{\EB}M_{\text{IS}}^{\EB}n_{\theta}^2.
\end{align*}
A sufficient condition is $M_{\tb}\lesssim n_{\theta}$ and $M_{\text{IS}}^{\text{B}}\lesssim M_{\EB}M_{\text{IS}}^{\EB}$. For other non-conjugate pairs of weighting and likelihood, computing $\hat{\bm\theta}^{\EB}(\hat{\bm\eta}_{\EB})$ often likewise requires repeated construction of $\bm\eta$-dependent proposals. Thus, similar computational advantages of $\hat{\bm\theta}^{\Bayes}(\pi_{\star})$ can also be expected in these settings.




\vspace{-2mm}
\section{Numerical Simulation}\label{sec: numerical simulation}

In this section, we consider the Gaussian and Student-$t$ EB weighting families studied in Section~\ref{sec: implementation and complexity}. For each of them, we compare $\hat{\bm\theta}^{\EB}(\hat{\bm\eta}_{\EB})$ and $\hat{\bm\theta}^{\Bayes}(\pi_{\star})$ in terms of their average performance, computation time, and local sensitivity.

\vspace{-3mm}

\subsection{Gaussian EB weighting family}\label{subsec: simulation Gaussian}

We first consider the Gaussian EB weighting family in \eqref{eq: Gaussian EB weighting families} under the parametrization in \eqref{eq: def of TC kernel}.

\subsubsection{Test data and simulation setup}

We use the test systems, test data, and simulation setup in Example~\ref{example: illustrating example}. We implement $\hat{\bm\theta}^{\EB}(\hat{\bm\eta}_{\EB})$ and $\hat{\bm\theta}^{\Bayes}(\pi_{\star})$ using the approaches in Section~\ref{subsec: Gaussian weighting with TC kernel}. For the EB hyper-parameter estimation, we first perform a coarse grid search over $c$ and $\alpha$ to obtain a good initial point. More specifically, we set $c=\check{c}\|\hat{\bm\theta}^{\ML}\|_{2}^2/n_{\theta}$ with $\check{c}$ selected by the MATLAB command \lstinline[style=Matlab-editor]{logspace(-3, 3, 10)}, and select $\alpha$ by \lstinline[style=Matlab-editor]{linspace(0.5, 0.95, 10)}. With the initial point, we then use
\lstinline[style=Matlab-editor]{fminsearch} for local refinement. Here, $M_{\EB}$ denotes the number of evaluations of $\mathscr{F}_{\EB}(\bm\eta)$ used in both the initial grid search and the \lstinline[style=Matlab-editor]{fminsearch} part. The former requires $100$ evaluations, and the latter uses at most $400$ evaluations by default. Therefore, we have $M_{\EB}\leq 500$. As for $\hat{\bm\theta}^{\Bayes}(\pi_{\star})$, we need to compute $\alpha_{\tb\star}(\bm\theta)$ in \eqref{eq: def of eta_b_star for theta} for each fixed $\bm\theta$, which is directly obtained by a grid search over $\{0.5,0.6,\cdots,0.9\}$. Hence, in this implementation, $M_{\tb}=5$. Moreover, to achieve a satisfactory balance between numerical accuracy and computational cost, we set $M_{\text{IS}}^{\text{B}}=7\times 10^3$.

Besides the shape hyper-parameter perturbation considered in Example~\ref{example: illustrating example}, we also perturb the scale hyper-parameter $c$. In this case, we keep $\hat{\alpha}_{\EB}$ and $\alpha_{\tb\star}(\bm\theta)$ fixed, and set $\log(\tilde{c}_{\EB})=\log(\hat{c}_{\EB})+\delta\log(c)$ and $\log(\tilde{c}_{\tb\star}(\bm\theta))=\log(c_{\tb\star}(\bm\theta))+\delta\log(c)$, where $\delta\log(c)=-1.5,-1.25,\cdots,1.5$. Here, the logarithm scale ensures that $\tilde{c}_{\EB}$ and $\tilde{c}_{\tb\star}(\bm\theta)$ remain positive. 

\subsubsection{Simulation results} 
\begin{figure}[!htbp]
\centering
\includegraphics[width=0.8\linewidth]{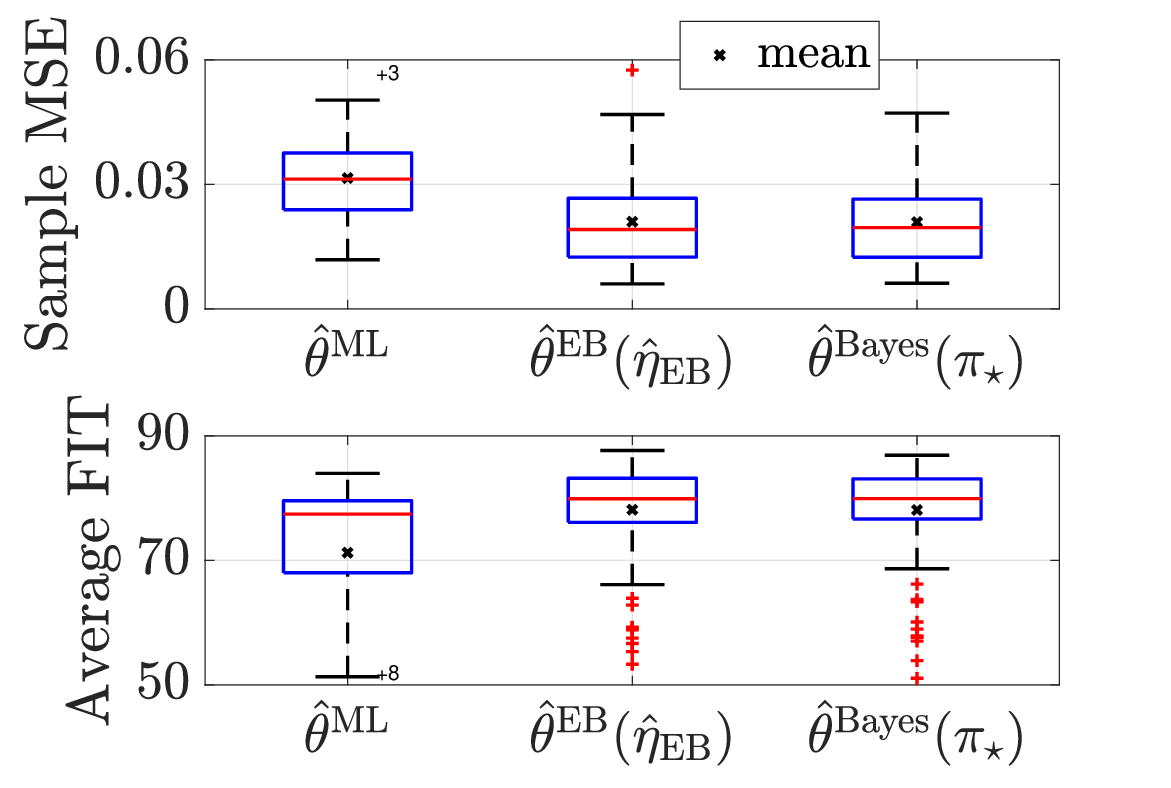}
\caption{Boxplots of the sample MSEs (top panel) and average FITs (bottom panel) of $\hat{\bm\theta}^{\ML}$, $\hat{\bm\theta}^{\EB}(\hat{\bm\eta}_{\EB})$ using the Gaussian EB weighting family in \eqref{eq: Gaussian EB weighting families} under the parametrization in \eqref{eq: def of TC kernel}, and $\hat{\bm\theta}^{\Bayes}(\pi_{\star})$ for $n_{\theta}=20$ and $N=80$.}
\label{fig: FIT_TC_n20N80}
\end{figure}

\begin{figure}[!htbp]
\centering
\begin{subfigure}[b]{0.82\linewidth}
 \includegraphics[width=\linewidth]{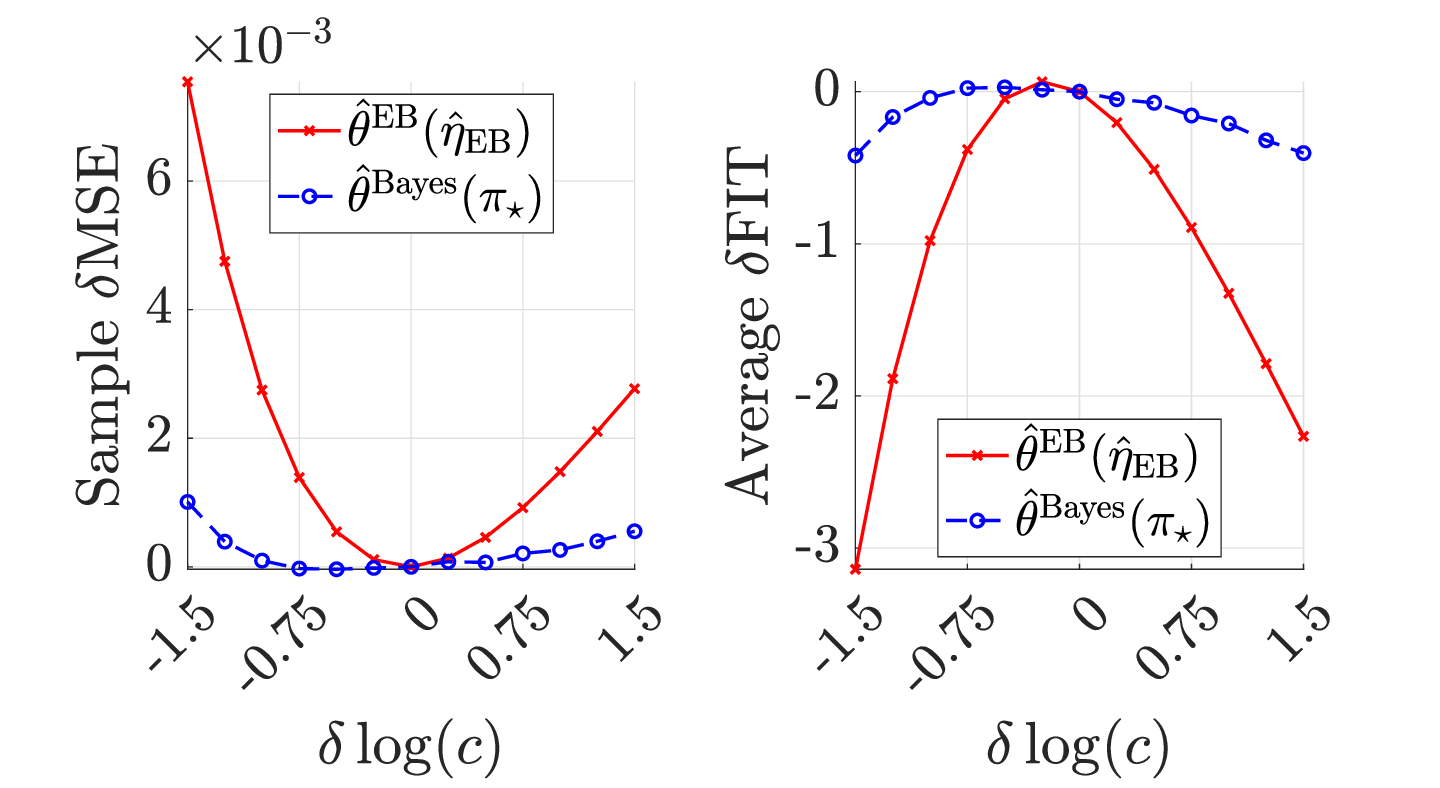}
\caption{}
\label{fig: delta_c TC}
\end{subfigure}
\begin{subfigure}[b]{0.82\linewidth}
 \includegraphics[width=\linewidth]{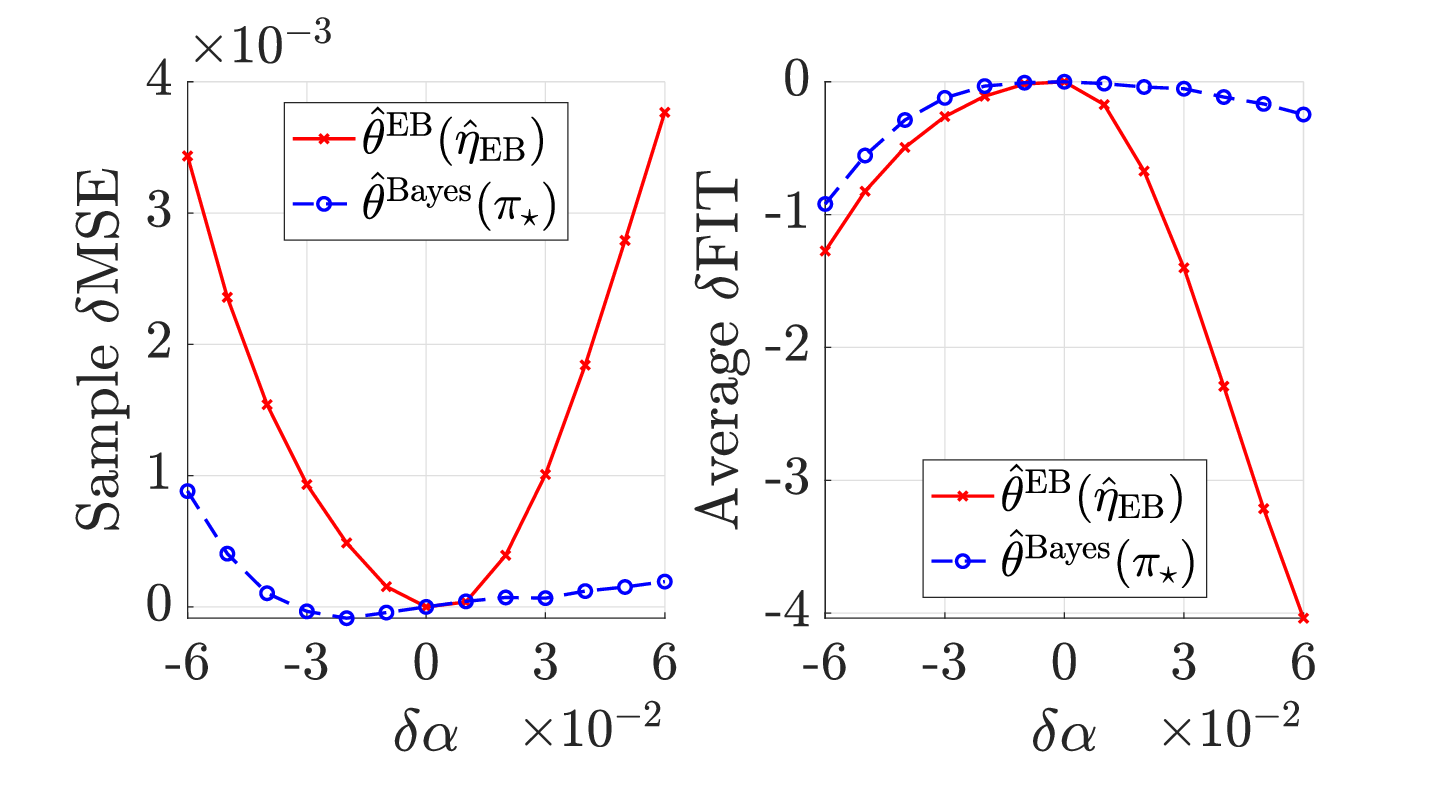}
\caption{}
\label{fig: delta_alpha TC}
\end{subfigure}
\caption{Sample means of sample $\delta$MSE and average $\delta$FIT of $\hat{\bm\theta}^{\EB}(\hat{\bm\eta}_{\EB})$ using the Gaussian EB weighting family in \eqref{eq: Gaussian EB weighting families} under the parametrization in \eqref{eq: def of TC kernel}, and $\hat{\bm\theta}^{\Bayes}(\pi_{\star})$ for $n_{\theta}=20$ and $N=80$. Panel (a): performance changes due to $\delta\log(c)$. Panel (b): performance changes due to $\delta\alpha$. }
\label{fig: perturbation_TC_n20N80}
\end{figure}

As shown in Fig.~\ref{fig: FIT_TC_n20N80} and Table~\ref{table: computation time of EB using Gaussian}, the estimation accuracies of $\hat{\bm\theta}^{\EB}(\hat{\bm\eta}_{\EB})$ and $\hat{\bm\theta}^{\Bayes}(\pi_{\star})$ are quite close, which is consistent with Theorem~\ref{thm: sufficient condition for general EB2 estimator}. The two estimators both perform better than $\hat{\bm\theta}^{\ML}$. Moreover, the computation time of $\hat{\bm\theta}^{\Bayes}(\pi_{\star})$ is comparable to that of $\hat{\bm\theta}^{\EB}(\hat{\bm\eta}_{\EB})$. Fig.~\ref{fig: perturbation_TC_n20N80} shows that, under both scale and shape hyper-parameter perturbations, $\hat{\bm\theta}^{\Bayes}(\pi_{\star})$ exhibits smaller performance degradation than $\hat{\bm\theta}^{\EB}(\hat{\bm\eta}_{\EB})$, which provides empirical support for the local performance robustness suggested by Theorem~\ref{thm: sensitivity analysis}.  

\vspace{-2mm}

\begin{table}[!htbp]
\centering
\caption{Sample means of sample MSEs and average FITs, and computation time of $\hat{\bm\theta}^{\ML}$, $\hat{\bm\theta}^{\EB}(\hat{\bm\eta}_{\EB})$ using the Gaussian EB weighting family in \eqref{eq: Gaussian EB weighting families} under the parametrization in \eqref{eq: def of TC kernel}, and $\hat{\bm\theta}^{\Bayes}(\pi_{\star})$ for $n_{\theta}=20$ and $N=80$.}
\label{table: computation time of EB using Gaussian}
\resizebox{0.86\hsize}{!}{
\begin{tabular}{cccc}
\hline
\centering
estimator & $\hat{\bm\theta}^{\ML}$ & $\hat{\bm\theta}^{\EB}(\hat{\bm\eta}_{\EB})$ & $\hat{\bm\theta}^{\Bayes}(\pi_{\star})$ \\
\hline
sample MSE & $3.15\times 10^{-2}$ & $2.10\times 10^{-2}$ & $2.09\times 10^{-2}$  \\
average FIT & $71.24$ & $78.15$ & $78.10$ \\
\makecell[c]{computation\\ time (s)} & - & $42.83$ & $40.6$ \\
\hline
\end{tabular}}
\end{table}

\vspace{-4mm}

\subsection{Student-t EB weighting family}

We next consider the Student-$t$ EB weighting family in \eqref{eq: student t weighting family}.

\subsubsection{Test data and simulation setup}

We first generate $100$ independent realizations from the distribution with density in \eqref{eq: student t weighting family}, where $\eta=2$, $\nu=3$ and $n_{\theta}=50$. These realizations are denoted by $\{\tilde{\bm\theta}_{0,k}\}_{k=1}^{100}$. For each $\tilde{\bm\theta}_{0,k}$, we generate $N=200$ independent realizations from a zero-mean Gaussian distribution with covariance matrix $\mathbf{\Sigma}$ with $[\mathbf{\Sigma}]_{k,l}=0.8^{|k-l|}$, which form the rows of the regression matrix $\mathbf{\Phi}$ in \eqref{eq: linear regresssion model}. We then set $\sigma^2=1$ and rescale $\bm\theta_{0,k}=m_{k}\tilde{\bm\theta}_{0,k}$ such that the sample SNR (see Footnote~\ref{footnote: sample snr}) is $10$. For each collection of $\bm\theta_{0,k}$ and $\mathbf{\Phi}$, we perform $100$ MC simulations. 

Given the test data collections, we implement $\hat{\bm\theta}^{\EB}(\hat{\eta}_{\EB})$ and $\hat{\bm\theta}^{\Bayes}(\pi_{\star})$ by using the approaches in Section~\ref{subsec: student t weighting family}. The hyper-parameter feasible set is $\mathcal{D}_{\eta}=[\underline{\eta},\overline{\eta}]$, where $\underline{\eta}=10^{-3}$ and $\overline{\eta}=20$. For the EB hyper-parameter estimation, we first perform a grid search over $\{0.1,2,4,\cdots,10\}$ to select a good initial point, and then use \lstinline[style=Matlab-editor]{fminsearch} to obtain $\hat{\eta}_{\EB}$. The former requires $6$ evaluations of $\mathscr{F}_{\EB}(\eta)$ in \eqref{eq: approximation of EB hp cost function using student t}, and the latter requires at most $200$ evaluations by default. Therefore, $M_{\EB}\leq 206$ in this implementation. In addition, we set $M_{\text{IS}}^{\EB}=200$. For the implementation of $\hat{\bm\theta}^{\Bayes}(\pi_{\star})$, $\eta_{\tb\star}(\bm\theta)$ is computed for each fixed $\bm\theta$ using the bisection method. The maximum number of evaluations of the left-hand side of \eqref{eq: student_t optimality condition} is set to $60$, i.e., $M_{\tb}\leq 60$. In addition, we set $M_{\text{IS}}^{\text{B}}=2000$.

For the perturbation simulation, we still apply the logarithm transformation to ensure that the perturbed hyper-parameter $\eta$ remains positive. Specifically, we set $\log(\tilde{\eta}_{\EB})=\log(\hat{\eta}_{\EB})+\delta\log(\eta)$ and $\log(\tilde{\eta}_{\tb\star}(\bm\theta))=\log({\eta}_{\tb\star}(\bm\theta))+\delta\log(\eta)$, where $\delta\log(\eta)=-0.6,-0.5,\cdots,0.6$. 

\subsubsection{Simulation results}

\begin{figure}[!htbp]
\centering
\includegraphics[width=0.8\linewidth]{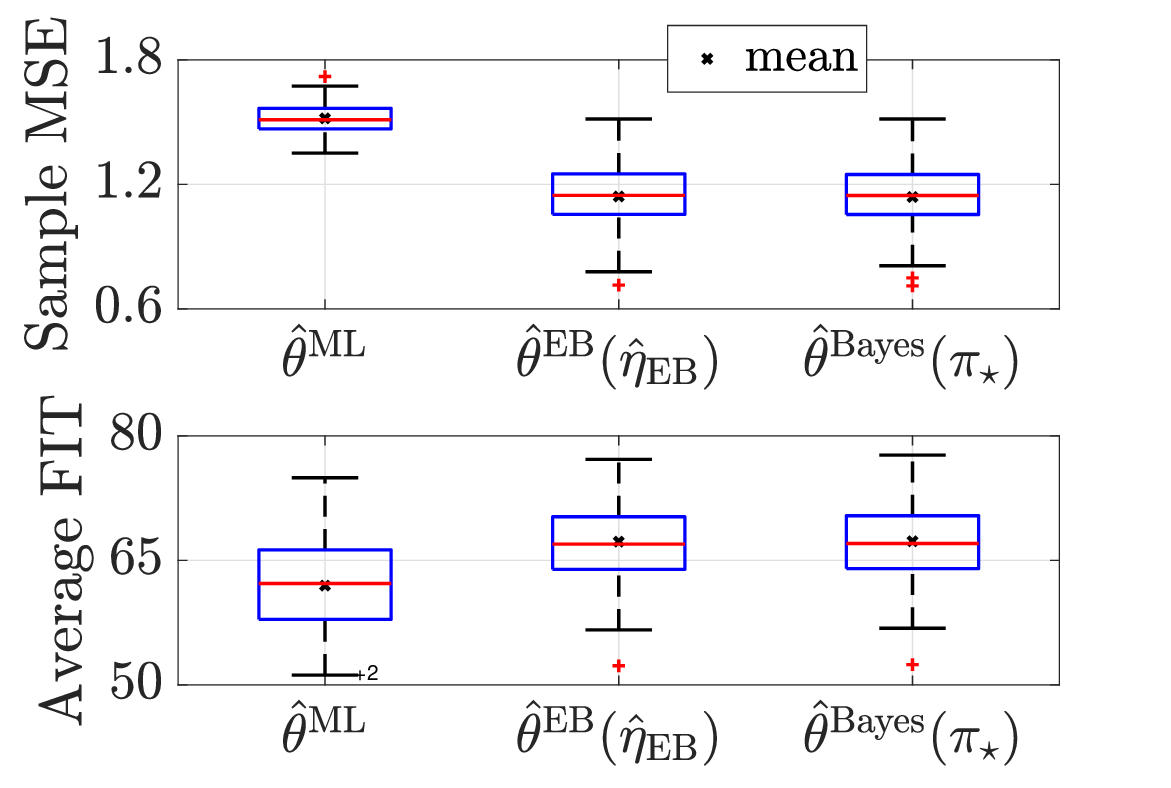}
\caption{Boxplots of the sample MSEs (top panel) and average FITs (bottom panel) of $\hat{\bm\theta}^{\ML}$, $\hat{\bm\theta}^{\EB}(\hat{\eta}_{\EB})$ using the Student-$t$ EB weighting family, and $\hat{\bm\theta}^{\Bayes}(\pi_{\star})$ for $n_{\theta}=50$ and $N=200$.}
\label{fig: FIT_studentt_n50N200}
\end{figure}

\begin{figure}[!htbp]
\centering
 \includegraphics[width=0.8\linewidth]{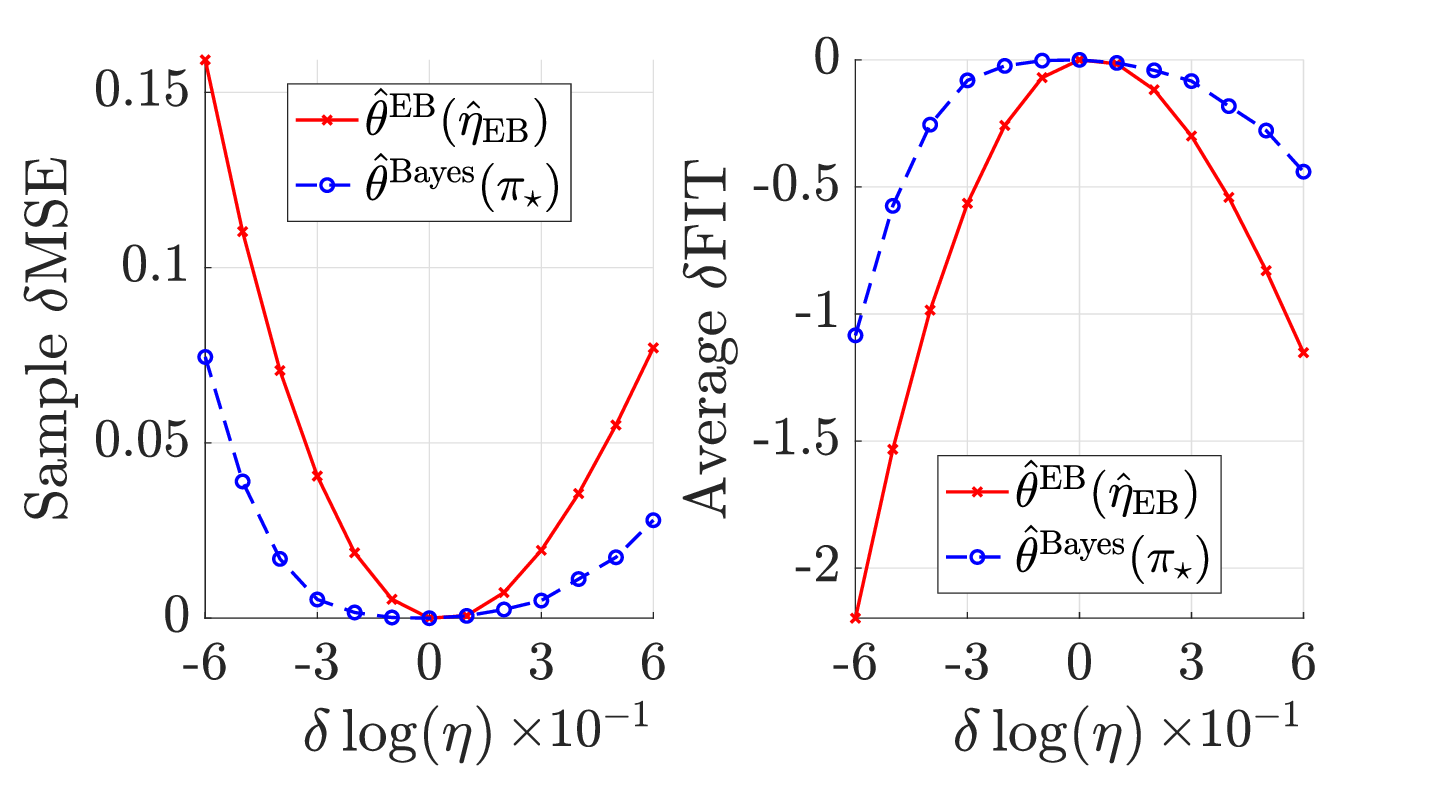}
\caption{Sample means of sample $\delta$MSE and average $\delta$FIT of $\hat{\bm\theta}^{\EB}(\hat{\eta}_{\EB})$ using the Student-$t$ EB weighting family in \eqref{eq: student t weighting family}, and $\hat{\bm\theta}^{\Bayes}(\pi_{\star})$ for $n_{\theta}=50$ and $N=200$.}
\label{fig: FIT_student_t_n50N200}
\end{figure}

As displayed in Fig.~\ref{fig: FIT_studentt_n50N200} and Table~\ref{table: computation time of EB using Student_t}, for the Student-$t$ EB weighting family in \eqref{eq: student t weighting family}, $\hat{\bm\theta}^{\EB}(\hat{\eta}_{\EB})$ and $\hat{\bm\theta}^{\Bayes}(\pi_{\star})$ are comparable in terms of estimation accuracy, and both outperform $\hat{\bm\theta}^{\ML}$. This provides numerical support for Theorem~\ref{thm: sufficient condition for general EB2 estimator}. Table~\ref{table: computation time of EB using Student_t} also shows that,  the computation time of $\hat{\bm\theta}^{\EB}(\hat{\eta}_{\EB})$ is about $12$ times that of $\hat{\bm\theta}^{\Bayes}(\pi_{\star})$, consistent with the analysis in Section~\ref{subsec: student t weighting family}. Moreover, Fig.~\ref{fig: FIT_student_t_n50N200} shows that the average performance of $\hat{\bm\theta}^{\Bayes}(\pi_{\star})$ is locally more robust to hyper-parameter perturbations than that of $\hat{\bm\theta}^{\EB}(\hat{\eta}_{\EB})$. This provides empirical evidence for the performance robustness of $\hat{\bm\theta}^{\Bayes}(\pi_{\star})$ suggested by Theorem~\ref{thm: sensitivity analysis}. 

\vspace{-2mm}

\begin{table}[!htbp]
\centering
\caption{Sample means of sample MSEs and average FITs, and computation time of $\hat{\bm\theta}^{\ML}$, $\hat{\bm\theta}^{\EB}(\hat{\eta}_{\EB})$ using the Student-$t$ EB weighting family in \eqref{eq: student t weighting family}, and $\hat{\bm\theta}^{\Bayes}(\pi_{\star})$ for $n_{\theta}=50$ and $N=200$.}
\label{table: computation time of EB using Student_t}
\resizebox{0.86\hsize}{!}{
\begin{tabular}{cccc}
\hline
\centering
estimator & $\hat{\bm\theta}^{\ML}$ & $\hat{\bm\theta}^{\EB}(\hat{\eta}_{\EB})$ & $\hat{\bm\theta}^{\Bayes}(\pi_{\star})$ \\
\hline
sample MSE & $1.52$ & $1.14$ & $1.14$  \\
average FIT & $61.96$ & $67.26$ & $67.30$ \\
computation time (s) & - & $1.43\times 10^3$ & $1.18\times 10^2$ \\
\hline
\end{tabular}}
\end{table}

\vspace{-2mm}

\section{Conclusion}\label{sec: conclusion}

This work considered the linear regression model and focused on the empirical Bayes (EB) estimator with the hyper-parameter estimated by maximizing the marginal likelihood, simply referred to as the EB estimator. This estimator is often appealing for its favorable estimation accuracy, but it can suffer from local sensitivity to hyper-parameter perturbations.


To reduce this sensitivity, a generalized Bayes estimator was constructed that is locally more robust than the EB estimator and exhibits comparable estimation accuracy for large sample sizes. This estimator was constructed based on the excess mean squared error (XMSE), a second-order asymptotic measure of the MSE difference between the estimator of interest and the maximum likelihood estimator. Specifically, the corresponding Bayes weighting function was explicitly characterized through a pointwise optimization problem such that the XMSE of the resulting generalized Bayes estimator matches that of the EB estimator. Furthermore, the constructed Bayes estimator and the EB estimator were shown to be at most second-order and first-order sensitive to hyper-parameter perturbations, respectively. Their computational complexities were also analyzed, especially for Gaussian and Student-$t$ EB weighting families. Under certain conditions, the constructed Bayes estimator can be computationally comparable to, or even more efficient than, the EB estimator. These theoretical findings were supported by numerical simulations.

A potential extension of this work is to construct generalized Bayes estimators that outperform the EB estimator. This was explored in \cite{JHWH26letter}, but only for ridge regression and the identity Gram matrix limit. Under general settings and following the XMSE-based idea of this work, requiring XMSE improvement over the EB estimator leads to a nonlinear partial differential equation. Generally, it does not admit an explicit solution and requires numerical methods. To make this construction more tractable, a possible strategy is to utilize the generalized Bayes estimator constructed in this work as a baseline and introduce additional shrinkage through its weighting function.

\renewcommand{\thesection}{Appendix}
\renewcommand{\thesubsection}{\Alph{subsection}}

\setcounter{subsection}{0}

\renewcommand{\theequation}{\thesubsection.\arabic{equation}}
\renewcommand{\thetheorem}{\thesubsection.\arabic{theorem}}

\setcounter{equation}{0}
\setcounter{theorem}{0}

\numberwithin{equation}{subsection}
\numberwithin{theorem}{subsection}

\section*{Appendix}\label{sec:Appendix A}


\subsection{Proof of Lemma~\ref{lemma: expression of EB estimator using MSS}}\label{subsec: proof of functions of MSS}

We first note that \eqref{eq: rewritten expression for EB estimator} is derived in \cite[(A.3)]{JCWH26}. We next recall from \cite[Proof of Theorem~2]{JCWH26} that
\begin{align}\label{eq: reformulation of likelihood}
p(\bm{Y}|\bm\theta)=C_{N}(\hat{\bm\theta}^{\ML})p(\bm\theta|\hat{\bm\theta}^{\ML}).
\end{align}
Substituting \eqref{eq: reformulation of likelihood} into $\mathscr{F}_{\EB}(\bm\eta)$ in \eqref{eq: def of EB hyperparameter estimator} yields \eqref{eq: rewritten expression for F_EB}. Therefore, $\hat{\bm\theta}^{\EB}(\hat{\bm\eta}_{\EB})$ can be expressed solely as a function of $\hat{\bm\theta}^{\ML}$.

\subsection{Proof of Theorem~\ref{thm: XMSE of EB2 estimator}}

To make the dependence of $\hat{\bm\eta}_{\EB}$, defined in \eqref{eq: def of EB hyperparameter estimator}, on $\hat{\bm\theta}^{\ML}$ more explicit, 
we write it as $\hat{\bm\eta}_{\EB}(\hat{\bm\theta}^{\ML})$ throughout this proof. For notational convenience, we let $\mathcal{C}>0$ denote a generic constant independent of $N$ and $\bm\eta$ in this proof.

As implied by \cite[Theorem~2]{JCWH26}, the XMSE expression for $\hat{\bm\theta}^{\EB}(\hat{\bm\eta}_{\EB})$ essentially relies on the two limits in \cite[Assumption~3]{JCWH26}, i.e., $\lim_{N\to\infty}\hat{\bm\eta}_{\EB}(\bm\theta_{0})$ and $\lim_{N\to\infty}\nabla_{\theta_{0}}\hat{\bm\eta}_{\EB}(\bm\theta_{0})$. So, our next step is to derive these two limits, respectively.

\paragraph{The limit of $\hat{\bm\eta}_{\EB}(\bm\theta_{0})$} Since the term $-\log[C_{N}(\hat{\bm\theta}^{\ML})]$ in \eqref{eq: rewritten expression for F_EB} is independent of $\bm\eta$, we have
\begin{align}\label{eq: rewritten minimizer}
\hat{\bm\eta}_{\EB}(\bm\theta_{0})={\argmin}_{\bm\eta\in\mathcal{D}_{\eta}}-\log\{\E_{\bm\theta|\bm\theta_{0}}[\pi(\bm\theta|\bm\eta)] \},
\end{align}
with $\E_{\bm\theta|\bm\theta_{0}}[\cdot]$ w.r.t. the density of $\bm\theta|\bm\theta_{0}\sim\mathcal{N}(\bm\theta_{0},\sigma^2(\mathbf{\Phi}^{\top}\mathbf{\Phi})^{-1})$. In what follows, we will first prove the uniform convergence of $-\log\{\E_{\bm\theta|\bm\theta_{0}}[\pi(\bm\theta|\bm\eta)] \}$ to $-\log[\pi(\bm\theta_{0}|\bm\eta)]$ for $\bm\eta\in\mathcal{D}_{\eta}$, and then apply the argmin consistency theorem, e.g., \cite[Theorem~8.2]{Ljung1999}, to prove the limit of minimizers
\begin{align}\label{eq: consistency of EB hyperparameter estimator}
\lim_{N\to\infty}\hat{\bm\eta}_{\EB}(\bm\theta_{0})=\bm\eta_{\tb\star}(\bm\theta_{0}),
\end{align}
where $\bm\eta_{\tb\star}(\bm\theta_{0})$ is defined in \eqref{eq: def of eta_b_star for theta}. Analogously to \cite[(A.4)]{JCWH26}, taking a second-order Taylor expansion of $\pi(\bm\theta|\bm\eta)$ around $\bm\theta=\bm\theta_{0}$ and then taking expectation w.r.t. $p(\bm\theta|\bm\theta_{0})$ yields
\begin{align}
&\pi(\bm\theta|\bm\eta)=\pi(\bm\theta_{0}|\bm\eta)+(\nabla_{\theta}\pi(\bm\theta|\bm\eta)|_{\bm\theta=\bm\theta_{0}})^{\top}(\bm\theta-\bm\theta_{0})\\
\label{eq: taylor expansion}
&+\tfrac{1}{2}(\bm\theta-\bm\theta_{0})^{\top}\nabla_{\theta\theta}^2\pi(\bm\theta|\bm\eta)|_{\bm\theta=\bm\theta_{0}}(\bm\theta-\bm\theta_{0})+\rho_{N}(\bm\theta,\bm\eta),\\
&\E_{\bm\theta|\bm\theta_{0}}[\pi(\bm\theta|\bm\eta)]=\pi(\bm\theta_{0}|\bm\eta)\\
&+\tfrac{\sigma^2}{2}\Tr[\nabla_{\theta\theta}^2\pi(\bm\theta|\bm\eta)|_{\bm\theta=\bm\theta_{0}}(\mathbf{\Phi}^{\top}\mathbf{\Phi})^{-1}]+\E_{\bm\theta|\bm\theta_{0}}[\rho_{N}(\bm\theta,\bm\eta)],\nonumber
\end{align}
where $\rho_{N}(\bm\theta,\bm\eta)$ denotes the remainder term. It follows that
\begin{align}
&\sup_{\bm\eta\in\mathcal{D}_{\eta}}\left|\E_{\bm\theta|\bm\theta_{0}}[\pi(\bm\theta|\bm\eta)]-\pi(\bm\theta_{0}|\bm\eta)\right|
\leq \sup_{\bm\eta\in\mathcal{D}_{\eta}}\left|\E_{\bm\theta|\bm\theta_{0}}[\rho_{N}(\bm\theta,\bm\eta)]\right|\\
\label{eq: uniform convergence of weighting difference}
& +({n_{\theta}\sigma^2}/{2})\|(\mathbf{\Phi}^{\top}\mathbf{\Phi})^{-1}\|_{2}\sup_{\bm\eta\in\mathcal{D}_{\eta}}\|\nabla_{\theta\theta}^2\pi(\bm\theta|\bm\eta)|_{\bm\theta=\bm\theta_{0}}\|_{2},
\end{align}
where we use the norm inequality $|\Tr(\mathbf{A}\mathbf{B})|\leq n_{\theta}\|\mathbf{A}\|_{2}\|\mathbf{B}\|_{2}$ for $\mathbf{A},\mathbf{B}\in\R^{n_{\theta}\times n_{\theta}}$. By the compactness of $\mathcal{D}_{\eta}$ and Item~$2)$ in Assumption~\ref{asp: regularity conditions for EB hyperparameter estimator}, there exists a neighborhood of $\{\bm\theta_{0}\}\times \mathcal{D}_{\eta}$ such that the third-order derivative of $\pi(\bm\theta|\bm\eta)$ w.r.t. $\bm\theta$ is uniformly bounded. Thus, there exists $\mathcal{C}>0$ such that for $\|\bm{z}_{N}\|_{2}\leq r$ with $\bm{z}_{N}\coloneqq \bm\theta-\bm\theta_{0}$ and sufficiently small $r>0$, we have
\begin{align}\label{eq: boundedness of local part}
{\sup}_{\bm\eta\in\mathcal{D}_{\eta}}|\rho_{N}(\bm\theta,\bm\eta)|\leq \mathcal{C}\|\bm{z}_{N}\|_{2}^3.
\end{align}
 Split $\E_{\bm\theta|\bm\theta_{0}}[\rho_{N}(\bm\theta,\bm\eta)]$ into a local part and a tail part,
\begin{align}
&\sup_{\bm\eta\in\mathcal{D}_{\eta}}|\E_{\bm\theta|\bm\theta_{0}}[\rho_{N}(\bm\theta,\bm\eta)]|
\leq \sup_{\bm\eta\in\mathcal{D}_{\eta}}\E_{\bm\theta|\bm\theta_{0}}[|\rho_{N}(\bm\theta,\bm\eta)|\bm{1}_{\{\|\bm{z}_{N}\|_{2}\leq r\}}]\nonumber\\
\label{eq: decomposition of remainder term}
&\qquad\qquad +{\sup}_{\bm\eta\in\mathcal{D}_{\eta}}\E_{\bm\theta|\bm\theta_{0}}[|\rho_{N}(\bm\theta,\bm\eta)|\bm{1}_{\{\|\bm{z}_{N}\|_{2}> r\}}].
\end{align}
For the local part, we derive from \eqref{eq: boundedness of local part} and Assumption~\ref{asp: limit of regression matrix} that, there exist constants $\mathcal{C},\tilde{\mathcal{C}}>0$ such that
\begin{align}
&{\sup}_{\bm\eta\in\mathcal{D}_{\eta}}\E_{\bm\theta|\bm\theta_{0}}[|\rho_{N}(\bm\theta,\bm\eta)|\bm{1}_{\{\|\bm{z}_{N}\|_{2}\leq r\}}]\\
\leq& \mathcal{C}\E_{\bm\theta|\bm\theta_{0}}[\|\bm{z}_{N}\|_{2}^3]
\leq \tilde{\mathcal{C}}\|(\mathbf{\Phi}^{\top}\mathbf{\Phi})^{-1}\|_{2}^{3/2}\to 0,\ \text{as}\ N\to\infty.
\end{align}
As for the tail part, we derive from \eqref{eq: taylor expansion} that 
\begin{align}
\rho_{N}(\bm\theta,\eta)=&\pi(\bm\theta|\bm\eta)-\pi(\bm\theta_{0}|\bm\eta)-(\nabla_{\theta}\pi(\bm\theta|\bm\eta)|_{\bm{z}_{N}=\bm{0}})^{\top}\bm{z}_{N}\\
&-(1/2)\bm{z}_{N}^{\top}\nabla_{\theta\theta}^2\pi(\bm\theta|\bm\eta)|_{\bm\theta=\bm\theta_{0}}\bm{z}_{N}.
\end{align}
By Items $2)$--$3)$ in Assumption~\ref{asp: regularity conditions for EB hyperparameter estimator}, there exists $\mathcal{C}>0$ such that
\begin{align}\label{eq: superium of remainder term}
{\sup}_{\bm\eta\in\mathcal{D}_{\eta}}|\rho_{N}(\bm\theta,\eta)|\leq {\mathcal{C}}\left[\exp({a}\|\bm\theta\|_{2}^2)+\|\bm{z}_{N}\|_{2}^2\right],
\end{align}
where we use $\sup_{\bm\eta\in\mathcal{D}_{\eta}}\pi(\bm\theta|\bm\eta)\leq \mathcal{C}\exp(a\|\bm\theta\|_{2}^2)$. Note that
\begin{align}
&\E_{\bm\theta|\bm\theta_{0}}[\exp({a}\|\bm\theta\|_{2}^2)\bm{1}_{\{\|\bm{z}_{N}\|_{2}> r\}}]\\
&\leq  \{\E_{\bm\theta|\bm\theta_{0}}[\exp(2a\|\bm\theta\|_{2}^2)]\}^{1/2}\{\E_{\bm\theta|\bm\theta_{0}}[\bm{1}_{\{\|\bm{z}_{N}\|_{2}>r\}}]\}^{1/2}\to 0,\\
&\E_{\bm\theta|\bm\theta_{0}}[\|\bm{z}_{N}\|_{2}^2 \bm{1}_{\{\|\bm{z}_{N}\|_{2}> r\}}]\\
&\leq \E_{\bm\theta|\bm\theta_{0}}[\|\bm{z}_{N}\|_{2}^2]=\sigma^2\Tr[(\mathbf{\Phi}^{\top}\mathbf{\Phi})^{-1}]\to 0,
\end{align}
where for the first inequality we utilize the Cauchy-Schwarz inequality, the boundedness of the Gaussian exponential moments of $\bm\theta$ for sufficiently large $N$, and $\E_{\bm\theta|\bm\theta_{0}}[\bm{1}_{\|\bm{z}_{N}\|_{2}>r}]\to 0$ as $N\to\infty$. For the second inequality, we use the property $\bm{1}_{\{\|\bm{z}_{N}\|_{2}>r\}}\leq 1$ and Assumption~\ref{asp: limit of regression matrix}. By applying these two inequalities and \eqref{eq: superium of remainder term}, we have
\begin{align}
&{\sup}_{\bm\eta\in\mathcal{D}_{\eta}}\E_{\bm\theta|\bm\theta_{0}}[|\rho_{N}(\bm\theta,\bm\eta)|\bm{1}_{\{\|\bm{z}_{N}\|_{2}> r\}}]\\
\leq & \E_{\bm\theta|\bm\theta_{0}}\left[ {\sup}_{\bm\eta\in\mathcal{D}_{\eta}}|\rho_{N}(\bm\theta,\eta)|\bm{1}_{\{\|\bm{z}_{N}\|_{2}>r\}} \right]\to 0.
\end{align}
Substituting the uniform convergence results of the local and tail parts into \eqref{eq: decomposition of remainder term} yields $\sup_{\bm\eta\in\mathcal{D}_{\eta}}\left| \E_{\bm\theta|\bm\theta_{0}}[\rho_{N}(\bm\theta,\bm\eta)] \right|\to 0$ as $N\to\infty$. We then combine this uniform convergence, $\|(\mathbf{\Phi}^{\top}\mathbf{\Phi})^{-1}\|_{2}\to0$ by Assumption~\ref{asp: limit of regression matrix}, the boundedness of $\sup_{\bm\eta\in\mathcal{D}_{\eta}}\|\nabla_{\theta\theta}^2\pi(\bm\theta|\bm\eta)|_{\bm\theta=\bm\theta_{0}}\|_{2}$ by Item~$2)$ in Assumption~\ref{asp: regularity conditions for EB hyperparameter estimator} and the inequality in \eqref{eq: uniform convergence of weighting difference} to derive
\begin{align}\label{eq: uniform convergence of pi}
{\sup}_{\bm\eta\in\mathcal{D}_{\eta}}\left|\E_{\bm\theta|\bm\theta_{0}}[\pi(\bm\theta|\bm\eta)]-\pi(\bm\theta_{0}|\bm\eta)\right|\to 0.
\end{align}
Based on the mean-value theorem for a logarithm function, we have $\log(x)-\log(y)=(x-y)/\zeta$, where $\zeta$ lies between $x,y>0$. It yields $|\log(x)-\log(y)|=|x-y|/\zeta\leq |x-y|/\min\{x,y\}$. We then use this logarithm inequality and \eqref{eq: uniform convergence of pi} to derive
\begin{gather}
{\sup}_{\bm\eta\in\mathcal{D}_{\eta}}\left|-\log\{\E_{\bm\theta|\bm\theta_{0}}[\pi(\bm\theta|\bm\eta)] \}+\log[\pi(\bm\theta_{0}|\bm\eta)] \right|\\
\label{eq: uniform convergence of log weighting}
\leq \frac{\sup_{\bm\eta\in\mathcal{D}_{\eta}}\left|\E_{\bm\theta|\bm\theta_{0}}[\pi(\bm\theta|\bm\eta)]-\pi(\bm\theta_{0}|\bm\eta)\right|}{\inf_{\bm\eta\in\mathcal{D}_{\eta}}\min\{\E_{\bm\theta|\bm\theta_{0}}[\pi(\bm\theta|\bm\eta)], \pi(\bm\theta_{0}|\bm\eta)\}}\to 0.
\end{gather}
Finally, we utilize the argmin consistency theorem, e.g., \cite[Theorem~8.2]{Ljung1999}, the uniform convergence in \eqref{eq: uniform convergence of log weighting}, together with the compactness of $\mathcal{D}_{\eta}$ and the uniqueness of $\bm\eta_{\tb\star}(\bm\theta_{0})$ in Assumption~\ref{asp: regularity conditions for EB hyperparameter estimator}, to obtain \eqref{eq: consistency of EB hyperparameter estimator}.

\paragraph{The limit of $\nabla_{\theta_{0}}\hat{\bm\eta}_{\EB}(\bm\theta_{0})$} As noted in Item~$1)$ of Assumption~\ref{asp: regularity conditions for EB hyperparameter estimator}, $\bm\eta_{\tb\star}(\bm\theta_{0})$ lies in the interior of $\mathcal{D}_{\eta}$. The consistency in \eqref{eq: consistency of EB hyperparameter estimator} implies that, for sufficiently large $N$, $\hat{\bm\eta}_{\EB}(\bm\theta_{0})$ is also an interior minimizer of $-\log\{\E_{\bm\theta|\bm\theta_{0}}[\pi(\bm\theta|\bm\eta)] \}$ in \eqref{eq: rewritten minimizer}. Hence, for sufficiently large sample sizes, it satisfies the first-order optimality condition, i.e., 
\begin{align}\label{eq: first order optimality condition}
\left.\nabla_{\eta}\log\{\E_{\bm\theta|\bm\theta_{0}}[\pi(\bm\theta|\bm\eta)]\}\right|_{\bm\eta=\hat{\bm\eta}_{\EB}(\bm\theta_{0})}=\bm{0}.
\end{align}
It follows that 
\begin{gather}
\nabla_{\eta\theta_{0}}^2\log\{\E_{\bm\theta|\bm\theta_{0}}[\pi(\bm\theta|\bm\eta)]\}=\tfrac{1}{\E_{\bm\theta|\bm\theta_{0}}[\pi(\bm\theta|\bm\eta)]}\nabla_{\eta\theta_{0}}^2\E_{\bm\theta|\bm\theta_{0}}[\pi(\bm\theta|\bm\eta)],\nonumber\\
\nabla_{\eta\eta}^2\log\{\E_{\bm\theta|\bm\theta_{0}}[\pi(\bm\theta|\bm\eta)]\}=\tfrac{1}{\E_{\bm\theta|\bm\theta_{0}}[\pi(\bm\theta|\bm\eta)]}\nabla_{\eta\eta}^2\E_{\bm\theta|\bm\theta_{0}}[\pi(\bm\theta|\bm\eta)]\nonumber
\end{gather}
with $\bm\eta=\hat{\bm\eta}_{\EB}(\bm\theta_{0})$. Taking the full derivative of \eqref{eq: first order optimality condition} w.r.t. $\bm\theta_{0}$ and using the two derivative equalities above yield
\begin{gather}
\left.\nabla_{\eta\theta_{0}}^2\E_{\bm\theta|\bm\theta_{0}}[\pi(\bm\theta|\bm\eta)]\right|_{\bm\eta=\hat{\bm\eta}_{\EB}(\bm\theta_{0})}\\
\label{eq: full derivative}
+\left.\nabla_{\eta\eta}^2\E_{\bm\theta|\bm\theta_{0}}[\pi(\bm\theta|\bm\eta)]\right|_{\bm\eta=\hat{\bm\eta}_{\EB}(\bm\theta_{0})}\nabla_{\theta_{0}}\hat{\bm\eta}_{\EB}(\bm\theta_{0})=\bm{0}.\quad
\end{gather}
Analogous to \eqref{eq: uniform convergence of weighting difference}, we can also derive
\begin{align}\label{eq: uniform convergence result of Hessian eta_theta}
&\sup_{\bm\eta\in\mathcal{D}_{\eta}}\left\|\nabla_{\eta\theta_{0}}^2\E_{\bm\theta|\bm\theta_{0}}[\pi(\bm\theta|\bm\eta)]-\nabla_{\eta\theta}^2\pi(\bm\theta|\bm\eta)|_{\bm\theta=\bm\theta_{0}}\right\|_{2}\to 0,\quad\ \  \\
\label{eq: uniform convergence result of Hessian eat_eta}
&\sup_{\bm\eta\in\mathcal{D}_{\eta}}\left\|\nabla_{\eta\eta}^2\E_{\bm\theta|\bm\theta_{0}}[\pi(\bm\theta|\bm\eta)]-\nabla_{\eta\eta}^2\pi(\bm\theta|\bm\eta)|_{\bm\theta=\bm\theta_{0}}\right\|_{2}\to 0,\quad
\end{align}
as $N\to\infty$. By the negative definiteness of $\nabla_{\eta\eta}^2\log[\pi(\bm\theta_{0}|\bm\eta)]$ with $\bm\eta$ evaluated at $\bm\eta_{\tb\star}(\bm\theta_{0})$ in Assumption~\ref{asp: regularity conditions for EB hyperparameter estimator}, and the uniform convergence in \eqref{eq: uniform convergence result of Hessian eat_eta}, we know that $\nabla_{\eta\eta}^2\log\{\E_{\bm\theta|\bm\theta_{0}}[\pi(\bm\theta|\bm\eta)]\}$ with $\bm\eta$ evaluated at $\hat{\bm\eta}_{\EB}(\bm\theta_{0})$ is nonsingular for sufficiently large $N$. Then, we rewrite \eqref{eq: full derivative} as
\begin{align}
\nabla_{\theta_{0}}\hat{\bm\eta}_{\EB}(\bm\theta_{0})=&-\{ \left.\nabla_{\eta\eta}^2\log\{\E_{\bm\theta|\bm\theta_{0}}[\pi(\bm\theta|\bm\eta)]\}\right|_{\bm\eta=\hat{\bm\eta}_{\EB}(\bm\theta_{0})}\}^{-1}\nonumber\\
\times &\left.\nabla_{\eta\theta_{0}}^2\log\{\E_{\bm\theta|\bm\theta_{0}}[\pi(\bm\theta|\bm\eta)]\}\right|_{\bm\eta=\hat{\bm\eta}_{\EB}(\bm\theta_{0})}.
\end{align}
Together with \eqref{eq: consistency of EB hyperparameter estimator} and \eqref{eq: uniform convergence result of Hessian eta_theta}--\eqref{eq: uniform convergence result of Hessian eat_eta}, we derive 
\begin{align}
\lim_{N\to\infty}&\nabla_{\bm\theta_{0}}\hat{\bm\eta}_{\EB}(\bm\theta_{0})=
-\left\{\nabla_{\eta\eta}^2\log[\pi(\bm\theta|\bm\eta)]\right\}^{-1}\\
\label{eq: limit of gradient of EB hyperparameter estimator}
&\times \left.\nabla_{\eta\theta}^2\log[\pi(\bm\theta|\bm\eta)]\right|_{\bm\theta=\bm\theta_{0},\bm\eta=\bm\eta_{\tb\star}(\bm\theta_{0})}.
\end{align}

Lastly, we substitute \eqref{eq: consistency of EB hyperparameter estimator} and \eqref{eq: limit of gradient of EB hyperparameter estimator} into \cite[Theorem~2]{JCWH26} to derive the expression for $\XMSE(\hat{\bm\theta}^{\EB}(\hat{\bm\eta}_{\EB}))$.

\subsection{Proof of Theorem~\ref{thm: sufficient condition for general EB2 estimator}}\label{subsec: derivation of weighting function}

We first establish the regularity of $\bm\eta_{\tb\star}(\bm\theta)$ defined in \eqref{eq: def of eta_b_star for theta}. By Item~$1)$ in Assumption~\ref{asp: regularity conditions for EB hyperparameter estimator}, $\bm\eta_{\tb\star}(\bm\theta_{0})$ is an interior minimizer of $-\log[\pi(\bm\theta_{0}|\bm\eta)]$ and thus satisfies the optimality condition $\nabla_{\eta}\log[\pi(\bm\theta_{0}|\bm\eta)]|_{\bm\eta=\bm\eta_{\tb\star}(\bm\theta_{0})}=\bm{0}$.
Moreover, we derive from Item~$1)$ in Assumption~\ref{asp: regularity conditions for EB hyperparameter estimator} that $\nabla_{\eta\eta}^2\log[\pi(\bm\theta_{0}|\bm\eta)]|_{\bm\eta=\bm\eta_{\tb\star}(\bm\theta_{0})}\prec 0$. By Item~$2)$ in Assumption~\ref{asp: regularity conditions for EB hyperparameter estimator}, $\log[\pi(\bm\theta|\bm\eta)]$ is jointly three times continuously differentiable and $\nabla_{\eta}\log[\pi(\bm\theta|\bm\eta)]$ is jointly twice continuously differentiable in an open neighborhood of $\{\bm\theta_{0}\}\times \mathcal{D}_{\eta}$. By using the implicit function theorem \cite[Theorem~9.28]{Rudin1976}, we then derive that there exist neighborhoods $\mathcal{U}(\bm\theta_{0})$ and $\mathcal{V}(\bm\eta_{\tb\star}(\bm\theta_{0}))$, together with a unique and twice continuously differentiable function $\bm{g}:\mathcal{U}(\bm\theta_{0})\to\mathcal{V}(\bm\eta_{\tb\star}(\bm\theta_{0}))$ such that $\bm{g}(\bm\theta_{0})=\bm\eta_{\tb\star}(\bm\theta_{0})$ and $\nabla_{\eta}\log[\pi(\bm\theta|\bm\eta)]|_{\bm\eta=\bm{g}(\bm\theta)}=\bm{0}$ for $\bm\theta\in\mathcal{U}(\bm\theta_{0})$.
Moreover, the uniqueness given by the implicit function theorem, together with Item~$1)$ in Assumption~\ref{asp: regularity conditions for EB hyperparameter estimator}, implies that $\bm{g}(\bm\theta)=\bm\eta_{\tb\star}(\bm\theta)\ \text{for}\ \bm\theta\in\mathcal{U}(\bm\theta_{0})$. Therefore, $\bm\eta_{\tb\star}(\bm\theta)$ is unique, interior, and twice continuously differentiable in $\mathcal{U}(\bm\theta_{0})$, and satisfies 
\begin{align}\label{eq: optimality condition for eta_star_theta}
\nabla_{\eta}\log[\pi(\bm\theta|\bm\eta)]|_{\bm\eta=\bm\eta_{\tb\star}(\bm\theta)}=\bm{0}\ \text{for}\ \bm\theta\in\mathcal{U}(\bm\theta_{0}).
\end{align}

We next show that $\pi_{\star}(\bm\theta)$ in \eqref{eq: characterization of weighting function} satisfies Assumption~\ref{asp: differentiablity and growth rate conditions} and the XMSE of $\hat{\bm\theta}^{\Bayes}(\pi_{\star})$ is well-defined. 

\paragraph{Verification of Item~$1)$ in Assumption~\ref{asp: differentiablity and growth rate conditions}} First, the positivity of $\pi_{\star}(\bm\theta)$ in \eqref{eq: characterization of weighting function} just follows from $\pi(\bm\theta|\bm\eta)>0$ in $\mathcal{U}(\bm\theta_{0})$ by Item~$2)$ in Assumption~\ref{asp: regularity conditions for EB hyperparameter estimator}. Then, we apply the chain rule
\begin{align}
\nabla_{\theta}\log\left[\pi_{\star}(\bm\theta)\right]=&\left.\left\{\nabla_{\theta}\log[\pi(\bm\theta|\bm\eta)]\right\}\right|_{\bm\eta=\bm\eta_{\tb\star}(\bm\theta)}\\
&+\left.[\nabla_{\theta}\bm\eta_{\tb\star}(\bm\theta)]^{\top}\nabla_{\eta}\log[\pi(\bm\theta|\bm\eta)]\right|_{\bm\eta=\bm\eta_{\tb\star}(\bm\theta)}.\nonumber
\end{align}
By the first-order optimality condition in \eqref{eq: optimality condition for eta_star_theta}, we have
\begin{align}
\nabla_{\theta}\log\left[\pi_{\star}(\bm\theta)\right]
\label{eq: sufficient condition on general EB2 estimator}
=&\left.\left\{\nabla_{\theta}\log\left[\pi(\bm\theta|\bm\eta) \right]\right\}\right|_{\bm\eta=\bm\eta_{\tb\star}(\bm\theta)}.
\end{align}
Because $\nabla_{\theta}\log[\pi(\bm\theta|\bm\eta)]$ is jointly twice continuously differentiable and $\bm\eta_{\tb\star}(\bm\theta)$ is twice continuously differentiable in $\mathcal{U}(\bm\theta_{0})$, the right-hand side of \eqref{eq: sufficient condition on general EB2 estimator} is also twice continuously differentiable in $\mathcal{U}(\bm\theta_{0})$. Therefore, we derive from \eqref{eq: sufficient condition on general EB2 estimator} that $\log\left[\pi_{\star}(\bm\theta)\right]$ is three times continuously differentiable in $\mathcal{U}(\bm\theta_{0})$.

\paragraph{Verification of Item~$2)$ in Assumption~\ref{asp: differentiablity and growth rate conditions}} By Item~$3)$ in Assumption~\ref{asp: regularity conditions for EB hyperparameter estimator}, there exist $\tilde{C},C,a>0$ such that $\pi_{\star}(\bm\theta)=\tilde{C}\pi(\bm\theta|\bm\eta_{\tb\star}(\bm\theta))\leq \tilde{C}\sup_{\bm\eta\in\mathcal{D}_{\eta}}\pi(\bm\theta|\bm\eta)\leq {C}\exp(a\|\bm\theta\|_{2}^2)$.

Finally, we prove \eqref{eq: XMSE matching condition} by matching the XMSE components of $\hat{\bm\theta}^{\EB}(\hat{\bm\eta}_{\EB})$ and $\hat{\bm\theta}^{\Bayes}(\pi_{\star})$, respectively. For $\XBias(\cdot)$, \eqref{eq: matching XBias} follows directly from the condition \eqref{eq: sufficient condition on general EB2 estimator}. For the remaining excess variance terms, we first apply the chain rule to derive
\begin{align}
&\nabla_{\theta}\left\{ \left.\left[\nabla_{\theta}\log\left[\pi(\bm\theta|\bm\eta) \right]\right]\right|_{\bm\eta=\bm\eta_{\tb\star}(\bm\theta)} \right\}\\
=&\left.\nabla_{\theta\theta}^2\log\left[\pi(\bm\theta|\bm\eta)\right]\right|_{\bm\eta=\bm\eta_{\tb\star}(\bm\theta)}\\
\label{eq: decomposition of derivative}
&+\left.\nabla_{\theta\eta}^2\log\left[\pi(\bm\theta|\bm\eta)\right]\right|_{\bm\eta=\bm\eta_{\tb\star}(\bm\theta)}\nabla_{\theta}\bm\eta_{\tb\star}(\bm\theta).
\end{align}
Then, take the derivative of the condition in \eqref{eq: optimality condition for eta_star_theta} w.r.t. $\bm\theta$, i.e., $\nabla_{\theta} \{\nabla_{\eta} \log[\pi(\bm\theta|\bm\eta)]|_{\bm\eta=\bm\eta_{\tb\star}(\bm\theta)} \}=\mathbf{0}$.
Apply the chain rule,
\begin{align}
&\left.\nabla_{\eta\theta}^2\log\left[\pi(\bm\theta|\bm\eta)\right]\right|_{\bm\eta=\bm\eta_{\tb\star}(\bm\theta)}\\
&+\left.\nabla_{\eta\eta}^2\log\left[\pi(\bm\theta|\bm\eta)\right]\right|_{\bm\eta=\bm\eta_{\tb\star}(\bm\theta)}\nabla_{\theta}\bm\eta_{\tb\star}(\bm\theta)=\mathbf{0}.
\end{align}
It follows that
\begin{align}\label{eq: reformulation of the gradient of eta_star}
\nabla_{\theta}\bm\eta_{\tb\star}(\bm\theta)=&-\left\{\left.\nabla_{\eta\eta}^2\log\left[\pi(\bm\theta|\bm\eta)\right]\right|_{\bm\eta=\bm\eta_{\tb\star}(\bm\theta)} \right\}^{-1}\\
&\times\left.\nabla_{\eta\theta}^2\log\left[\pi(\bm\theta|\bm\eta)\right]\right|_{\bm\eta=\bm\eta_{\tb\star}(\bm\theta)}.
\end{align}
Substituting this into \eqref{eq: decomposition of derivative}, we have
\begin{align}
&\nabla_{\theta}\left\{ \left.\left[\nabla_{\theta}\log\left[\pi(\bm\theta|\bm\eta) \right]\right]\right|_{\bm\eta=\bm\eta_{\tb\star}(\bm\theta)} \right\}\\
=&\nabla_{\theta\theta}^2\log\left[\pi(\bm\theta|\bm\eta)\right]
-\nabla_{\theta\eta}^2\log\left[\pi(\bm\theta|\bm\eta)\right]\\
\label{eq: final decomposition of derivative}
&\times\left\{\nabla_{\eta\eta}^2\log\left[\pi(\bm\theta|\bm\eta)\right] \right\}^{-1}
\nabla_{\eta\theta}^2\log\left[\pi(\bm\theta|\bm\eta)\right]\quad\
\end{align}
with $\bm\eta$ evaluated at $\bm\eta_{\tb\star}(\bm\theta)$. Combining \eqref{eq: final decomposition of derivative}, Theorem~\ref{thm: XMSE of EB2 estimator} and Lemma~\ref{lemma: XMSE expression for Bayes estimator}, we then obtain \eqref{eq: matching XVar} with $\XVar(\hat{\bm\theta}^{\Bayes}(\pi_{\star}))=
2(\sigma^2)^2\mathbf{\Sigma}^{-1}\nabla_{\theta\theta}^2\log\left[\pi(\bm\theta|\bm\eta_{\tb\star}(\bm\theta))\right]\mathbf{\Sigma}^{-1}$.
Therefore, $\hat{\bm\theta}^{\EB}(\hat{\bm\eta}_{\EB})$ and $\hat{\bm\theta}^{\Bayes}(\pi_{\star})$ share the same XMSE.


\subsection{Proof of Lemma~\ref{lemma: pde for Gaussian weighting families}}

We first show that the regularity of $\mathbf{P}(\bm\eta)$ guarantees the Gaussian weighting families to satisfy Items~$2)$--$3)$ in Assumption~\ref{asp: regularity conditions for EB hyperparameter estimator}. Its regularity implies that $\pi(\bm\theta|\bm\eta)$ in \eqref{eq: Gaussian EB weighting families} is positive and three times continuously differentiable jointly in $(\bm\theta,\bm\eta)$, which corresponds to Item~$2)$ in Assumption~\ref{asp: regularity conditions for EB hyperparameter estimator}. The compactness of $\mathcal{D}_{\eta}$ in Item~$1)$ and the regularity of $\mathbf{P}(\bm\eta)$ also yield the uniform boundedness of $\mathbf{P}(\bm\eta)^{-1}$ and the derivatives of $\mathbf{P}(\bm\eta)$ on $\mathcal{D}_{\eta}$. Thus, there exist constants $0<\underline{m}<\overline{m}<\infty$ such that
\begin{align}
\underline{m}\mathbf{I}_{n_{\theta}}\preceq \mathbf{P}(\bm\eta)\preceq \overline{m}\mathbf{I}_{n_{\theta}},\ \text{for}\ \bm\eta\in\mathcal{D}_{\eta}.
\end{align}
We have $\det(\mathbf{P}(\bm\eta))\geq \underline{m}^{n_{\theta}}$ and $\bm\theta^{\top}\mathbf{P}(\bm\eta)^{-1}\bm\theta\geq {1}/{\overline{m}}\|\bm\theta\|_{2}^2$. Thus, $\pi(\bm\theta|\bm\eta)=(2\pi)^{-\frac{n_{\theta}}{2}}\det(\mathbf{P}(\bm\eta))^{-\frac{1}{2}}\exp[-\frac{1}{2}\bm\theta^{\top}\mathbf{P}(\bm\eta)^{-1}\bm\theta]$ satisfies Item~$3)$. Analogously, the uniform boundedness of the first and second order derivatives of $\pi(\bm\theta|\bm\eta)$ w.r.t. $\bm\eta$ also holds.

We then recall that for a general EB weighting family, the nonlinear PDE in \eqref{eq: sufficient condition on general EB2 estimator} ensures the matching of the XMSE components in \eqref{eq: matching XBias}--\eqref{eq: matching XVar}. For the Gaussian weighting families in \eqref{eq: Gaussian EB weighting families}, \eqref{eq: sufficient condition on general EB2 estimator} specializes to \eqref{eq: EB PDE}. It completes the proof.

\subsection{Proof of Corollary~\ref{corollary: characterization of weighting function}}

When we adopt the Gaussian EB weighting families in \eqref{eq: Gaussian EB weighting families}, $\bm\eta_{\tb\star}(\bm\theta)$ in \eqref{eq: def of eta_b_star for theta} coincides with that in \eqref{eq: def of eta_b_star for Gaussian weighting}. The representation in \eqref{eq: analytic form} follows directly by substituting \eqref{eq: def of eta_b_star for Gaussian weighting} into \eqref{eq: characterization of weighting function}. 

Moreover, $\pi_{\star}(\bm\theta)$ in \eqref{eq: characterization of weighting function} can be shown to characterize the complete family of solutions to the nonlinear PDE in \eqref{eq: sufficient condition on general EB2 estimator}. To see this, let $\check{\pi}_{\star}(\bm\theta)>0$ satisfy \eqref{eq: sufficient condition on general EB2 estimator}. By using the chain rule and \eqref{eq: optimality condition for eta_star_theta}, we obtain 
\begin{align}
\nabla_{\theta}\left\{\log[\check{\pi}_{\star}(\bm\theta)]-\log[\pi(\bm\theta|\bm\eta_{\tb\star}(\bm\theta))]\right\}=\bm{0}.
\end{align}
It means that the expression in the brackets is constant, i.e., $\check{\pi}_{\star}(\bm\theta)=C\pi(\bm\theta|\bm\eta_{\tb\star}(\bm\theta))$ for some $C>0$, which is exactly the form in \eqref{eq: characterization of weighting function}. Therefore, \eqref{eq: characterization of weighting function} gives all the solutions to \eqref{eq: sufficient condition on general EB2 estimator}. For the Gaussian weighting families in \eqref{eq: Gaussian EB weighting families}, substituting \eqref{eq: def of eta_b_star for Gaussian weighting} into \eqref{eq: sufficient condition on general EB2 estimator} reduces it to the PDE in \eqref{eq: EB PDE}. Consequently, \eqref{eq: analytic form} gives the complete family of solutions to \eqref{eq: EB PDE}.

\subsection{Proof of Corollary~\ref{corollary: characterization with kernel structure}}\label{subsec: derivation of weighting for parametrized kernel}

Under \eqref{eq: kernel structure}, $W_{\tb}(\bm\eta;\bm\theta)$ in \eqref{eq: def of Wb for Gaussian weighting} takes the form
\begin{align*}
W_{\tb}(c,\bm\alpha;\bm\theta)
&= \tfrac{1}{c}\bm\theta^\top \mathbf{K}(\bm\alpha)^{-1}\bm\theta + n_{\theta}\log c + \log\det(\mathbf{K}(\bm\alpha)).
\end{align*}
For fixed $\bm\alpha$, minimizing $W_{\tb}(c,\bm\alpha;\bm\theta)$ w.r.t. $c$ leads to the first-order optimality condition $-({1}/{c^2})\bm\theta^\top \mathbf{K}(\bm\alpha)^{-1}\bm\theta + ({n_{\theta}}/{c}) = 0$ and hence \eqref{eq: optimal scale hyperparameter}. Substituting this into $W_{\tb}(c,\bm\alpha;\bm\theta)$ gives 
\begin{align}
W_{\tb}(c_{\tb\star},\bm\alpha;\bm\theta)=n_{\theta}-n_{\theta}\log(n_{\theta})+\widetilde{W}_{\tb}(\bm\alpha;\bm\theta),
\end{align}
where $\widetilde{W}_{\tb}(\bm\alpha;\bm\theta)$ is defined in \eqref{eq: def of widetilde_Wb}. Thus, we obtain \eqref{eq: more explicit characterization} with
\begin{align}
\bm\alpha_{\tb\star}\coloneqq \argmin_{\bm\alpha\in\mathcal{D}_{\alpha}}W_{\tb}(c_{\tb\star},\bm\alpha;\bm\theta)=\argmin_{\bm\alpha\in\mathcal{D}_{\alpha}}\widetilde{W}_{\tb}(\bm\alpha;\bm\theta).
\end{align}


\subsection{Proof of Corollary~\ref{corollary: characterization with diagnal kernel}}\label{subsec: proof of diagonal parametrization}

For the Gaussian EB weighting family in \eqref{eq: Gaussian EB weighting families} with \eqref{eq: diagonal kernel}, the cost function in \eqref{eq: def of Wb for Gaussian weighting} specializes to
\begin{align}\label{eq: def of Wb for diagonal P}
W_{\tb}(\bm\eta;\bm\theta)={\textstyle\sum}_{k=1}^{n_{\eta}}\left\{{[\bm\theta]_{k}^2}/{[\bm\eta]_{k}}+\log([\bm\eta]_{k})\right\},
\end{align}
and the feasible set is $\mathcal{D}_{\eta}=[\underline{\eta},\infty)^{n_{\eta}}$, which is closed but not compact. Given any fixed $\bm\theta$, $W_{\tb}(\bm\eta;\bm\theta)$ in \eqref{eq: def of Wb for diagonal P} is continuous on $\mathcal{D}_{\eta}$. Moreover, as $\|\bm\eta\|_{2}\to\infty$, we have $W_{\tb}(\bm\eta;\bm\theta)\to \infty$, and $W_{\tb}(\bm\eta;\bm\theta)$ is therefore coercive on $\mathcal{D}_{\eta}$. Its coercivity leads to the level-boundedness of $W_{\tb}(\bm\eta;\bm\theta)$ and the existence of its minimizer on $\mathcal{D}_{\eta}$, see \cite[Definition~1.8 and Theorem~1.9]{RoW:98}. 

We then calculate the first-order derivative of $W_{\tb}(\bm\eta;\bm\theta)$ in \eqref{eq: def of Wb for diagonal P} w.r.t. $[\bm\eta]_{k}$, i.e., $[\nabla_{\eta}W_{\tb}(\bm\eta;\bm\theta)]_{k}=\left\{[\bm\eta]_{k}-[\bm\theta]_{k}^2\right\}/{[\bm\eta]_{k}^2}$,
where $[\bm\eta]_{k}\geq \underline{\eta}$. It follows that the minimizer of $W_{\tb}(\bm\eta;\bm\theta)$ is unique, i.e., $[\bm\eta_{\tb\star}(\bm\theta)]_{k}=\max\{[\bm\theta]_{k}^2,\underline{\eta}\}$, where $k=1,\cdots,n_{\eta}$. Substituting this into $W_{\tb}(\bm\eta_{\tb\star}(\bm\theta);\bm\theta)$ yields \eqref{eq: weighting function for diagonal kernel}.

Moreover, since $\mathcal{D}_{\eta}=[\underline{\eta},\infty)^{n_{\eta}}$ is not compact, we need to verify some additional conditions to ensure that the XMSEs of EB and Bayes estimators are well-defined. 

\paragraph{Verification of Item~$1)$ in Assumption~\ref{asp: regularity conditions for EB hyperparameter estimator} (except for the compactness of $\mathcal{D}_{\eta}$)} By $\underline{\eta}<\min_{k}[\bm\theta_{0}]_{k}^2$, we derive $[\bm\eta_{\tb\star}(\bm\theta_{0})]_{k}=[\bm\theta_{0}]_{k}^2$ and hence $\bm\eta_{\tb\star}(\bm\theta_{0})$ is in the interior of $\mathcal{D}_{\eta}$. In addition, 
\begin{align*}
&\nabla_{\eta\eta}^2\log(\pi(\bm\theta_{0}|\bm\eta))|_{\bm\eta=\bm\eta_{\tb\star}(\bm\theta_{0})}\\
=&-(1/2)\diag\{[\bm\theta_{0}]_{1}^{-4},\cdots,[\bm\theta_{0}]_{n_{\theta}}^{-4}\}\prec 0.
\end{align*}

\paragraph{Verification of Items~$2)$--$3)$ in Assumption~\ref{asp: regularity conditions for EB hyperparameter estimator}} It follows directly from the Gaussian EB weighting family in \eqref{eq: Gaussian EB weighting families} with \eqref{eq: diagonal kernel} that $\pi(\bm\theta|\bm\eta)=(2\pi)^{-n_{\eta}/2}\prod_{k=1}^{n_{\eta}}[\bm\eta]_{k}^{-1/2}\exp\{-[\bm\theta]_{k}^2/(2[\bm\eta]_{k})\}$. It is positive and infinitely differentiable in $(\bm\theta,\bm\eta)$. In addition, the uniform boundedness of $\pi(\bm\theta|\bm\eta)$ and its derivatives in Item~$3)$ follows from its specific form and $\mathcal{D}_{\eta}=[\underline{\eta},\infty)^{n_{\eta}}$.

\paragraph{Verification of Assumption~\ref{asp: differentiablity and growth rate conditions}} Notice that $\underline{\eta}<\min_{k}[\bm\theta_{0}]_{k}^2$ also implies that there exists an open neighborhood $\mathcal{U}(\bm\theta_{0})$ of $\bm\theta_{0}$ such that $[\bm\theta]_{k}^2>\underline{\eta}$, and hence $[\bm\eta_{\tb\star}(\bm\theta)]_{k}=[\bm\theta]_{k}^2$. The resulting weighting function is $\pi_{\star}(\bm\theta)=C\prod_{k}|[\bm\theta]_{k}|^{-1}$, which is positive and infinitely differentiable  on $\mathcal{U}(\bm\theta_{0})$. In addition, \eqref{eq: weighting function for diagonal kernel} satisfies $\pi_{\star}(\bm\theta)\leq C\exp(a\|\bm\theta\|_{2}^2)$ for some $C,a>0$.

\paragraph{Verification of the coercivity of the EB hyper-parameter cost function} The compactness of $\mathcal{D}_{\eta}$ in Assumption~\ref{asp: regularity conditions for EB hyperparameter estimator} ensures not only the existence of the minimizer of $W_{\tb}(\bm\eta;\bm\theta)$ but also that of the minimizer of $\mathscr{F}_{\EB}(\bm\eta)$. Hence, for the noncompact $\mathcal{D}_{\eta}$, we also need the coercivity of the EB hyper-parameter cost function. Note from \eqref{eq: expression for F_EB} that $\mathscr{F}_{\EB}(\bm\eta)$ in \eqref{eq: expression for F_EB} can be rewritten as $\mathscr{F}_{\tb}(\bm\eta)=(\hat{\bm\theta}^{\ML})^{\top}\mathbf{S}(\bm\eta)^{-1}\hat{\bm\theta}^{\ML}+\log\det(\mathbf{S}(\bm\eta))$,
up to constant terms independent of $\bm\eta$, where $\mathbf{S}(\bm\eta)=\mathbf{P}(\bm\eta)+\sigma^2(\mathbf{\Phi}^{\top}\mathbf{\Phi})^{-1}\succ 0$. It follows that $\mathscr{F}_{\tb}(\bm\eta)\geq \log\det(\diag\{\bm\eta\})$. Therefore, as $\|\bm\eta\|_{2}\to \infty$, $\mathscr{F}_{\tb}(\bm\eta)\to\infty$, which yields the coercivity of $\mathscr{F}_{\tb}(\bm\eta)$ and the existence of the minimizer $\hat{\bm\eta}_{\EB}=\argmin_{\bm\eta\in\mathcal{D}_{\eta}}\mathscr{F}_{\EB}(\bm\eta)=\argmin_{\bm\eta\in\mathcal{D}_{\eta}}\mathscr{F}_{\tb}(\bm\eta)$.

\subsection{Proof of Theorem~\ref{thm: sensitivity analysis}}\label{subsec: proof of sensitivity analysis}

We first establish an integrable function that will uniformly dominate the integrands, up to some constants, used in this proof. Let $a>0$ be the constant in Item~$3)$ of Assumption~\ref{asp: regularity conditions for EB hyperparameter estimator}. Define
\begin{align}\label{eq: def of g_N function}
g_{N}(\bm\theta)\coloneqq (1+\|\bm\theta\|_{2})p(\bm{Y}|\bm\theta)\exp(a\|\bm\theta\|_{2}^2).
\end{align}
By Assumption~\ref{asp: limit of regression matrix}, for sufficiently large sample sizes, we have $\mathbf{\Phi}^{\top}\mathbf{\Phi}/(2\sigma^2)- a\mathbf{I}_{n_{\theta}}\succ 0$. Note that $p(\bm{Y}|\bm\theta)\exp(a\|\bm\theta\|_{2}^2)\propto\exp\left\{-\bm\theta^{\top}[\mathbf{\Phi}^{\top}\mathbf{\Phi}/(2\sigma^2)- a\mathbf{I}_{n_{\theta}}]\bm\theta+\bm{Y}^{\top}\mathbf{\Phi}\bm\theta/\sigma^2\right\}$. Then, there exists $C_{N},c_{N}>0$ independent of $\bm\theta$ such that
\begin{align}
p(\bm{Y}|\bm\theta)\exp(a\|\bm\theta\|_{2}^2)\leq C_{N}\exp(-c_{N}\|\bm\theta\|_{2}^2).
\end{align}
The Gaussian-type exponential tail can dominate any polynomial growth and therefore, 
\begin{align}\label{eq: integrable condition}
{\textstyle\int} g_{N}(\bm\theta) d\bm\theta <\infty.
\end{align}

Next, we analyze the local sensitivity of $\hat{\bm\theta}^{\EB}(\hat{\bm\eta}_{\EB})$ to hyper-parameter perturbation $\delta\bm\eta$. For $\bm\eta\in\mathcal{D}_{\eta}$, we define $\bm{f}_{num}(\bm\eta)\coloneqq \int \bm\theta p(\bm{Y}|\bm\theta)\pi(\bm\theta|\bm\eta)d\bm\theta$ and $f_{den}(\bm\eta)\coloneqq \int p(\bm{Y}|\bm\theta)\pi(\bm\theta|\bm\eta)d\bm\theta$. By \eqref{eq: integrable condition} and Item~$3)$ of Assumption~\ref{asp: regularity conditions for EB hyperparameter estimator}, there exists a constant $C>0$, independent of $\bm\theta$, such that $\|\bm{f}_{num}(\bm\eta)\|_{2}\leq C\int g_{N}(\bm\theta) d\bm\theta<\infty$ and $0<f_{den}(\bm\eta)\leq C \int g_{N}(\bm\theta) d\bm\theta<\infty$. It follows that the EB estimate $\hat{\bm\theta}^{\EB}(\bm\eta)=\bm{f}_{num}(\bm\eta)/f_{den}(\bm\eta)$ is well-defined. By Item~$3)$ of Assumption~\ref{asp: regularity conditions for EB hyperparameter estimator} and Items~$1)$--$2)$ of Assumption~\ref{asp: admissible perturbations}, the dominated convergence theorem permits differentiation under the integral sign in $\bm{f}_{num}(\bm\eta)$ and $f_{den}(\bm\eta)$. We then have
\begin{align}
\nabla_{\eta}\bm{f}_{num}(\bm\eta)=&{\textstyle\int} \bm\theta p(\bm{Y}|\bm\theta)\nabla_{\eta}\pi(\bm\theta|\bm\eta)^{\top}d\bm\theta,\\
\nabla_{\eta}f_{den}(\bm\eta)=&{\textstyle\int} p(\bm{Y}|\bm\theta)\nabla_{\eta}\pi(\bm\theta|\bm\eta)d\bm\theta.
\end{align}
By \eqref{eq: integrable condition} and Item~$3)$ of Assumption~\ref{asp: regularity conditions for EB hyperparameter estimator}, the dominated convergence theorem also gives that the derivatives $\nabla_{\eta}\bm{f}_{num}(\bm\eta)$ and $\nabla_{\eta}f_{den}(\bm\eta)$ are continuous in $\bm\eta$. Therefore, 
\begin{align*}
\nabla_{\eta}\hat{\bm\theta}^{\EB}(\bm\eta)=\frac{f_{den}(\bm\eta)\nabla_{\eta}\bm{f}_{num}(\bm\eta)-\bm{f}_{num}(\bm\eta)\nabla_{\eta}f_{den}(\bm\eta)^{\top}}{[f_{den}(\bm\eta)]^2}
\end{align*}
is also continuous in $\bm\eta$. By Item~$1)$ of Assumption~\ref{asp: admissible perturbations} and the fundamental theorem of calculus, we obtain
\begin{align}
\hat{\bm\theta}^{\EB}(\tilde{\bm\eta}_{\EB})-\hat{\bm\theta}^{\EB}(\hat{\bm\eta}_{\EB})=&{\textstyle\int}_{0}^{1}\nabla_{\eta}\hat{\bm\theta}^{\EB}(\bm\eta)|_{\bm\eta=\hat{\bm\eta}_{\EB}+t\delta\bm\eta}\delta\bm\eta dt.
\end{align}
It follows that
\begin{align}
&\|\hat{\bm\theta}^{\EB}(\tilde{\bm\eta}_{\EB})-\hat{\bm\theta}^{\EB}(\hat{\bm\eta}_{\EB})\|_{2}\\
\leq & {\textstyle\sup_{t\in [0,1]}} \|\nabla_{\eta}\hat{\bm\theta}^{\EB}(\bm\eta)|_{\bm\eta=\hat{\bm\eta}_{\EB}+t\delta\bm\eta}\|_{F}\|\delta\bm\eta\|_{2}
\end{align}
Since $\nabla_{\eta}\hat{\bm\theta}^{\EB}(\bm\eta)$ is continuous, it is locally bounded around $\bm\eta=\hat{\bm\eta}_{\EB}$. Therefore, there exists $C_{N}>0$, independent of $\delta\bm\eta$, such that for sufficiently small $\|\delta\bm\eta\|_{2}$, we have $\|\hat{\bm\theta}^{\EB}(\tilde{\bm\eta}_{\EB})-\hat{\bm\theta}^{\EB}(\hat{\bm\eta}_{\EB})\|_{2}\leq C_{N}\|\delta\bm\eta\|_{2}$. This leads to \eqref{eq: sensitivity of EB2 estimator}.

Finally, we analyze local sensitivity of $\hat{\bm\theta}^{\Bayes}(\pi_{\star})$ to hyper-parameter perturbation $\delta\bm\eta(\bm\theta)$. Using Items~$1)$--$2)$ of Assumption~\ref{asp: admissible perturbations}, we apply the second-order Taylor expansion to $\tilde{\pi}_{\star}(\bm\theta)$ around $\tilde{\bm\eta}_{\tb\star}(\bm\theta)=\bm\eta_{\tb\star}(\bm\theta)$ with integral remainder,
\begin{align*}
&\tilde{\pi}_{\star}(\bm\theta)=\pi_{\star}(\bm\theta)+C[\nabla_{\eta}\pi(\bm\theta|\bm\eta)|_{\bm\eta=\bm\eta_{\tb\star}(\bm\theta)}]^{\top}\delta\bm\eta(\bm\theta)\\
&+C\textstyle{\int_{0}^{1}}(1-t)\delta\bm\eta(\bm\theta)^{\top}[\nabla_{\eta\eta}^2\pi(\bm\theta|\bm\eta)|_{\bm\eta=\bm\eta_{\tb\star}(\bm\theta)+t\delta\bm\eta(\bm\theta)}]\delta\bm\eta(\bm\theta)dt.
\end{align*}
By using \eqref{eq: directional stationarity}, $\int_{0}^{1}(1-t)dt=1/2$ and the norm inequalities $|\bm{a}^{\top}\mathbf{A}\bm{a}|\leq \|\mathbf{A}\|_{F}\|\bm{a}\|_{2}^2$ for $\bm{a}$ and $\mathbf{A}$ with suitable dimensions, we then derive
\begin{align}
|\tilde{\pi}_{\star}(\bm\theta)-\pi_{\star}(\bm\theta)|\leq \tfrac{C}{2}\|\delta\bm\eta(\bm\theta)\|_{2}^2{\sup}_{\bm\eta\in\mathcal{D}_{\eta}}\|\nabla_{\eta\eta}^2\pi(\bm\theta|\bm\eta)\|_{F}.
\end{align}
Using Item~$3)$ of Assumption~\ref{asp: regularity conditions for EB hyperparameter estimator} and the definition of $\|\delta\bm\eta\|_{\infty}$ in \eqref{eq: uniform norm}, there exists $\tilde{C}>0$, independent of $\bm\theta$, such that
\begin{align}\label{eq: bound of diff of pi_star}
|\tilde{\pi}_{\star}(\bm\theta)-\pi_{\star}(\bm\theta)|\leq \tilde{C}\|\delta\bm\eta\|_{\infty}^2\exp(a\|\bm\theta\|_{2}^2)
\end{align}
Define 
\begin{align*}
\bm{f}_{num\star}\coloneqq \textstyle{\int} \bm\theta p(\bm{Y}|\bm\theta)\pi_{\star}(\bm\theta)d\bm\theta,\ f_{den\star}\coloneqq \textstyle{\int} p(\bm{Y}|\bm\theta)\pi_{\star}(\bm\theta)d\bm\theta,\\
\tilde{\bm{f}}_{num\star}\coloneqq \textstyle{\int} \bm\theta p(\bm{Y}|\bm\theta)\tilde{\pi}_{\star}(\bm\theta)d\bm\theta,\ \tilde{f}_{den\star}\coloneqq \textstyle{\int} p(\bm{Y}|\bm\theta)\tilde{\pi}_{\star}(\bm\theta)d\bm\theta.
\end{align*}
Analogously, we utilize \eqref{eq: integrable condition} and Item~$3)$ in Assumption~\ref{asp: regularity conditions for EB hyperparameter estimator} to show that the four quantities above are all finite, and therefore $\hat{\bm\theta}^{\Bayes}(\pi_{\star})=\bm{f}_{num\star}/f_{den\star}$ and $\hat{\bm\theta}^{\Bayes}(\tilde{\pi}_{\star})=\tilde{\bm{f}}_{num\star}/\tilde{f}_{den\star}$ are well-defined. We can also derive from \eqref{eq: integrable condition}--\eqref{eq: bound of diff of pi_star} that, there exists $\tilde{C}_{\star}>0$, independent of $\bm\theta$, such that
\begin{align}
\|\tilde{\bm{f}}_{num\star}-\bm{f}_{num\star}\|_{2}\leq& \tilde{C}\|\delta\bm\eta\|_{\infty}^2\left[\textstyle{\int}\|\bm\theta\|_{2}p(\bm{Y}|\bm\theta)\exp(a\|\bm\theta\|_{2}^2)d\bm\theta\right]\nonumber\\
\leq& \tilde{C}_{\star} \|\delta\bm\eta\|_{\infty}^2,\\
|\tilde{{f}}_{den\star}-{f}_{den\star}|\leq& \tilde{C}\|\delta\bm\eta\|_{\infty}^2\left[\textstyle{\int}p(\bm{Y}|\bm\theta)\exp(a\|\bm\theta\|_{2}^2)d\bm\theta\right]\nonumber\\
\label{eq: bound of f_den_star difference}
\leq& \tilde{C}_{\star} \|\delta\bm\eta\|_{\infty}^2.
\end{align}
From \eqref{eq: bound of f_den_star difference}, for sufficiently small $\|\delta\bm\eta\|_{\infty}$, $|\tilde{f}_{den\star}-f_{den\star}|\leq f_{den\star}/2$. Hence, $\tilde{f}_{den\star}\geq f_{den\star}/2$. We then decompose the difference between the two Bayes estimates as
\begin{align}
&\hat{\bm\theta}^{\Bayes}(\tilde{\pi}_{\star})-\hat{\bm\theta}^{\Bayes}({\pi}_{\star})\\
=&\tfrac{\tilde{\bm{f}}_{num\star}}{\tilde{f}_{den\star}}-\tfrac{\bm{f}_{num\star}}{f_{den\star}}
=\tfrac{\tilde{\bm{f}}_{num\star}-\bm{f}_{num\star}}{\tilde{f}_{den\star}}
-\tfrac{\bm{f}_{num\star}[\tilde{f}_{den\star}-f_{den\star}]}{\tilde{f}_{den\star}f_{den\star}}.
\end{align}
It follows that
\begin{align*}
\scalebox{0.95}{$\|\hat{\bm\theta}^{\Bayes}(\tilde{\pi}_{\star})-\hat{\bm\theta}^{\Bayes}({\pi}_{\star})\|_{2}
\leq \left(\tfrac{2\tilde{C}_{\star}}{f_{den\star}}+\tfrac{2\|\bm{f}_{num\star}\|_{2}\tilde{C}_{\star}}{f_{den\star}^2} \right)\|\delta\bm\eta\|_{\infty}^2.$}
\end{align*}
As a result, the finiteness of $f_{num\star}$ and $\bm{f}_{den\star}$ yields \eqref{eq: sensitivity of Bayes estimator}.



\subsection{Complexity analysis of $\hat{\bm\theta}^{\Bayes}(\pi_{\star})$ in Section~\ref{subsec: Gaussian weighting with TC kernel}} \label{subsec: computational analysis of Bayes under Gaussian weighting families}

We first discuss the computation of the basic data-dependent quantities. Given $\mathbf{\Phi}$, $\bm{Y}$ and $\sigma^2$, the computational complexities of $\mathbf{\Phi}^{\top}\mathbf{\Phi}$ and $\mathbf{\Phi}^{\top}\bm{Y}$ are $\mathcal{O}(Nn_{\theta}^2)$ and $\mathcal{O}(Nn_{\theta})$, respectively. Then, to compute $\hat{\bm\theta}^{\ML}$ in \eqref{eq: def of ML estimator}, we first take a Cholesky factorization $\mathbf{\Phi}^{\top}\mathbf{\Phi}=\mathbf{L}\mathbf{L}^{\top}$, where $\mathbf{L}\in\R^{n_{\theta}\times n_{\theta}}$ denotes a lower triangular matrix, which costs $\mathcal{O}(n_{\theta}^3)$. We then use two triangular solves to compute $\hat{\bm\theta}^{\ML}=\mathbf{L}^{-\top}\mathbf{L}^{-1}\mathbf{\Phi}^{\top}\bm{Y}$, which costs $\mathcal{O}(n_{\theta}^2)$. To compute $\pi(\bm\theta|\bm\eta_{\tb\star}(\hat{\bm\theta}^{\ML}))$, we evaluate $\widetilde{W}_{\tb}(\alpha;\bm\theta)$ in \eqref{eq: rewritten formulation of Wb_tilde_TC} $M_{\tb}$ times, and each time costs $\mathcal{O}(n_{\theta})$. Hence, the computation of $\pi(\bm\theta|\bm\eta_{\tb\star}(\hat{\bm\theta}^{\ML}))$ costs $\mathcal{O}(M_{\tb}n_{\theta})$. We then apply the Cholesky factorization $\mathbf{\Phi}^{\top}\mathbf{\Phi}+\sigma^2\mathbf{P}(\bm\eta_{\tb\star}(\hat{\bm\theta}^{\ML}))^{-1}=\mathbf{L}_{q}\mathbf{L}_{q}^{\top}$, together with $\bm{\mu}_{G}$ and $\mathbf{\Sigma}_{G}$ in \eqref{eq: def of mu_q and Sigma_q}, which costs $\mathcal{O}(n_{\theta}^3)$. Thus, the total computational complexity is $\mathcal{O}(Nn_{\theta}^2 + n_{\theta}^3 + M_{\tb}n_{\theta})$.

The computation of $\pi_{\star}(\bm\theta)$ essentially depends on the computation of the importance weight $w_{\star}(\bm\theta)$. We first generate an IS sample by $\bm\theta^{(k)}=\bm{\mu}_{q}+\sigma\mathbf{L}_{q}^{-\top}\bm{z}^{(k)}$ with $\bm{z}^{(k)}\sim\mathcal{N}(\bm{0},\mathbf{I}_{n_{\theta}})$, and its computational complexity is $\mathcal{O}(n_{\theta}^2)$. For each $\bm\theta^{(k)}$, we compute $\bm\eta_{\tb\star}(\bm\theta^{(k)})$ by evaluating $\widetilde{W}_{\tb}$ in \eqref{eq: rewritten formulation of Wb_tilde_TC} $M_{\tb}$ times, which costs $\mathcal{O}(M_{\tb}n_{\theta})$. With $\bm\eta_{\tb\star}(\bm\theta^{(k)})$, we then compute the weights $w_{\star}(\bm\theta^{(k)})$ in $\mathcal{O}(n_{\theta})$ operations. Thus, evaluating $\pi_{\star}$ for all $M_{\text{IS}}^{\text{B}}$ samples costs $\mathcal{O}(M_{\text{IS}}^{\text{B}}n_{\theta}^2+M_{\text{IS}}^{\text{B}}M_{\tb}n_{\theta})$.

Given the IS samples and importance weights, we directly use \eqref{eq: Bayes estimate approximation} to approximate $\hat{\bm\theta}^{\Bayes}(\pi_{\star})$, which costs $\mathcal{O}(M_{\text{IS}}^{\text{B}}n_{\theta})$.



\subsection{Complexity analysis of $\hat{\bm\theta}^{\EB}(\hat{\eta}_{\EB})$ in Section~\ref{subsec: student t weighting family}}\label{subsec: computational complexity of EB using student t}

For the basic data-dependent quantities, their computational complexity analysis is the same as that in Appendix~\ref{subsec: computational analysis of Bayes under Gaussian weighting families}.



Before computing $\mathscr{F}_{\EB}(\eta)$ and $\hat{\bm\theta}^{\EB}(\eta)$, we first construct $\bm{\mu}_{S}(\eta)$ in \eqref{eq: def of mu_S} and $\mathbf{\Sigma}_{S}(\eta)$ in \eqref{eq: def of Sigma_S} for a fixed $\eta$. The construction of $\bm{\mu}_{S}(\eta)$ relies on the evaluation of $\nabla_{\theta}\mathcal{J}$ in \eqref{eq: explicit expression for mu_S}. Given $\mathbf{\Phi}^{\top}\mathbf{\Phi}$ and $\mathbf{\Phi}^{\top}\bm{Y}$, $M_{L}$ evaluations of \eqref{eq: explicit expression for mu_S} cost $\mathcal{O}(M_{L}n_{\theta}^2)$. We then construct the Hessian matrix in \eqref{eq: explicit expression for Hessian} and compute its inverse, which costs $\mathcal{O}(n_{\theta}^3+n_{\theta}^2)$. Therefore, the computational complexity of constructing the proposal $q_{S}(\bm\theta;\eta)$ is $\mathcal{O}(M_{L}n_{\theta}^2+n_{\theta}^3)$. With $\bm{\mu}_{S}(\eta)$ and $\mathbf{\Sigma}_{S}(\eta)$ computed, the generation of $M_{\text{IS}}^{\EB}$ IS samples $\{\bm\theta^{(k)}\}$ costs $\mathcal{O}(M_{\text{IS}}^{\EB}n_{\theta}^2)$. We then compute the IS weights $\{w_{\EB}(\bm\theta^{(k)};\eta)\}$, whose complexity is $\mathcal{O}(M_{\text{IS}}^{\EB}n_{\theta}^2)$. Repeating this procedure $M_{\EB}$ times gives a total complexity of $\mathcal{O}(M_{\EB}[M_{L}n_{\theta}^2+n_{\theta}^3+M_{\text{IS}}^{\EB}n_{\theta}^2])$.

By reusing the IS samples and IS weights, $\hat{\bm\theta}^{\EB}(\hat{\eta}_{\EB})$ only requires a weighted average in \eqref{eq: approximation of EB estimate using student t}, which costs $\mathcal{O}(M_{\text{IS}}^{\EB}n_{\theta})$.

\subsection{Complexity analysis of $\hat{\bm\theta}^{\Bayes}(\pi_{\star})$ in Section~\ref{subsec: student t weighting family}} \label{subsec: computational analysis of Bayes under student_t weighting families}

For the basic data-dependent quantities, given $\mathbf{\Phi}$ and $\bm{Y}$, the computation of $\mathbf{\Phi}^{\top}\mathbf{\Phi}$, $\mathbf{\Phi}^{\top}\bm{Y}$ and $\hat{\bm\theta}^{\ML}$ still costs $\mathcal{O}(Nn_{\theta}^2+n_{\theta}^3)$. To construct the proposal using $\eta_{\tb\star}(\hat{\bm\theta}^{\ML})$, we evaluate the left-hand side of \eqref{eq: student_t optimality condition} $M_{\tb}$ times, which has complexity $\mathcal{O}(M_{\tb}n_{\theta})$. Given $\eta=\eta_{\tb\star}(\hat{\bm\theta}^{\ML})$, the computation of $\bm{\mu}_{S}(\eta)$ and $\mathbf{\Sigma}_{S}(\eta)$ in \eqref{eq: def of mu_S}--\eqref{eq: def of Sigma_S} is analogous to that in Appendix~\ref{subsec: computational complexity of EB using student t}, which has complexity $\mathcal{O}(M_{L}n_{\theta}^2+n_{\theta}^3)$. Thus, the total complexity of this step is $\mathcal{O}(M_{\tb}n_{\theta}+[N+M_{L}]n_{\theta}^2+n_{\theta}^3)$.

As for the generation of $\{w_{\star}(\bm\theta^{(k)})\}$ and the approximation of $\hat{\bm\theta}^{\Bayes}(\pi_{\star})$, the complexity analyses are analogous to those in Appendix~\ref{subsec: computational analysis of Bayes under Gaussian weighting families}, and are therefore omitted here.

\bibliographystyle{abbrv}
\bibliography{database} 


\vskip 0pt plus -1fil

\begin{IEEEbiography}[{\includegraphics[width=1in,height=1.25in,clip,keepaspectratio]{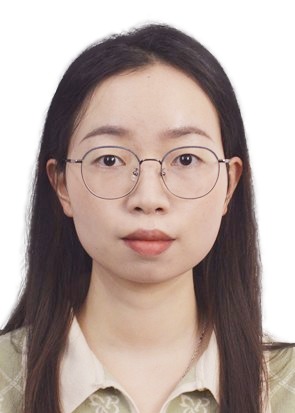}}] {Yue Ju} received the B.E. degree in automation from the Nanjing University of Science and Technology, Nanjing, China, in 2017 and the Ph.D. degree in computer and information engineering from the Chinese University of Hong Kong, Shenzhen, China, in 2022.
	
She is currently a postdoc at the KTH Royal Institute of Technology. She has been mainly working in the area of system identification.
\end{IEEEbiography}

\begin{IEEEbiography}[{\includegraphics[width=1in,height=1.25in,clip,keepaspectratio]{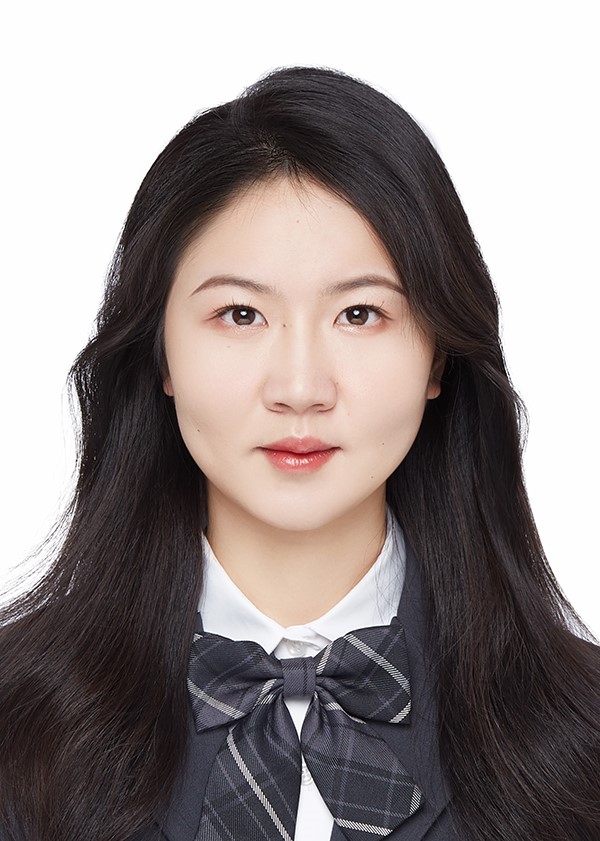}}] {Ying Wang} received the B.S. degree in automation from Northwestern Polytechnical University, Xi'an, China in 2020, and the M.Sc. degree in control science and engineering from Shanghai Jiao Tong University, Shanghai, China, in 2023.
	
She is currently pursuing a Ph.D. degree in electrical engineering at KTH Royal Institute of Technology, Stockholm, Sweden. Her research interests include dual control, system identification and input design.
\end{IEEEbiography}

\begin{IEEEbiography}
		[{\includegraphics[width=1in,height=1.25in,clip,keepaspectratio]{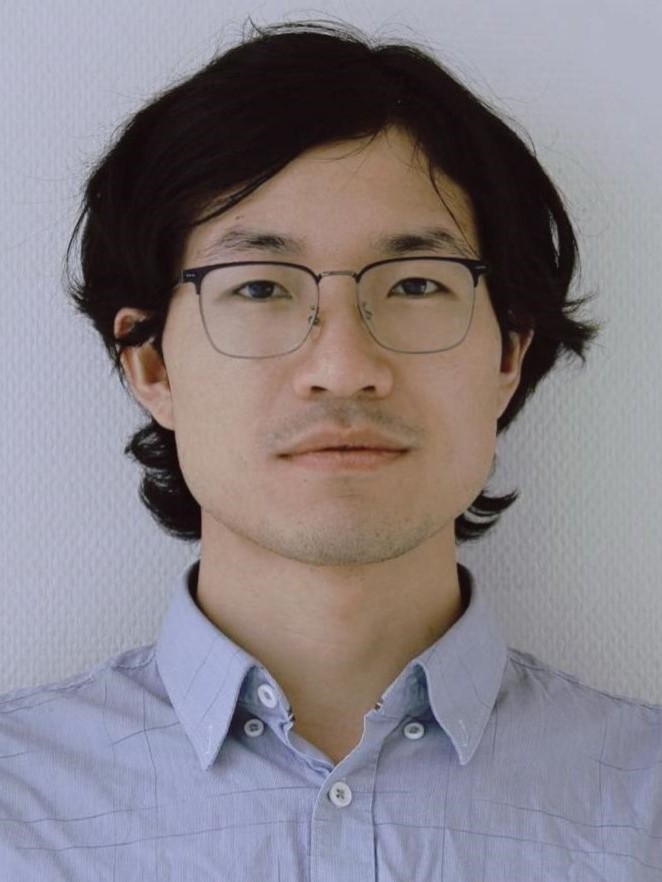}}]{Jiabao He}
		received the B.E. and M.E. degrees in mechanical engineering and control engineering from Tsinghua University, Beijing, China, in 2018 and 2021, respectively. He is currently working toward the Ph.D. degree with the Division of Decision and Control Systems, School of Electrical Engineering and Computer Science (EECS), KTH Royal Institute of Technology, Stockholm, Sweden. 
		
		His research interests include system identification and data driven control.	
		
		Mr. He received the Outstanding Graduate Award in Tsinghua University, and 2025 IEEE TC SIAC Student Paper Award.
\end{IEEEbiography}

\vskip 0pt plus -1fil

\begin{IEEEbiography}[{\includegraphics[width=1in,height=1.25in,clip,keepaspectratio]{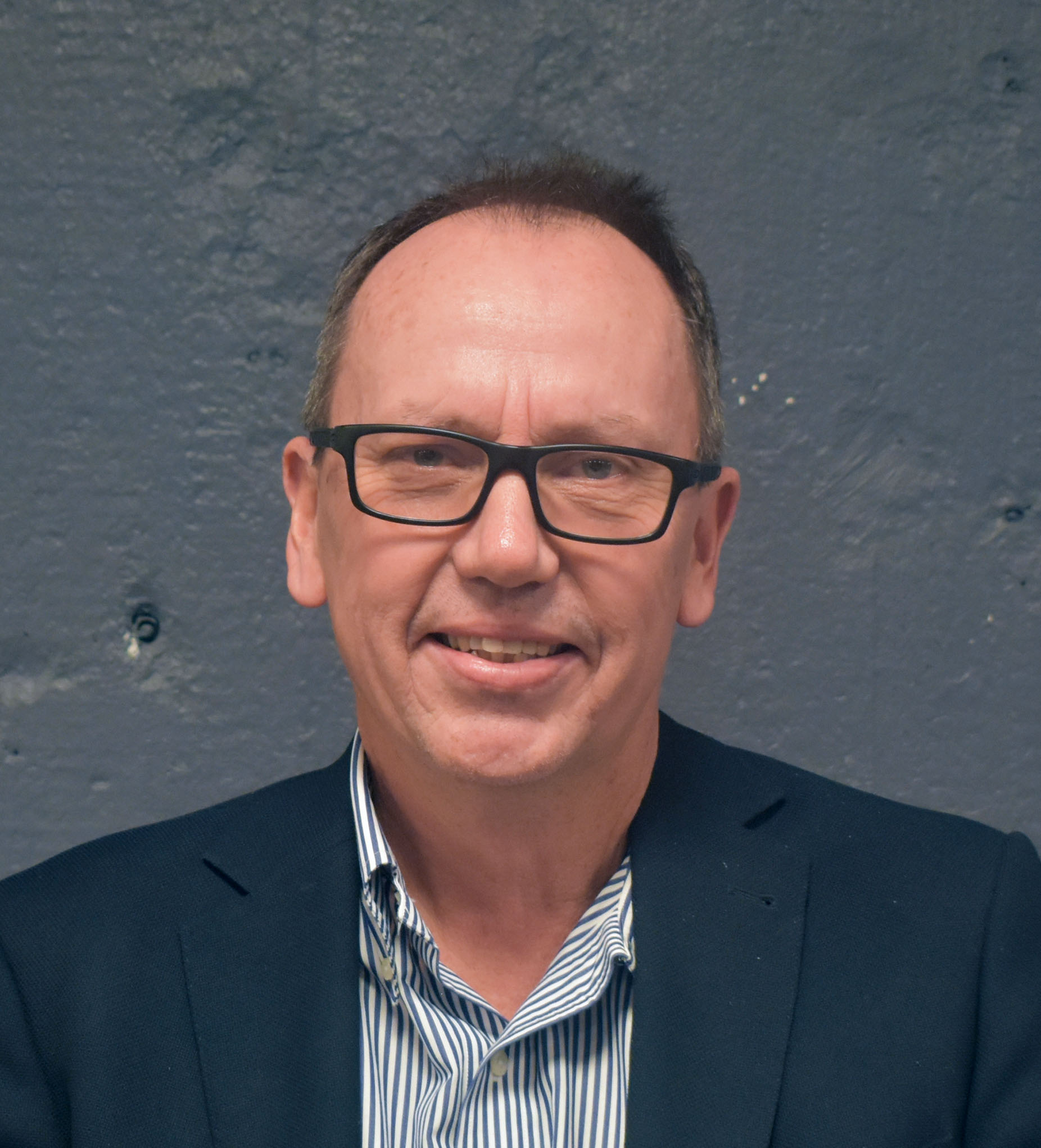}}]{Bo Wahlberg} has been the Professor of Chair in Automatic Control at KTH Royal Institute of Technology since 1991. He was elected an IEEE Fellow in 2007 for his contributions to system identification using orthonormal basis functions and a Fellow of IFAC in 2019 for his contributions to system identification and the development of orthonormal basis function models. Bo Wahlberg is since 2023 a Fellow of the Royal Swedish Academy of Engineering Sciences (IVA), Electrical Engineering Division.
 
Bo Wahlberg has received several awards, including the IEEE Transactions on Automation Science and Engineering Best New Application Paper Award in 2016. He is the author of over 250 scientific publications and has been the supervisor of more than 130 master’s students and 25 PhD students. His main research interest is in estimation and optimization in system identification, decision and control systems, and signal processing with applications in process industry and transportation.
\end{IEEEbiography}

\vskip 0pt plus -1fil

\begin{IEEEbiography}[{\includegraphics[width=1in,height=1.25in,clip,keepaspectratio]{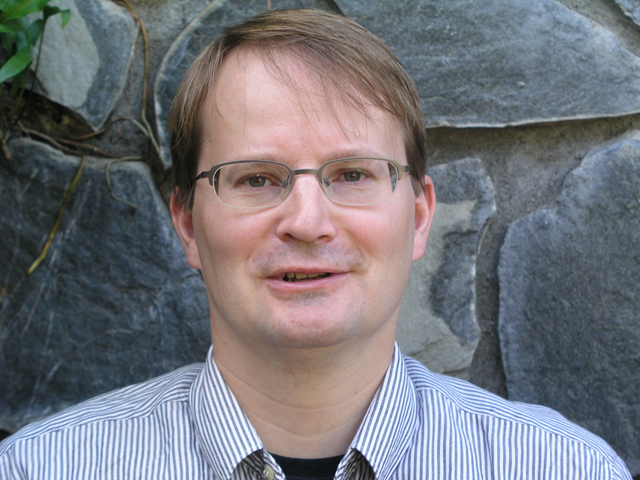}}] {H\r{a}kan Hjalmarsson} was born in 1962. He received the M.S. degree in Electrical Engineering in 1988, and the Licentiate degree and the Ph.D. degree in Automatic Control in 1990 and 1993, respectively, all from Link{\"o}ping University, Sweden. He has held visiting research positions at California Institute of Technology, Louvain University and at the University of Newcastle, Australia. He has served as an Associate Editor for Automatica (1996-2001), and IEEE Transactions on Automatic Control (2005-2007) and been Guest Editor for European Journal of Control and Control Engineering Practice. He is Professor at the Division of Decision and Control Systems, School of Electrical Engineering and Computer Science, KTH, Stockholm, Sweden and also affiliated with the Competence Centre for Advanced BioProduction by Continuous Processing, AdBIOPRO. He is an IEEE Fellow and past Chair of the IFAC Coordinating Committee CC1 Systems and Signals. In 2001, he received the KTH award for outstanding contribution to undergraduate education. He was General Chair for the IFAC Symposium on System Identification held in 2018. His research interests include system identification, learning of dynamical systems for control, process modeling control and also estimation in communication networks. 
\end{IEEEbiography}

\end{document}